\documentclass{article}
\usepackage[utf8]{inputenc}
\usepackage{authblk}
\usepackage[margin=1.25in]{geometry}
\usepackage{graphicx}
\usepackage{subcaption}
\usepackage{amsmath}
\usepackage{amssymb} 
\usepackage{booktabs}
\usepackage{doi}
\usepackage{setspace}
\usepackage[T1]{fontenc}

\usepackage[round,authoryear]{natbib}
\usepackage{aas_macros}
\title{Star Clusters in the Ultraviolet}

\author[1]{Annapurni Subramaniam}
\author[2,3]{Samyaday Choudhury}
\author[4,5,6]{Richard de Grijs}
\author[7,8]{\\Vikrant V. Jadhav}
\author[9,10]{Chengyuan Li}
\author[11,*]{Snehalata Sahu}
\author[12]{Kaushar Vaidya}
\author[9,10]{Li Wang}

\affil[1]{Indian Institute of Astrophysics, II Block, Koramangala, Bangalore-560034, India.}
\affil[2]{School of Arts and Sciences, Ahmedabad University, Ahmedabad-380009, India.}
\affil[3]{International Centre for Space and Cosmology, Ahmedabad University, Ahmedabad-380009, India.} 
\affil[4]{School of Mathematical and Physical Sciences, Macquarie University, Balaclava Road, Sydney, NSW 2109, Australia.}
\affil[5]{Astrophysics and Space Technologies Research Centre, Macquarie University, Balaclava Road, Sydney, NSW 2109, Australia.}
\affil[6]{International Space Science Institute--Beijing, 1 Nanertiao, Zhongguancun, Hai Dian District, Beijing 100190, China.}
\affil[7]{Astronomical Institute, Faculty of Mathematics and Physics, Charles University, V Hole\v{s}ovi\v{c}k\'ach 2, CZ-180 00 Praha 8, Czechia.}
\affil[8]{Helmholtz-Institut f\"ur Strahlen- und Kernphysik, Universit\"at Bonn, Nussallee 14--16, D-53115 Bonn, Germany.}
\affil[9]{School of Physics and Astronomy, Sun Yat-sen University, Daxue Road, Zhuhai, 519082, China.}
\affil[10]{CSST Science Center for the Guangdong-Hong Kong-Macau Greater Bay Area, Zhuhai, 519082, China.}
\affil[11]{Department of Physics, University of Warwick, Coventry, CV4 7AL, UK.}
\affil[12]{Department of Physics, Birla Institute of Technology and Science, Pilani-333031, India.}
\affil[*]{Address correspondence to: Snehalata.Sahu@warwick.ac.uk}

\date{2026}

\begin{document}

\maketitle

\begin{abstract}
Ultraviolet (UV) observations provide a uniquely powerful window into the hot and evolved stellar populations that shape the structure, evolution and integrated light of star clusters. Because UV wavelengths are highly sensitive to massive main-sequence stars, blue straggler stars (BSS), extreme-horizontal branch (HB) stars, post-AGB objects, interacting binaries and compact remnants, they probe key evolutionary processes that are inaccessible at optical and infrared wavelengths. This review synthesises five decades of UV studies of star clusters across the Milky Way, the Magellanic Clouds (MCs) and nearby galaxies, drawing on results from early space missions, wide-field surveys, high-resolution imaging and the recent capabilities of {\sl AstroSat}/UVIT and {\sl Swift}/UVOT. In Galactic open clusters, UV studies have revealed diverse compact companions---including white dwarfs, hot subdwarfs and stripped stars---and established the mass-transfer origins of BSS and related populations such as Blue Lurkers and yellow stragglers. In globular clusters, UV imaging has been instrumental in identifying multiple stellar populations through UV-sensitive molecular bands, probing helium enrichment and mapping HB morphologies. Wide-field UVIT surveys have extended earlier Hubble Space Telescope studies by providing homogeneous catalogues of HB and post-HB stars across entire clusters. In the MCs, UV observations have transformed our understanding of multiple populations, rotation-driven extended main-sequence turn-offs and the recently identified UV-dim phenomenon, while spectroscopic surveys have constrained massive-star evolution, stellar winds and binarity at low metallicity. UV mapping of the Magellanic Bridge has further demonstrated ongoing massive-star formation in low-density tidal environments. Beyond the Local Group, UV studies of extragalactic clusters have revealed star-formation histories, stellar feedback and population synthesis constraints across diverse galactic environments. Collectively, UV observations now form a cornerstone of star cluster astrophysics and will continue to do so with upcoming missions.
\end{abstract}

\section{Introduction}
Star clusters are fundamental laboratories for testing stellar evolution and probing the formation history of galaxies. As gravitationally bound groups of stars originating from a common progenitor molecular cloud, cluster members share age, distance and initial chemical composition, allowing stellar evolutionary models to be evaluated under controlled conditions \citep[e.g.,][]{1993A&AS...98..477M, 2013ApJ...775..134V}. Across the Milky Way and the nearby Universe, clusters span a wide range of physical properties---from the relatively young and low-mass open clusters (OCs) of the Galactic disc to the ancient, metal-poor globular clusters (GCs) in the Galactic halo \citep{1996AJ....112.1487H, 2010arXiv1012.3224H} and the rich cluster populations of the Magellanic Clouds \citep[MCs;][]{2011AJ....142...36G, 2016A&A...586A.148N}. Together, these systems trace the star-formation and chemical-enrichment histories of their host galaxies \citep{2002ARA&A..40..487F, 2005ARA&A..43..387G}.

Ultraviolet (UV) observations have profoundly advanced cluster astrophysics by isolating evolutionary phases that dominate at short wavelengths but are faint, blended or difficult to disentangle at optical and infrared (IR) wavelengths. The UV is particularly sensitive to hot and evolved stars---including massive main-sequence stars, classical Be stars, blue straggler stars (BSS), blue-horizontal-branch stars (BHB), extreme-horizontal-branch (extreme-HB) stars, post-asymptotic-giant-branch (post-AGB) stars, interacting binaries and compact remnants \citep{1990ApJ...364...35G, Brown2001, Schiavon2012AJ....143..121S}. Analysis of these populations has enabled precise measurements of cluster ages, helium enrichment, stellar mass-loss processes, binary evolution and dynamical interactions in dense environments \citep{Ferraro2012Natur.492..393F, 2018ApJ...864...33D}. UV diagnostics provide complementary leverage of multiple stellar populations (MPs) in GCs \citep{Piotto2015AJ....149...91P} and of low-level recent star formation in young and intermediate-age systems \citep{2011Ap&SS.335...51B}, especially in unresolved extragalactic clusters. 

\begin{figure}
    \centering
    \includegraphics[width=1\linewidth]{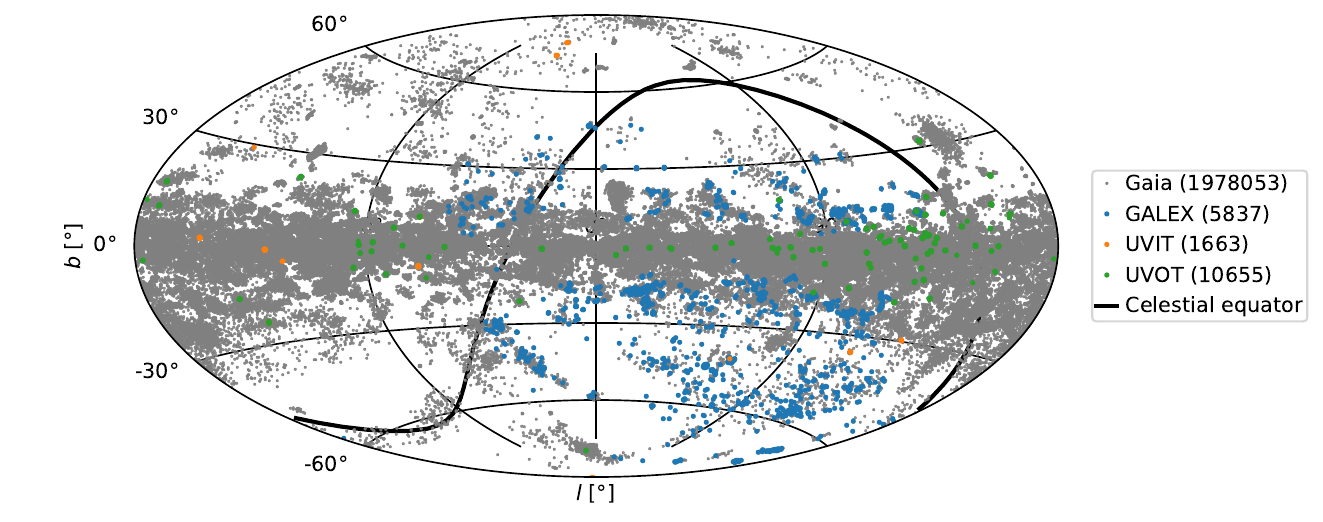}
    \caption{Distribution of {\sl Gaia}-detected OC and GC members in the Galactic coordinate plane \citep{2024A&A...686A..42H}. The members detected in {\sl Gaia} DR3 \citep{2023A&A...674A...1G}, {\sl GALEX} \citep{2017ApJS..230...24B}, UVIT \citep{2021MNRAS.503..236J, Sahu2022, Jadhav2023A&A...676A..47J, Jadhav2024A&A...688A.152J, 2024ApJS..275...34P} and UVOT \citep{2019AJ....158...35S} are coloured in grey, blue, orange, and green, respectively. The total number of cluster members detected in each survey is given in the legend.}
    \label{fig:uv_detection}
\end{figure}

More than 7000 clusters have been parametrised using the recent {\sl Gaia} Data Release 3 \citep[DR3;][]{2024A&A...686A..42H}. These clusters are distributed across the sky, with the majority located within the Galactic disc (Figure~\ref{fig:uv_detection}). However, the Milky Way disc remains poorly explored in the UV, primarily owing to instrumental limitations. The major UV observatories such as GAlaxy Evolution eXplorer ({\sl GALEX}) and {\sl AstroSat}/UltraViolet Imaging Telescope (UVIT) are photon-counting instruments for which saturation leads to detector degradation. As such, they generally avoid targeting regions containing very bright UV sources ($\lesssim$12 mag), including the majority of the Galactic disc. This is demonstrated in Figure~\ref{fig:uv_detection}, which shows that only $\approx$400 clusters (6 per cent) have UV data. The upcoming Ultraviolet Explorer ({\sl UVEX}) mission plans to perform the first all-sky survey in the UV bands to fill this gap \citep{Kulkarni2021UVEX}.
Nevertheless, even with this limited set of observations, UV observations of clusters have provided unique insights about stellar and star cluster evolution.

Recent studies using the UVIT have further demonstrated the power of UV colours to characterise BSS, hot HB stars, post-AGB objects and other UV-bright populations in Galactic and MC clusters \citep{Subramaniam2016ApJ...833L..27S, Sahu2019, Rani2021}.
The progress achieved in recent decades builds on a long heritage of space-based UV astronomy. Early missions including NASA’s Orbiting Astronomical Observatory-2 ({\sl OAO}-2), the Astronomical Netherlands Satellite ({\sl ANS}) and the International Ultraviolet Explorer ({\sl IUE}) laid the foundation by obtaining the first UV photometry and spectroscopy of star clusters \citep{1970ApJ...161..377C, 1989ARA&A..27..397K}.
The Ultraviolet Imaging Telescope ({\sl UIT}) and the Far Ultraviolet Spectroscopic Explorer ({\sl FUSE}) expanded this work, enabling wide-field imaging and high-resolution spectroscopy of hot stellar populations in the Milky Way and MCs \citep{1994A&A...289..715D, 1997PASP..109..584S, 2000ApJ...538L...1M}. 

The increased coverage of the sky and sensitivity of the {\sl GALEX} revolutionised studies of young and extragalactic clusters \citep{2005mmgf.conf..197M, 2014AdSpR..53..900B}, while the Hubble Space Telescope's ({\sl HST}) Space Telescope Imaging Spectrograph (STIS), the Advanced camera for Surveys (ACS)/Solar Blind Channel and the Wide-Field Camera-3 (WFC3)/UV and Visible light channel (UVIS) enabled direct investigations of exotic populations in dense cluster cores \citep{Brown2010ApJ...723.1072B, 2017MNRAS.469..267D}. More recently, {\sl AstroSat}/UVIT has provided  UV imaging with superior resolution that has been instrumental in studying BSS, extreme-HB stars and white-dwarf (WD) cooling sequences in OCs, GCs and the MCs \citep{Subramaniam2017, Jadhav2021JApA...42...89J, Sahu2022, 2023JApA...44...82M}. Time-domain capabilities from {\sl Swift}'s UV--Optical Telescope (UVOT) have further expanded the scope of UV investigations in nearby clusters \citep{2014AJ....148..131S}.

Early missions such as the {\sl IUE} established UV spectroscopy as a fundamental astrophysical tool, revealing the ubiquity of stellar winds in massive stars and providing critical insights into stellar evolution, mass loss, and the interaction between stars and the interstellar medium \citep{1987ApJS...64...83C, 1987ASSL..129.....K}. {\sl FUSE} further revolutionised the field through the detection of highly ionised gas in the warm--hot intergalactic medium, providing key evidence for the location of a significant fraction of the `missing baryons' in the local Universe \citep{2002ApJ...564..631S,2005ApJ...624..555D}.

Subsequent missions expanded UV astronomy into studies of galaxy evolution and stellar populations. {\sl GALEX} mapped star formation across the nearby and distant Universe, establishing the recent cosmic star-formation history and discovering extended UV discs that demonstrated ongoing star formation far beyond the optical extents of galaxies \citep{Martin2005ApJ...619L...1M,2007ApJS..173..538T}. UV instruments onboard the {\sl HST} enabled high-resolution investigations of active galactic nuclei, the intergalactic medium, and the cosmic web \citep{2012ApJ...744...60G}. More recently, UVIT onboard {\sl AstroSat} has combined wide-field imaging with arcsecond-scale spatial resolution, leading to major breakthroughs in understanding BSS and UV-bright stellar populations in clusters and galaxies \citep{Subramaniam2016ApJ...833L..27S,2018MNRAS.481..226S}. Together, these missions have established UV astronomy as an indispensable probe of stellar evolution, galaxy evolution, and the baryonic structure of the Universe.

Despite these advances, several key questions remain open. Which physical mechanisms drive the formation and evolution of extreme UV-bright stars in dense stellar environments \citep{2021ApJ...908..102P, Sahu2022}? What roles do binaries, stellar rotation, internal dynamics and helium enrichment play in shaping the UV properties of clusters across environments \citep{Milone2018MNRAS.481.5098M, Subramaniam2020JApA...41...45S}? How do UV evolutionary pathways vary with metallicity, particularly among Galactic GCs, MC clusters and extragalactic cluster systems \citep{2021ApJ...920..105C, Rani2023ApJ...945...11R}? And to what extent can integrated UV diagnostics be used to derive accurate mass and age estimates for unresolved extragalactic star clusters, particularly in the presence of stochastic sampling of the upper stellar initial mass function \citep[IMF;][]{2010A&A...521A..22F, 2015MNRAS.452.1447K}?

This review covers UV studies of OCs, GCs, MC clusters and extragalactic star clusters, starting from section 2 with emphasis on their UV-bright stellar constituents, the scientific insights derived to date and the outstanding problems that motivate future theoretical and observational work. The review progresses from nearby systems (e.g., OCs) to increasingly distant environments, culminating in clusters in external galaxies. Stellar populations that predominantly contribute to the UV flux and are studied are covered in the respective sections. By integrating results across diverse cluster environments, we identify the unifying physical processes revealed in the UV and highlight a number of outstanding questions that can be addressed by upcoming UV missions.

\section{Galactic Star Clusters}

\begin{figure}
    \centering
    \includegraphics[width=0.9\linewidth]{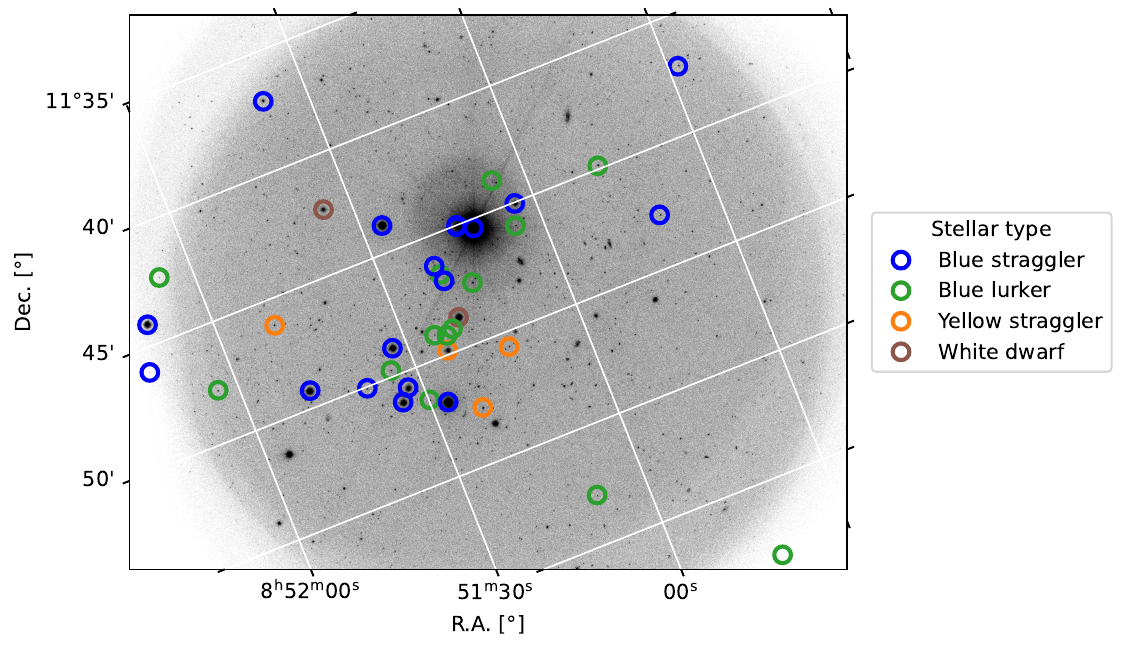}
    \caption{UVIT image of cluster M67 in the F148W filter. The positions of known BSS, yellow stragglers, white dwarfs and blue lurkers are highlighted.}
    \label{fig:M67_stragglers}
\end{figure}
\subsection{Open Clusters}

Figure~\ref{fig:UOCS_OC} shows the variety of compact companions detected in OCs using spectral energy distribution (SED) analysis \citep{Subramaniam2016ApJ...833L..27S, Jadhav2019ApJ...886...13J, Sindhu2019ApJ...882...43S, Subramaniam2020JApA...41...45S, Jadhav2021JApA...42...89J, Panthi2022MNRAS.516.5318P, Rao2022MNRAS.516.2444R, Vaidya2022MNRAS.511.2274V, Jadhav2023A&A...676A..47J, Rani2023ApJ...945...11R, Chand2024AJ....168..278C, Jadhav2024A&A...688A.152J, Pal2024ApJ...970L..39P, Panthi2024MNRAS.52710335P, Panthi2024MNRAS.527.8325P, Nedhath2025A&A...699L...1N}. It has only been possible to characterise optically subluminous companions thanks to the availability of UV photometry. The majority of compact objects are likely WDs, including extremely low-mass WDs. However, there is a significant population of hot subdwarfs (e.g., sdO/B stars) or stripped-star candidates present in the clusters, alongside main-sequence or giant stars. In the following subsections, we discuss the specific stellar populations found in OCs.

\subsubsection{Blue stragglers} 
\label{OC.BSS.sec}

The bluest and brightest stars in open clusters are BSS, apart from the scanty subdwarf population in older clusters \citep[see][for more discussion on the UV properties of hot subdwarf populations]{2026SSRv..222...42F}. They are formed through mass accretion via either mergers or mass transfer \citep{Hills1976ApL....17...87H, Naoz2014ApJ...793..137N} in binary or higher-order multiple systems \citep{McCrea1964MNRAS.128..147M}. These rejuvenated stars are UV-bright and the best targets for UV-based studies to understand their formation pathways. 

The first comprehensive catalogues of BSS in OCs were compiled by \citet{1995A&AS..109..375A, 2007A&A...463..789A}. Their catalogues comprised a few thousand BSS in approximately 400 OCs. These studies examined the dependence of the BSS frequency and cluster properties, including age, richness, stellar spatial distributions and the ratio of BSS to main-sequence stars; however, a major limitation of these catalogues was the lack of membership information of the identified BSS. In the {\sl Gaia} era, BSS have been identified more comprehensively in a large number of OCs with reliable and homogeneously determined cluster membership information \citep{2021MNRAS.507.1699J, 2021A&A...650A..67R}. \citet{2021MNRAS.507.1699J} reported that most OCs older than $\sim$1 Gyr and masses greater than $\sim$1000 $M_{\odot}$ contain BSS populations. A similar correlation between cluster age and the presence of BSS was reported by \citet{2021A&A...650A..67R}, who found that BSS begin to appear in OCs with ages of approximately 500 Myr and older. Owing to their low-density environments and close proximity (for some), OCs allow for a detailed investigation of their members, including BSS. The WIYN Open Cluster Survey, which used the 3.5m WIYN telescope, thoroughly investigated selected OCs using photometric and spectroscopic observations. 

Another interesting aspect pertaining to BSS relates to their binary companions. In particular, using high-resolution spectroscopic observations, \citet{2011Natur.478..356G} discovered that a majority of the BSS in NGC 188 have companions with masses close to $0.5 M_\odot$, consistent with WD masses. Such companions may also be hot owing to their stripped atmospheres, resulting in significant UV emission. Characterising both components of a binary can provide better constraints on the formation history of the system.

In fact, UV observations have been crucial in revealing WD companions to BSS. Early {\sl HST} UV studies of 47~Tucanae (47 Tuc) combined far-UV spectroscopy and near-UV imaging to construct composite SEDs of BSS, uncovering hot WD companions with $T_{\mathrm{eff}} \sim 12,000$--30,000~K that contribute significantly to the far-UV flux \citep[e.g.,][]{Knigge2008,Raso2019}. This strongly suggested Case C mass transfer\footnote{Case C mass transfer refers to Roche-lobe overflow that begins after the donor star has completed core helium burning, typically when it has expanded to an AGB configuration with a deep convective envelope.} to be the dominant formation mechanism for these systems in star clusters. 

These results provided impetus for UV-based studies to search for such unresolved hot BSS companions using multi-wavelength SEDs by means of the detection of UV excesses. Using high-resolution UV-imaging observations with the {\sl HST}, \citet{2014ApJ...783L...8G, Gosnell2015ApJ...814..163G} discovered WD companions of several BSS in NGC 188. This was the first observational evidence of a mass-transfer origin of BSS in an OC. Hot WD companions were also found to be associated with BSS in M67 \citep{2019ApJ...885...45G}. Building on these results, high-resolution far- and near-UV imaging  with {\sl AstroSat}/UVIT has recently enabled the  photometric detection of WD companions through multi-band SED analyses \citep{sahu2019b, Dattarey2023ApJ...943..130D}. These findings demonstrate the UV domain as the most effective window for detecting and characterising such systems. Binary systems that host UV-bright companions can potentially evolve to become cataclysmic variables \citep{2022MNRAS.517.2867H}. Although such systems are expected to be present in clusters, they are rarely detected. 

Multi-wavelength observations, particularly UV imaging and spectroscopy, have been crucial in identifying and characterising binary BSS. 
Figure~\ref{fig:M67_stragglers} shows the UVIT image of  M67, highlighting the position of known BSS and other UV-bright sources.
Numerous BSS+WD \citep{Gosnell2015ApJ...814..163G, Sindhu2019ApJ...882...43S}, BSS+sdA \citep[A-type hot subdwarf;][]{Jadhav2023A&A...676A..47J}, BSS+sdB \citep{Jadhav2021JApA...42...89J} and BSS+post-AGB \citep{Subramaniam2016ApJ...833L..27S} systems have been identified based on UV imaging from {\sl HST}, {\sl AstroSat}/UVIT and {\sl Swift}/UVOT (see fig. \ref{fig:UOCS_OC}). The properties of BSS and their companions show that mass transfer is ubiquitous in star clusters and that there are multiple complex evolutionary scenarios for binary systems.

\subsubsection{Blue Lurkers and Yellow Stragglers} 
\label{OC.lurker.sec}

The main-sequence counterparts of BSS are called `Blue Lurkers'. They are identified using signatures of rapid rotation \citep{Leiner2019ApJ...881...47L} or by demonstrating the presence of a stellar companion that can only be formed through mass donation \citep{Jadhav2019ApJ...886...13J}. As demonstrated by UV studies of BSS, characterisation of the compact companion is useful to confirm the system's mass-transfer history. Detection of a similar companion to a main-sequence star can be used to confirm the mass-transfer history of the main-sequence system while simultaneously leading to the classification of the main-sequence component as a Blue Lurker. Recent studies find the presence of Blue Lurkers in many OCs, including M67, NGC 362 and NGC 752 \citep{Leiner2019ApJ...881...47L, 2023ApJ...944..145N, Jadhav2024A&A...688A.152J}, and in one GC, NGC 362 \citep{Dattarey2023ApJ...943..130D}. 

Yellow stragglers are the cooler, bloated counterparts of BSS lying above the sub-giant branch (SGB) in the Hertzsprung--Russell (HR) diagram and between the zero-age main sequence and the red-giant branch (RGB). Similar to BSS, their presence indicates ongoing or (recent) past stellar interactions. Colour--magnitude diagram (CMD)-based searches have resulted in a combined total of $\approx240$ and $\approx1230$ yellow stragglers in all known Galactic OCs and GCs, respectively \citep{Rain2024A&A...685A..33R, Carrasco2025A&A...699A.142C}. 
However, the yellow straggler catalogues are known to be contaminated by unresolved binaries.
Nevertheless, the median ratio of yellow stragglers to BSS is 0.27 for clusters that contain both stellar species \citep{Carrasco2025A&A...699A.142C}.
Two main formation scenarios have been proposed for yellow stragglers: (i) they may represent BSS evolving towards the RGB, making them analogous to SGB stars \citep{Landsman1997ApJ...481L..93L}; or (ii) they may be close-to-MSTO binary systems that appear bloated owing to ongoing mass transfer \citep{Rain2024A&A...685A..33R, Wang2025ApJ...984...52W}.

UV spectroscopy with the Goddard High-Resolution Spectrograph on the {\sl HST} and imaging with {\sl UIT} enabled the first detailed study of a yellow straggler \citep{Landsman1997ApJ...481L..93L}. Chromospheric activity in this system was revealed through the Mg\,{\sc ii} $\lambda2800$ {\AA} doublet. Since the flux of yellow stragglers dominates at optical wavelengths, their companions are best studied in the far-UV regime, where the flux contribution from a hot, compact companion can be significant and the contribution from the chromospheric activity, if present, is minimal. The formation and evolutionary pathways of these systems are still to be explored. A massive CO WD would point to an AGB donor, whereas a low-mass ($<0.5 M_{\odot}$) WD would suggest an RGB donor. Thus, constraining the properties of the companion is essential to understanding the evolutionary pathways of these systems, with UV observations providing the most effective means to achieve this.

\subsubsection{Classical Be Stars}

Classical Be stars are identified by the presence of Balmer emission lines in their optical spectra. Such emission lines originate from a Keplerian gaseous disc formed from the matter ejected from the star subject to rapid rotation. The rapid rotation of these stars is thought to have been acquired through either binary interactions or internal stellar evolution, and the material may also be ejected through non-radial pulsations. 

Observations probing the binary channel for the formation of Be stars have resulted in the detection of several Be--sdO/B binaries (hot O/B-type subdwarfs). The detection of a hot sub-dwarf O/B companion to a classical Be star is challenging and UV data has greatly helped their detection. Studies of classical Be stars in the field with subdwarf companions have progressed in the recent years, whereas such studies in young star clusters are still in their infancy. \citet{2008ApJ...686.1280P} detected a hot companion with an effective temperature of $T_\mathrm{eff} =45\pm5$ kK to FY Canis Majoris (FY CMa), and \citet{2013ApJ...765....2P} detected a hot companion ($T_\mathrm{eff} = 52.1 \pm 4.8$ kK) to 59 Cygni (59 Cyg), using a large number of {\sl IUE} spectra. Similarly, \citet{2017ApJ...843...60W} found a hot subdwarf companion to 60 Cyg and \citet{2016ApJ...828...47P} found a faint signal of a hot companion in HR 2142, again using many {\sl IUE} spectra. \citet{2018ApJ...853..156W} found 12 new Be+sdO systems from a sample of 264 stars from {\sl IUE} spectra and suggested that there could be many more Be+sdO systems whereas many more might be undetectable owing to the faint luminosities of the companions.

\begin{figure}
    \centering
    \includegraphics[width=0.99\linewidth]{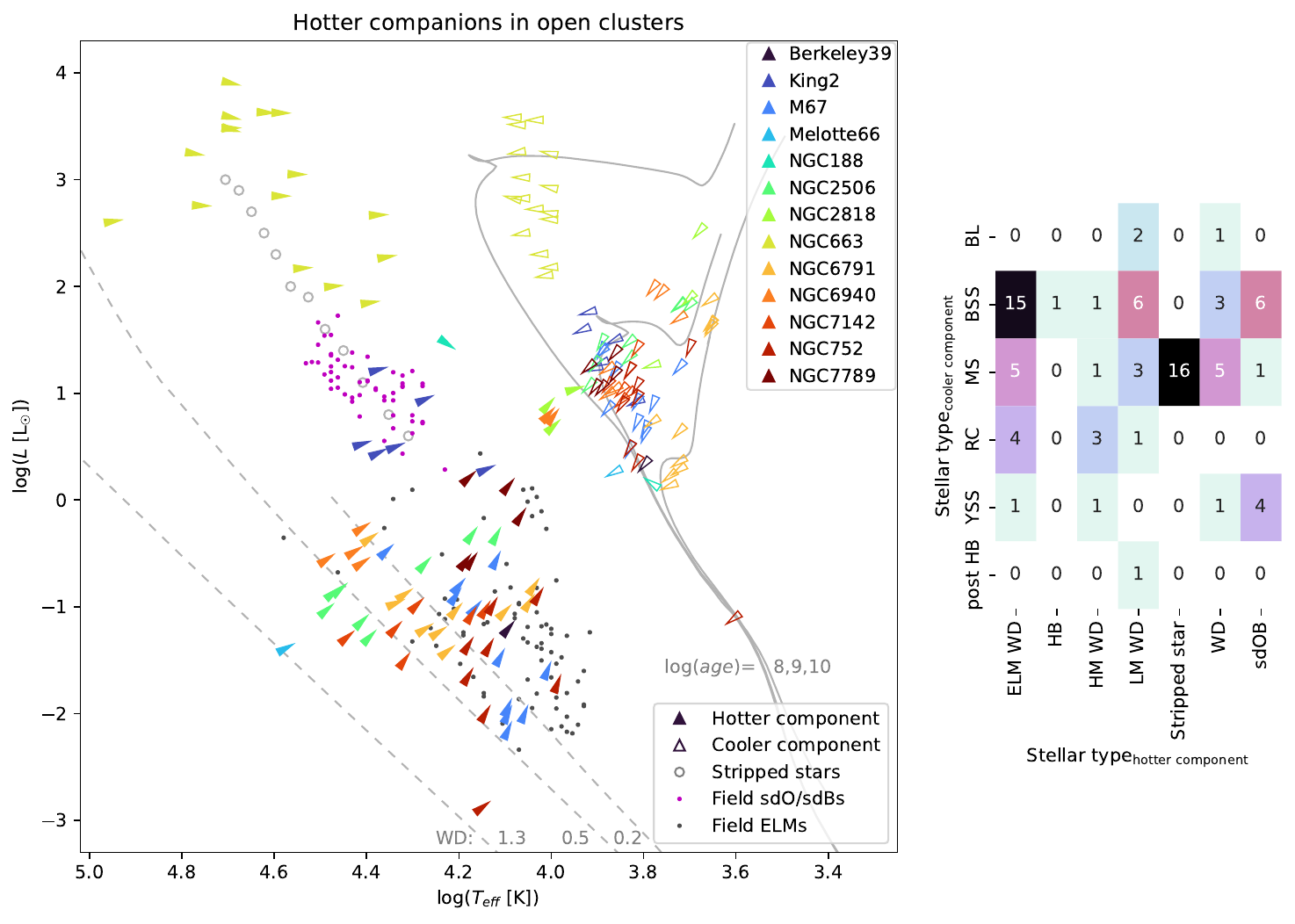}
    \caption{Loci in the Hertzsprung--Russell diagram of binary companions in 13 OCs identified using {\sl GALEX}, UVIT and UVOT observations. The hotter compact companions (filled triangles) and cooler components (open triangles) are coloured by reference to the host cluster, and they point towards their companion in the diagram. The positions of extremely low-mass field WDs \citep[grey dots;][]{Brown2010ApJ...723.1072B}, field sdO/sdB stars \citep[hot subdwarf O and B stars: magenta dots][]{Geier2020A&A...635A.193G}, models of stripped stars \citep[grey circles;][]{Gotberg2018A&A...615A..78G}, WD cooling curves \citep[$M_\mathrm{WD} = 1.3, 0.5, 0.2 M_{\odot}$, grey dashed lines;][]{Bedard2020ApJ...901...93B} and solar-metallicity {\sc parsec} isochrones \citep{Bressan2012MNRAS.427..127B} ($\log(t \mbox{ [yr}^{-1}]) = 8, 9, 10$; grey lines) are also shown, for reference. The right-hand panel shows the statistics of the Hertzsprung--Russell diagram-based stellar types (and their pair-wise combinations) in the binary systems. LM, HM: low-mass, high-mass.}
    \label{fig:UOCS_OC}
\end{figure}

Classical Be stars in OCs have been studied in detail by \citet{2005ApJS..161..118M,2005ApJ...622.1052M}, \citet{2008MNRAS.388.1879M}, \citet{2010A&A...509A..11M} and \citet{2021JApA...42..109J}. These OCs are ideal systems to probe the binary mode of evolution, since clusters contain a significant fraction of binary stars \citep{2021AJ....162..264J, 2024ApJ...971...71J}. {\sl IUE} spectra of Be stars in nearby clusters are yet to be fully explored. Recently, \citet{Nedhath2025A&A...699L...1N} explored UV photometry to probe binarity among Be stars in the young star cluster NGC 663. Based on SEDs constructed using 20 data points from the far-UV to the mid-IR (including five data points in the UV), they identified hot companions for 16 out of 23 Be stars, with properties similar to stripped sdOB-type stars (see Figure~\ref{fig:UOCS_OC}). The authors further suggest that access to UV spectroscopic data is essential to confirm the evolutionary status of the companion stars.  The above result is indicative of the significant contribution of binary-initiated formation of classical Be stars in the cluster. More clusters need to be studied to confirm the above and to assess the relative contribution of single-star versus binary pathways for the formation of classical Be stars.

\subsubsection {Reflections}

Figure 2 shows the location in the CMD of the components of binary companions detected in 13 OCs studied based on UV photometry from UVIT and UVOT. While most of the binaries found in clusters are $\ge 1$ Gyr old, with NGC 6791 being the oldest cluster, the figure also shows the binary Be+sdOB systems detected in the young 25 Myr-old OC NGC 663. The location of extremely low-mass (ELM) WDs and sdOBs found in the field and the theoretically estimated location of stripped stars are also shown. The BSS, yellow stragglers, Blue Lurkers and Be stars are shown as open symbols, whereas the filled symbols indicate their compact companions. Note that the compact companions fall in the categories of WD, low-mass WD, ELM WDs, sdOBs and stripped stars. Except for the WDs, the other types of the companions will likely be end product of binary mass transfer. Low-luminosity sdBs are found in the binary systems of King 2 (age $\sim 6$ Gyr). It will be important to model these binary systems to understand the details of mass transfer and its link with the end product, as a function of binary and cluster properties.

As for the individual populations, as more Be+sdOB systems are being discovered, more theoretical simulations to explore their formation pathways are needed. Yellow stragglers are relatively little explored, and more studies are needed to understand their formation pathways. Although Blue Lurkers are the latest entries in the landscape of post-mass-transfer products, it has become evident that they are present in many old OCs. Blue Lurkers are also ideal targets to understand mass-transfer details in low-mass-ratio binary systems. 

\subsection{Globular Clusters}

\subsubsection{Multiple stellar populations}
Star clusters have long been considered typical `simple' stellar populations (SSPs), because the star-formation process within them occurs extremely rapidly \citep[on timescales of only 1--3 times the free-fall time of the progenitor molecular cloud;][]{2019MNRAS.487..364L}, and stellar feedback processes are intense. The combined effect of photoionisation, stellar winds, and supernovae from the first massive stars injects sufficient energy and momentum to expel the residual gas on timescales of a few Myr---much shorter than the duration required for a second burst of star formation. Consequently, the supply of fresh gas is truncated almost immediately, enforcing the assumption that all cluster stars share a common age and metallicity. All of these factors result in stars in the cluster forming from the same molecular cloud within an extremely short temporal window. Therefore, it would be reasonable to assume that all stars in the cluster have nearly identical metallicities and ages. 

However, the classical SSP paradigm has been challenged by the discovery of multiple populations (MPs): distinct groups of stars co-existing within individual clusters that exhibit correlated abundance variations in light elements (e.g., the Na--O, C--N, and Mg--Al anti-correlations) and in helium. Although the physical origin of MPs, whether resulting from multiple star-formation episodes, early internal pollution by massive stars, or external accretion, remains actively debated, the observational signature is robust and remarkably uniform across the Galactic GC system. Specifically, in nearly all GCs and the vast majority of old, massive stellar clusters, significant chemical abundance variations have been observed among the member stars \citep{2012A&ARv..20...50G,Piotto2015AJ....149...91P}. These variations involve elements ranging from helium to carbon, nitrogen, oxygen, sodium, magnesium, aluminium and other light elements, and in some massive GCs, even iron-peak elements exhibit inhomogeneities. This phenomenon, which is at odds with the traditional SSP picture, is referred to as the `multiple populations' problem. The MP problem in star clusters has been extensively reviewed in recent years \citep{Bastian2018ARA&A..56...83B,Gratton2019A&ARv..27....8G,Milone2022Univ....8..359M} and thus will not be reiterated here. Instead, we focus on the critical role of UV observations in advancing our understanding of the MP problem. 

Traditional spectroscopic studies of MPs face two fundamental, intertwined observational limitations: (i) the faintness of key stellar tracers at typical GC distances, which demands prohibitively long exposures for high-resolution, high-signal-to-noise spectroscopy; and (ii) extreme stellar crowding in cluster cores, where surface densities can exceed $10^5$ stars pc$^{-2}$ and overlapping point-spread functions render single-star abundance measurements unreliable. For distant clusters, it becomes challenging to obtain medium- to high-resolution spectra with sufficient signal-to-noise ratios for low-luminosity stars, such as main-sequence stars, which are precisely the key stellar population needed for inferring the presence of MPs in a cluster. Meanwhile, the high stellar densities in cluster cores lead to overlapping stellar images, thus causing severe contamination from crowded fields that heavily affects traditional spectroscopic observations. As a result, the use of spectroscopy to study MPs is often restricted to the outer regions of GCs, where stellar (number) densities are relatively low. This, in turn, greatly limits investigations of the dynamical origins of MPs, since the densest central regions of clusters, which typically hold crucial kinematic information about MPs, are often inaccessible with conventional spectroscopic techniques. {\it Nevertheless, within these restricted domains---the less crowded outskirts and the intrinsically bright evolved sequences (e.g., RGB and HB stars)---high-resolution spectroscopy remains feasible and has yielded the definitive chemical abundance measurements (e.g., the Na--O and Mg--Al anti-correlations) that serve as the critical benchmarks for validating photometric MP classifications.}

UV imaging has been pivotal in transforming our understanding of MPs in star clusters, especially in Galactic GCs. UV passbands sample strong molecular features---OH in F275W, NH in F336W and CN/CH in F438W---whose sensitivities to light-element variations (CNO) translate into large photometric separations at a fixed effective temperature. Leveraging this, the {\sl HST}’s WFC3/UVIS filters enabled the construction of pseudo-colours such as $C$(F275W, F336W, F438W) = (F275W$-$F336W)$-$(F336W$-$F438W) that cleanly disentangle populations along the main sequence, SGB and RGB, even in extremely crowded cluster cores. The {\sl HST} UV Legacy Survey of Galactic GCs \citep{Milone2014,Piotto2015AJ....149...91P} provided homogeneous, deep imaging for dozens of GCs with WFC3/UVIS and the ACS, demonstrating that MPs are ubiquitous and tightly linked to variations in He and the Na--O anti-correlation, as confirmed by spectroscopy; the resulting `chromosome maps' offered an empirical, cluster-by-cluster census of population complexity and its correlation with GC parameters \citep[e.g., mass, concentration, HB morphology;][]{Milone2017MNRAS.464.3636M}.

UV-selected first-population (1P) and second-population (2P) samples, combined with high-precision proper motions and spectroscopy, reveal spatial and kinematic segregation. In many systems, 2P stars are more centrally concentrated than 1P stars \citep{Lardo2011, Milone2012}. In some clusters, 2P stars show more radially anisotropic velocity distributions than 1P stars, consistent with formation scenarios followed by long-term dynamical evolution \citep{Cordoni2020}. However, diversity abounds: the mapping between UV colours and abundance spreads, as well as the degree of spatial/kinematic segregation, varies from cluster to cluster \citep{Bastian2018ARA&A..56...83B, Gratton2019A&ARv..27....8G}. Studies of young/intermediate-age massive clusters in the MCs show MPs in light elements (and sometimes He) with little Fe spread, indicating dependencies on age, mass and environment \citep{Martocchia2018, Li2019, Saracino2020}. Further insights into MPs were gained from studies in the MCs, which are discussed in Section~\ref{MPs in MCs}.

While the above discussion focuses on MP separation along the main sequence, subgiant branch, and red-giant branch, UV observations are equally powerful for characterising intrinsically UV-bright evolutionary phases---specifically the horizontal branch (HB), white dwarf (WD), and BSS populations. It is important to emphasise that these are stellar evolutionary stages, not population labels: both 1P and 2P stars evolve through the HB and WD cooling sequences. Nevertheless, their UV properties differ systematically because of initial abundance variations. For instance, He-enhanced 2P stars tend to populate bluer and more extended HB morphologies \citep{Milone2018MNRAS.481.5098M}, and their WD descendants may retain distinct kinematic signatures tied to dynamical heating. BSS, whose formation channels (collisional versus binary) may depend sensitively on local density and primordial abundances, also show complex spatial and kinematic associations with the underlying MP structure. MPs are therefore not restricted to the unevolved stellar phases; they manifest across the full Hertzsprung--Russell diagram, from the main sequence to the WD cooling track.

Ultimately, high-resolution UV imaging resolves individual 1P/2P stars in the densest cores where spectroscopy fails; wide-field astrometric missions provide the global kinematic framework, revealing large-scale segregation; and ground-based spectrographs deliver precise radial velocities and detailed chemical patterns in the less crowded outer regions. Only by integrating these complementary approaches can one reconstruct the complete lifecycle of MPs from their formation environment to their long-term dynamical diffusion.

These advances in MP characterisation point to a broader principle. Star clusters are not merely passive laboratories for testing stellar evolution models; they are chemically tagged, dynamically evolving systems whose internal substructures preserve the imprint of their birth environments and subsequent Galactic orbits. The same UV-based techniques that disentangle 1P and 2P populations in crowded cores also provide the precise astrometry and chemodynamical tags needed to use clusters as probes of the Milky Way's gravitational potential, assembly history, and large-scale stellar populations. We now turn to this wider perspective, examining how both globular and open clusters serve as tracers of Galactic structure and dynamics beyond their internal MP complexity.

\subsubsection{Horizontal Branch and Post-HB Stars}

\begin{figure}
    \centering
    \includegraphics[width=\textwidth]{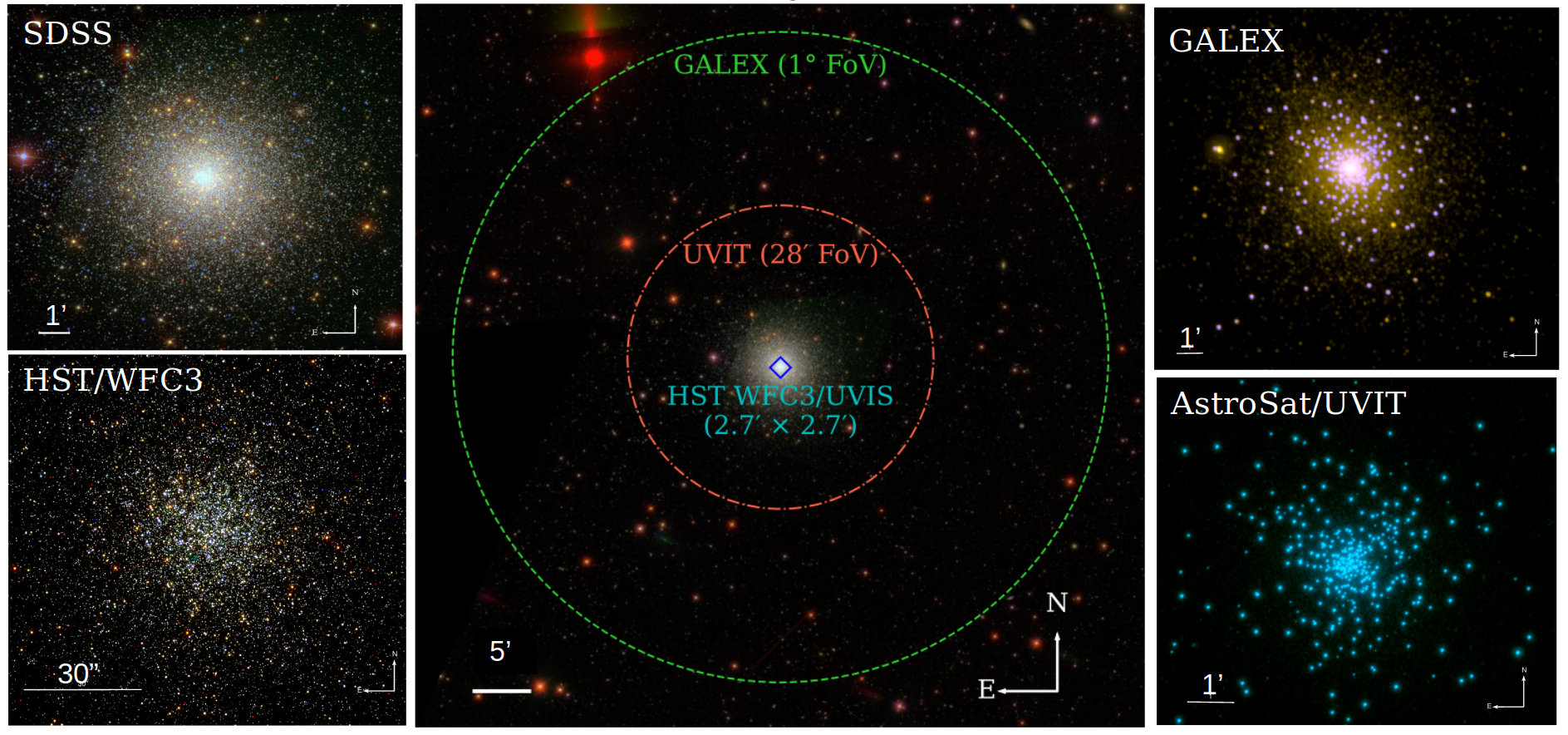}
    \caption{UV images of a dense GC M92 illustrating the differences in spatial resolution and field of view (FoV) among major UV instruments. {\sl HST}/WFC3, with its small FoV of $2.7^\prime \times2.7^\prime$ and high spatial resolution of $\sim0.1^{\prime\prime}$ (solid blue line), resolves even the crowded cluster core \citep{Piotto2015AJ....149...91P}. {\sl GALEX}, offering a much larger $\sim 1^\circ$ FoV but coarser 4--$5^{\prime\prime}$ resolution (dashed green line), enables mapping of the entire cluster out to its tidal radius in both the far-UV (blue dots) and the near-UV \citep[yellow dots;][]{Schiavon2012AJ....143..121S}. {\sl AstroSat}/UVIT provides an intermediate capability in FUV (blue dots) with $\sim1.4^{\prime\prime}$ resolution and a 28$^{\prime}$ circular FoV (dash-dotted red line); its multiple far-UV filters help in the spectral sampling of hot stellar populations \citep{Sahu2022}.}
    \label{fig:gc_images}
    \vspace{-0.5cm}
\end{figure}

Prior to the launch of the {\sl HST}, UV investigations of HB stars in GCs relied on early space missions, most notably {\sl IUE} (1978--1990). Pioneering {\sl IUE} UV photometry of individual hot stars in clusters such as NGC~6752 provided some of the first direct estimates of the effective temperatures of blue and extreme-HB stars \citep[e.g.,][]{Caloi1983,Cacciari1995}. These observations confirmed the presence of HB stars with $T_{\mathrm{eff}} \geq 20,000$~K, significantly hotter than predicted by canonical zero-age HB models at the time. In parallel, theoretical studies \citep{Caloi1989, Castellani1994} developed evolutionary tracks for extreme-HB and post-HB stars, predicting the formation of AGB-manqu\'e and UV-bright post-HB populations as natural outcomes of enhanced mass loss and helium enrichment on the RGB. Together, these observational and theoretical efforts established the physical basis for interpreting the UV output of GCs in terms of their hot HB populations, laying the groundwork for the high-resolution imaging and detailed stellar population analyses later achieved with {\sl HST}.

{\sl HST} revolutionised UV studies of HB stars in GCs by overcoming the crowding limitations and sensitivity constraints that plagued earlier missions. With its superb spatial resolution ($\sim$0.1$^{\prime\prime}$) and stable photometric performance, {\sl HST} enabled, for the first time, direct imaging and photometry of individual HB stars across entire cluster cores (see Figure~\ref{fig:gc_images}), even in dense, metal-rich systems where ground-based optical and earlier UV telescopes could not resolve individual stars. Early {\sl HST} UV instruments, particularly the Wide Field and Planetary Camera-2 (WFPC2) and STIS, provided deep UV CMDs that revealed the full temperature distribution of HB stars, from the red HB ($T_{\rm eff} \sim 5,000$--8,000~K) through the blue HB ($T_{\rm eff} \sim 10,000$--20,000~K) to the extreme-HB stars with $T_{\rm eff} \geq 20,000$~K \citep{Ferraro+1998ApJ}. These early {\sl HST} studies uncovered complex morphological features such as extended blue tails, gaps, and `blue-hook' stars, the latter reaching $T_{\rm eff}\,\approx 35,000-40,000~K$ \citep[for example, see][]{Brown+2010ApJ, Moehler+2011A&A} and believed to originate from late He flashes or He-enriched subpopulations. UV observations also revealed distinct discontinuities along the HB \citep{Dalessandro2011, Brown2016}, including the `Grundahl jump' at $T_{\rm eff} \sim 11,500$~K and the `second U jump' near $T_{\rm eff} \sim 23,000$~K, which are attributed to radiative levitation and diffusion effects in the atmospheres of hot HB stars, respectively \citep{Grundahl1999,Momany2004}. Observations of $\omega$~Centauri \citep[$\omega$~Cen;][]{DCruz2000} and NGC~2808 \citep{Brown2001} established the presence of these hot, subluminous blue-hook objects, whereas \citet{Moehler2004} confirmed their helium and calcium enrichment through UV spectroscopy, supporting the `late hot flasher' scenario. More broadly, using WFPC2 \citet{Piotto1999} demonstrated that the HB morphology varies systematically with cluster parameters beyond metallicity, pointing to MPs and differential mass loss as key drivers. 

Building on these results, the {\sl HST} UV Legacy Survey of Galactic GCs established precise HB temperature calibrations and quantified how He enhancement ($\Delta Y \sim 0.03$--$0.10$) systematically shifts HB stars to higher effective temperatures and bluer UV colours. Subsequent analyses \citep[e.g.,][]{Dalessandro2011,Lagioia2015,Milone2015} confirmed that clusters with extended blue HBs and large internal He spreads are among the most massive systems, revealing a direct link between cluster mass, helium enrichment and HB extent.

\begin{figure}
    \centering
    \includegraphics[width=\linewidth]{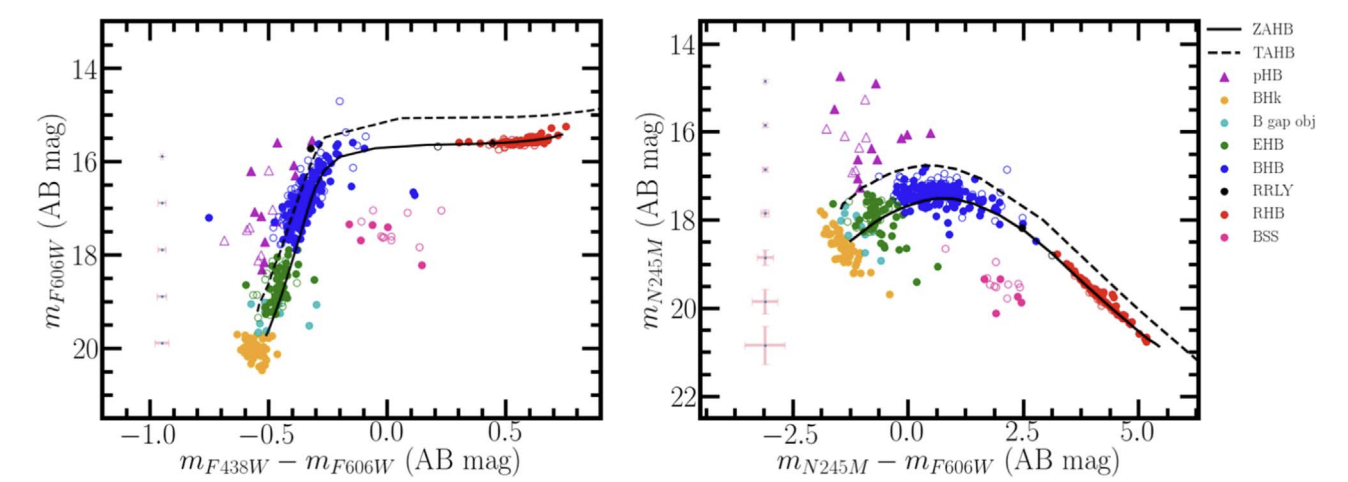}
    \caption{Observed optical and NUV (N245M filter of UVIT) CMDs of a massive GC NGC 2808 \citep{Prabhu2021}, constructed by combining {\sl Astrosat}/UVIT photometry with {\sl HST} (dots) and {\sl Gaia} (open circles) observations. Several UV-bright stellar populations are marked in the figure, including the different HB populations spanning the RHB, BHB to RHB and blue hook stars, as well as BSSs. Owing to their higher effective temperatures and SEDs peaking at shorter wavelengths, these stars appear significantly brighter in the near-UV, making UV observations particularly effective for identifying and characterising hot stellar populations in GCs.}
    \label{fig:gc_uv_cmd}
\end{figure}

Expanding on the relatively limited field of view studies from {\sl HST}, the {\sl GALEX} mission (2003--2013) extended HB research into the wide-field regime, providing the first homogeneous UV survey of Galactic GCs. With its simultaneous far-UV ($\lambda_{\mathrm{eff}} \simeq 1528$~\text{\AA}) and near-UV ($\lambda_{\mathrm{eff}} \simeq 2271$~\text{\AA}) imaging over a $1.2^{\circ}$ field of view, {\sl GALEX} could capture entire clusters, including their extended haloes and tidal structures, in a single exposure (see Figure\,\ref{fig:gc_images}). This capability resulted in the first systematic UV CMDs for a large sample of 44 Galactic GCs \citep{Schiavon2012AJ....143..121S}, revealing the full range of blue, extreme- and post-HB stars spanning effective temperatures from $T_{\mathrm{eff}} \sim 8,000$~K up to 35,000~K. The {\sl GALEX} data showed that clusters with bluer HB morphologies and lower metallicities ($\mathrm{[Fe/H]} \leq -1.5$ dex), such as M13 and NGC~6752, exhibit much stronger far-UV fluxes than metal-rich systems like 47~Tuc ($\mathrm{[Fe/H]} \simeq -0.7$ dex), quantitatively strengthening the long-standing metallicity--UV relation \citep{1995ApJ...442..105D}. Integrated (far-UV$-$near-UV) colours measured by {\sl GALEX} correlated tightly with the fraction of blue HB and extreme-HB stars \citep{Dalessandro2012}, providing a statistical diagnostic for He enhancement and mass-loss efficiency across the Galactic cluster system. Although its moderate spatial resolution ($\sim$5$^{\prime\prime}$) limited its ability to resolve crowded cluster cores, the wide-field and uniform sensitivity of {\sl GALEX} enabled robust population-level analyses, identification of hot sub-dwarfs and UV-bright post-HB stars, and comparisons with integrated UV light from extragalactic systems exhibiting the `UV upturn' phenomenon \citep{Rey2007}.

The subsequent Indian {\sl AstroSat}/UVIT mission, launched in 2015, ushered in a new phase, combining high spatial resolution ($\sim$1.5--1.8$^{\prime\prime}$) with a wide $28^{\prime}$ field of view. In contrast to the narrow but deep imaging of {\sl HST} and the all-sky coverage of {\sl GALEX}, UVIT bridges these regimes, enabling resolved UV photometry of entire clusters from core to halo (see Figure\,\ref{fig:gc_images}). Multi-band far- and near-UV imaging with UVIT has revealed well-defined red, blue and extreme-HB sequences in several clusters, including NGC~1851, NCG~288, NGC~4590, NGC~1261 and NGC~2808 \citep{Subramaniam2017, Sahu2019, Prabhu2021, Rani2021, Kumar2022} as shown in Figure \ref{fig:gc_uv_cmd} for example. These studies identified HB stars with temperatures extending up to $T_{\mathrm{eff}} \approx 35,000$--40,000~K and detected numerous UV-bright post-HB and blue-hook objects. When combined with optical and {\sl Gaia} photometry, UVIT data has enabled precise luminosity and temperature estimates for individual stars, revealing clear signatures of MPs with He enrichment up to $\Delta Y \sim 0.05$--0.10.

\begin{figure}
    \centering
    \includegraphics[width=\linewidth]{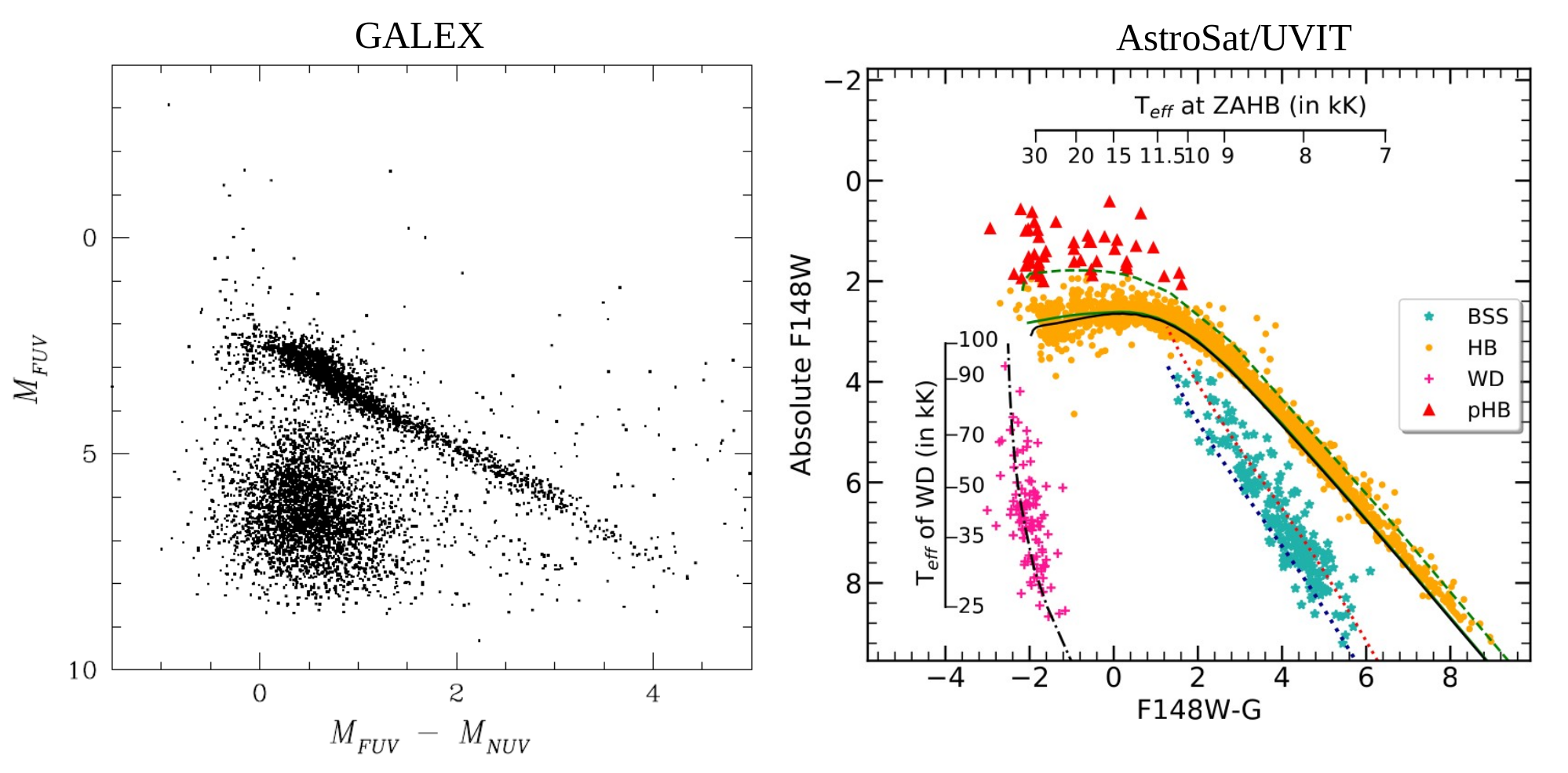}
    \caption{(left) Stacked far-UV versus (far-UV$-$near-UV) CMD constructed from {\sl GALEX} photometric observations of 44 Galactic GCs, illustrating the HB and BSS populations. Stacking multiple clusters enhances the visibility of these sparse UV-bright populations, which are often difficult to identify in individual clusters; however, the CMD remains affected by background contaminants, particularly towards bluer colours and fainter far-UV magnitudes \citep{Schiavon2012AJ....143..121S}. (right) Stacked far-UV--optical CMD derived from {\sl AstroSat}/UVIT far-UV photometry of 11 GCs, combined with optical photometry and proper-motion memberships from {\sl HST} (inner regions) and {\sl Gaia} (outer regions) \citep{Sahu2022}. Membership information has been used to remove background contaminants and the UV data clearly delineate the HB, BSS, and WD sequences, demonstrating the power of UV photometry for studying hot stellar populations in GCs.}
    \label{fig:stack_cmd}
\end{figure}

The Globular Cluster UVIT Legacy Survey (GlobULeS) systematically extended UV studies to a larger sample of Galactic GCs \citep{Sahu2022}, presenting combined far-UV--optical CMDs for eleven clusters covering the entire cluster region, i.e. the full spatial extent of each GC covered by the UVIT field of view. The survey identified more than 2800 HB stars, including about 190 extreme-HB candidates, across a wide metallicity range ($-2.3 \leq \mathrm{[Fe/H]} \leq -0.5$ dex) and detected HB discontinuities consistent with those identified through {\sl HST} observations in the cluster cores. By combining UVIT photometry with proper-motion memberships from {\sl HST} and {\sl Gaia}, the stacked far-UV--optical CMDs provided clean sequences of hot stellar populations revealing HB, extreme-HB, and post-HB stars over entire cluster fields, thus providing an improvement over {\sl GALEX} as shown in Figure~\ref{fig:stack_cmd}. The uniform UVIT data set also enabled a statistical study of HB colour extensions, showing that the extent of the blue HB correlates with both metallicity and cluster mass, consistent with an earlier {\sl HST} study \citep{Milone2014}. The catalogue further allowed estimates of He variations ($\Delta Y \approx 0.05$--0.06) and supported the interpretation that differences in mass loss and He abundance drive the classical second-parameter effect between M3 and M13 \citep{Kumar2023}. Deep far-UV imaging of $\omega$~Cen identified more than 5000 hot sources, including about 1000 extreme-HB and blue-hook stars \citep{Prabhu2022}, spanning $T_{\mathrm{eff}} \simeq 8,000$--45,000~K and confirming that the hottest stars belong to the He-rich subpopulations previously found with {\sl HST}. Overall, UVIT provided a powerful complement to {\sl HST} by enabling wide-field, population-resolved studies of HB morphology in the UV and offering essential constraints to models of hot HB and post-HB evolution.

GCs provide a uniquely powerful laboratory for studying HB evolution because all HB stars in a given cluster share a common distance and, to first order, a similar age and metallicity. This greatly reduces the degeneracies that affect field-star samples, where uncertainties in distance, evolutionary history and chemical composition complicate interpretation. At the same time, comparisons among clusters spanning a wide range of metallicities, masses and helium abundances enable direct tests of the relative importance of these parameters in shaping HB morphology. In contrast, MC clusters offer a complementary perspective by extending such studies to different age and metallicity regimes, particularly among intermediate-age populations. Together, Galactic and MC clusters provide a framework for disentangling the effects of metallicity, helium enrichment, mass loss and MPs on HB evolution in ways that are difficult to achieve using field stars alone. This is particularly important for understanding the long-standing second-parameter problem, for which UV observations provide some of the most sensitive diagnostics of helium-rich and extreme-HB populations.

Future UV facilities will extend these capabilities further. The proposed Cosmological Advanced Survey Telescope for Optical and ultraviolet Research ({\sl CASTOR}) combines near-{\sl HST} spatial resolution ($\sim0.15''$ in the UV) with a field of view almost two orders of magnitude larger than that of the {\sl HST}. Such a facility would enable homogeneous UV imaging of entire Galactic GCs, from their crowded cores to their tidal radii, while simultaneously resolving HB, extreme-HB, blue-hook and post-HB populations. {\sl CASTOR} would therefore provide an important bridge between the detailed cluster-core studies enabled by the {\sl HST} and the population-level surveys pioneered by {\sl GALEX} and UVIT, allowing systematic investigations of MPs, helium enrichment and the UV properties of hot stellar populations across large cluster samples \citep{2019BAAS...51g.209M,2023PASP..135c5002M}.

\subsubsection{White Dwarfs}

OCs and GCs provide clean laboratories to uncover WD binaries, because their well-constrained distances, reddening values and ages---combined with an empirically calibrated initial--final mass relation \citep[e.g.,][]{2018ApJ...866...21C}---enable robust estimates of a WD’s total age, \(t_{\mathrm{tot}} = t_{\mathrm{prog}}(M_{\mathrm{init}}, Z) + t_{\mathrm{cool}}(M_{\mathrm{WD}})\). When a cluster WD would appear `too old' under a single-star assumption---i.e. its inferred \(t_{\mathrm{tot}}\) exceeds the cluster age---this strongly signals past binary interaction (stable mass transfer or common-envelope evolution), often producing undermassive He-core WDs that single-star evolution cannot create at the observed cluster age. Deep {\sl HST} studies of NGC~6397 established this diagnostic in practice: the WD cooling sequence and a centrally concentrated population of low-mass, likely He-core WDs require binary-evolution pathways rather than isolated evolution \citep{2007ApJ...671..380H,2009ApJ...699...40S}. 

UV observations are pivotal for discovering, identifying and characterising compact WD binaries in clusters, because they isolate the hottest emission and expose accretion and magnetic diagnostics that are weak or ambiguous in the optical. Time-tagged {\sl HST}/Cosmic Origins Spectrograph (COS) and STIS spectroscopy of ZTF~J2008+4449 revealed (i) a large-amplitude, phase-coherent far-UV continuum modulation ($\sim 30$ per cent bluewards of 1300\,{\AA}), (ii) a spin-dependent, Zeeman-split Ly$\alpha$ component near 1344\,{\AA} that disappears for half a cycle, and (iii) narrow C/Si absorption consistent with inter- or circumstellar material. Combined with a UV SED break characteristic of highly magnetised atmospheres, these UV signatures trace magnetically confined, ionised gas and hot spots that mirror the observables expected from interacting binary-channel remnants. They also efficiently flag WD+WD systems even in crowded cluster fields, while joint optical/IR constraints help reject non-degenerate donors \citep{2025arXiv250713850C}. Moreover, cluster-anchored age constraints for ZTF~J2008+4449 (a confirmed member of the RSG~5 cluster) indicate a total age of only $\sim 35$ Myr and support a recent binary-origin pathway; in this context, ZTF~J2008+4449 may be the youngest cluster-identified binary-channel WD to date, reinforcing the unique role of phase-resolved UV photometry/spectroscopy in both identifying and physically interpreting such systems \citep{Yan2025}. Recently, the detection of massive WDs in the cluster NGC~362 was found by \citet{2025ApJ...990L..62Y}. As NGC~362 is a core-collapsed cluster, the authors argue that it may host massive WDs through mergers or collisions.
Very recently, \citet{Gupta2025ApJ} investigated deep {\sl HST} UV CMDs of NGC~2808 and found an excess (60--70 per cent) of luminous WDs compared with predictions from evolutionary models. NGC~2808 is only the third GC known to show such a surplus of WDs, after M13 and NGC~6752 \citep{Chen2021NatAs,Chen2022ApJ}. The authors suggest that this excess might be explained by the presence of either a group of slowly cooling CO-core WDs that are embedded in a hot hydrogen envelope \citep{Chen2021NatAs} or He-core WDs \citep{Calamida2008ApJ}. The origin of these luminous WDs is likely linked to NGC~2808’s extended blue HB and its complex stellar chemistry. In addition, binary interactions could have played a role. 

\subsubsection{Blue stragglers and blue lurkers}
\label{lurkers.sec}

Unlike in OCs, the formation of BSS in GCs is largely driven by the combined effects of binary evolution and dynamical interactions. The dense and highly evolved environment of GCs enhances both binary and collisional formation channels, but increasing observational and theoretical evidence suggests that mass transfer in binaries is the dominant mechanism producing BSS and Blue Lurkers (see also Sections \ref{OC.BSS.sec} and \ref{OC.lurker.sec}), a conclusion that holds across a wide range of cluster parameters. 
A recent study showed that the number of BSS normalised to the sampled luminosity in 48 Galactic GCs exhibits environmental dependences remarkably similar to those of the cluster binary fraction, decreasing with increasing central density, collision rate, and dynamical age, while showing no significant correlation with other cluster parameters such as metallicity, age, distance, or reddening. These correlations indicate that low-density regions, possibly owing to a higher binary production and survival rate, are the natural habitat of both BSS and binary systems, and that most BSS have a binary-related origin mediated by environmental conditions \citep{Ferraro2026}.

Earlier {\sl HST}/ACS imaging of dynamically evolved clusters, such as M30 and NGC~362, revealed two distinct BSS sequences \citep{2009Natur.462.1028F, Dalessandro2013b}. The appearance of such bifurcated BSS sequences in GC CMDs was initially interpreted as the simultaneous enhancement of two distinct formation mechanisms during core collapse, with each sequence corresponding to a different mechanism \citep{2009Natur.462.1028F}. In this scenario, the narrow blue sequence is attributed to BSS formed through direct stellar collisions during core collapse, while the broad red sequence represents BSS originating from binary evolution. However, the discovery of double BSS sequences in young clusters suggests a different explanation instead: the bifurcation may reflect different mass-transfer efficiencies and evolutionary stages \citep{Wang2025ApJ...984...52W}. Ongoing mass transfer in close binaries produces the red sequence, while post-mass-transfer BSS+WD systems form a bluer, UV-brighter branch. Binary population synthesis shows that binary evolution can explain the diversity of BSS sequences across clusters of different ages and dynamical states, with UV data providing key constraints on the contributions of each formation channel \citep{Jiang2017, Jiang2022, Wang2025ApJ...984...52W}. In this unified binary-evolution scenario, Blue Lurkers can also be reasonably explained as BSS+WD systems formed via highly non-conservative mass-transfer processes \citep{Leiner2019ApJ...881...47L, Wang2025ApJ...984...52W}. A number of Blue Lurkers and BSS+WD systems are discovered in NGC 362 \citep{2023MNRAS.523L..58D, Dattarey2023ApJ...943..130D}, a GC that is believed to have undergone a recent core-collapse.

UV observations, especially far- and near-UV photometry and spectroscopy, provide crucial diagnostic tools for unifying the mass-transfer formation mechanisms of BSS and Blue Lurkers in GCs. Far- and near-UV observations can directly detect the hot remnants left by mass transfer, such as WDs and sdB stars, and, through UV--optical SEDs, quantitatively separate hot companions and determine their cooling ages. This enables both individual objects and populations to be mapped to specific formation channels. Within this binary framework, BSS and Blue Lurkers are part of a continuous mass-transfer sequence: BSS appear above the MSTO with higher luminosity, while Blue Lurkers are hidden at or just below the MSTO but can be identified by subtle far-UV excess and rotational properties. UV observations integrate far-UV excesses, chemical fingerprints, and WD cooling timescales into a comprehensive toolkit, connecting binary-evolution physics (mass-transfer evolutionary stages and efficiency distributions) with cluster dynamics \citep{Reggiani2025}.

\begin{figure}
    \centering
    \includegraphics[width=0.9\linewidth]{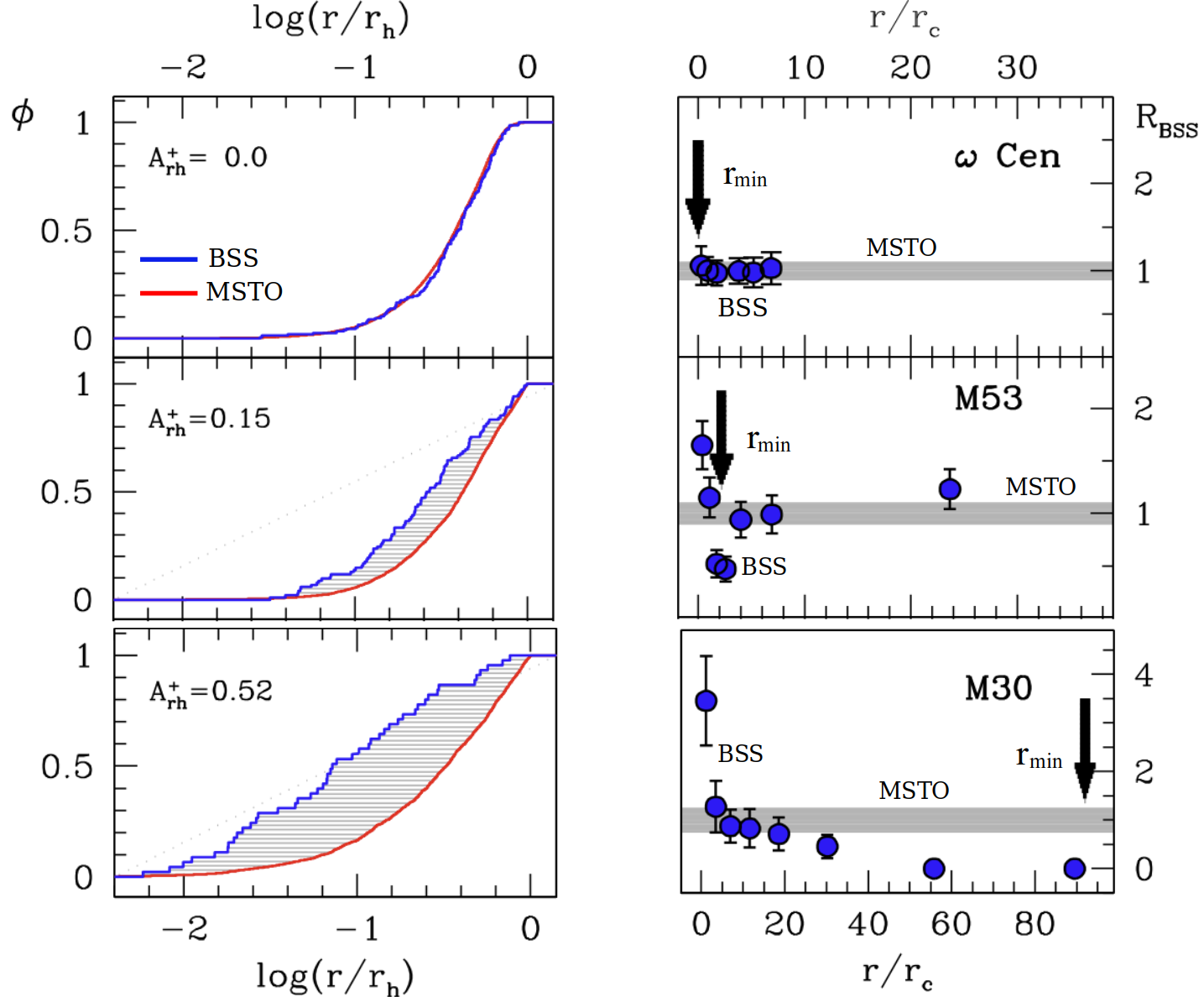}
    \caption{(Left) Cumulative radial distribution functions (CDFs; $\phi$) of BSSs (solid blue line) compared to MSTO stars (solid red line) within the half-light radius ($r_\mathrm{h}$) of GCs. The clusters are arranged in order of increasing area between the two CDFs, defined as $A^{+}{r\mathrm{h}}$, which quantifies the degree of dynamical evolution driven by mass segregation; larger $A^{+}_{r\mathrm{h}}$ values correspond to more dynamically evolved systems. (Right) Normalised radial distribution of BSSs ($R_{\mathrm{BSS}}$; blue circles) compared with that of MSTO stars \citep[grey shading, shown for reference;][]{2016ApJ...833L..29L}. The black arrows mark the location of the minimum in the BSS radial distribution ($r_{\rm min}$), which shifts outwards as dynamical friction causes the more massive BSS to segregate towards the cluster centre. Together, $A^{+}_{r\mathrm{h}}$ and $r_{\rm min}$ serve as complementary diagnostics of cluster dynamical evolution, forming the basis of the BSS `dynamical clock'.}
    \label{fig:bss_dyn}
    \vspace{-0.5cm}
\end{figure}

The use of UV imaging observations to study BSS dates back to the study of the Galactic GC M3 with {\sl OAO-2} \citep{1974ApJS...28..157G}. Subsequently, {\sl HST} UV imaging of the same GC \citep{1997A&A...324..915F} led to the discovery of a normalised bimodal radial distribution of the cluster's BSS, which is indicative of an intermediate dynamical cluster age. Normalised BSS radial distributions with respect to another massive reference population, such as the HB in GCs and subgiant branch (SGB), red-giant branch (RGB), and/or main-sequence turn-off (MSTO) stars in OCs, have been used to probe the dynamical ages of both OCs and GCs \citep{2009Natur.462.1028F, 2015ASSL..413...99F, 2016ApJ...833L..29L, 2018ApJ...860...36F, 2021MNRAS.508.4919R, 2023MNRAS.526.1057R}.

{\sl HST} established BSS as precise dynamical tracers of GC evolution by enabling accurate measurements of their spatial distributions and luminosity functions in dense cluster environments. In dynamically evolved (relaxed) clusters (where the half-mass relaxation time $t_{\mathrm{rh}} \ll t_{\mathrm{age}} \sim 12$~Gyr), repeated two-body encounters drive mass segregation, causing more massive stars to sink towards the centre through dynamical friction. Since BSS are typically $\sim$1.2--1.5 times more massive than MSTO stars, they experience this effect more strongly and thus provide an excellent probe of internal dynamical evolution. High-resolution near-UV--optical imaging with {\sl HST}/WFPC2 and ACS enabled the construction of cumulative radial distribution functions for BSS in dozens of GCs, revealing systematic variations in their central concentration relative to normal stars \citep[e.g.,][]{ferraro2003,Ferraro2012Natur.492..393F,2016ApJ...833L..29L}. 

These observations led to the development of the BSS `dynamical clock' \citep{Ferraro2012Natur.492..393F}, an empirical framework in which the degree of BSS mass segregation provides a measure of a cluster's dynamical age. As dynamical evolution proceeds, the BSS radial distribution transitions from a nearly flat profile in dynamically young clusters to a bimodal distribution at intermediate stages, and eventually to a centrally concentrated profile in dynamically old systems. As shown in Figure~\ref{fig:bss_dyn}, dynamically young clusters such as $\omega$~Cen ($t_{\mathrm{rh}} > 10^{10}$~yr) exhibit nearly flat BSS radial distributions, while clusters such as M53 and 47~Tucanae ($t_{\mathrm{rh}} \sim 10^8$--$10^9$~yr) display bimodal profiles characterised by a prominent central peak and an outer rise \citep{2016ApJ...833L..29L}. By contrast, dynamically old clusters such as M30 ($t_{\mathrm{rh}} \simeq 10^{7}$~yr) show a single central peak, consistent with advanced mass segregation and a dynamically evolved state.

Subsequent UV observations from the {\sl HST} UV Legacy Survey \citep{Raso2017,2018ApJ...860...36F}, using WFC3/UVIS imaging, refined this framework by providing homogeneous near-UV photometry of BSSs across a large sample of clusters. These data revealed that clusters with higher central BSS concentrations systematically host hotter and more massive BSS populations, linking their photometric properties to the dynamical state of the host cluster. UV observations therefore not only refined the empirical dynamical clock but also firmly established BSSs as key tracers of stellar interactions, binary evolution, and energy redistribution processes within GCs.

\subsubsection{Reflections}
UV observations of GCs are crucial to decipher the details of the end-state evolutionary stages of low-mass stars. The exotic populations in the central regions of GCs are easily detected even with the moderate spatial resolution of UVIT, leading to the detection of the hot HB population, BSS, WDs, and other species. The spectroscopic data of these populations in the UV are lacking, which is essential to decipher their surface chemistry. The flattening of the HB as seen in the UV CMD (Right panel of fig. \ref{fig:stack_cmd} ) is attributed to atomic diffusion resulting in the increased surface opacity. Tracing the surface chemistry to understanding diffusion requires UV spectroscopy. 

The integrated UV luminosity of GCs can put a limit on the UV flux that can be generated by the old stellar population and connect it with the UV-upturn in elliptical galaxies. The upcoming UV missions, such as ULTRASAT and UVEX, will be able to provide UV flux of more GCs, whereas the HST/UVIT data will still be required to resolve the inner stellar population, highlighting the importance of good spatial resolution in UV imaging.

\section{Star clusters in the Magellanic Clouds}

Star clusters in MCs are located in a metal-poor, gas-rich and shallow gravitational potential environment compared with the Milky Way. The proximity and location of MCs at high Galactic latitudes allow us to resolve faint stellar populations in their star cluster systems, thus enabling us to study their stellar IMF, whereas their metallicity and dynamical state provide a contrast to Galactic conditions. In particular, the UV regime offers a critical window into these systems, since UV radiation traces the hot, young and massive (O- and B-type) stellar populations, as well as evolved stellar populations with strong UV fluxes, such as post-AGB and HB stars. Even the signatures of interactions in the inter-Cloud region (the Magellanic Bridge) are ideal sites for investigating the origin of UV-bright stellar populations in a metal-poor, low gas density and tidally distorted environment. Such conditions are apparently not conducive to star formation or, at least, the formation of massive stars.

Identification and parameterisation of star clusters in the MC have been the subject of extensive investigation for several decades. Over 4000 star clusters in the MCs have been identified, particularly by \citet{Bica+1999AJ, Bica_2008, Bica2020}, which has led to the creation of an updated MC cluster catalogue. With the advent of optical and near-IR sky surveys over the last three decades, the number of newly discovered star clusters has increased substantially. Starting from 1~m-class facilities \citep[Optical Gravitational Lensing Experiment--II, Magellanic Cloud Photometric Survey/MCPS;][]{Udalski+1997, Zaritsky+2002AJ, Zaritsky+2004} since the start of the year 2000, through deep near-IR surveys like the Vista Survey of the Magellanic Clouds \citep[VMC;][]{Cioni2011} and deep optical surveys like the Survey of the MAgellanic Stellar History \citep[SMASH;][]{SMASH2017} using 4~m-class facilities in more recent times. Each generation of surveys has expanded the breadth of cluster catalogues. Of all these surveys, only the MCPS included the $U$ band, as one of the closest filters to the UV regime. 

The goals of using star clusters as tools to study the MCs have been manifold: from reconstructing the star-formation history of the MCs \citep[e.g.,][]{PU2000, Nayak2016, Nayak2018, Piatti+2018MNRAS, dhanushp1}, discovering low-mass open cluster-like clusters in the MCs \citep[e.g.,][]{Choudhury2015, Nayak2016, Nayak2018} to decipher the IMF, derivation of age--metallicity relations and estimations of chemical enrichment and metallicity gradients in the MCs. The only dedicated recent survey focused on star clusters in the MCs is the the VIsible Soar photometry of star Clusters in tApii and Coxi HuguA \citep[VISCACHA;][]{VISCACHA+2019}, a survey undertaken in the $V$ and $I$ bands using the 4~m SOAR telescope. Their studies have helped decipher {\it in situ} star formation in the Magellanic Bridge, discovered distance variations across the Bridge and uncovered metallicity variations. UV studies of star clusters in the MCs, both photometric and spectroscopic, have focused on a few major areas: unravelling the phenomenon of MPs outside our Galaxy, constraining the IMF and estimating properties of massive (O-type) stars (e.g., their masses, mass-loss and/or rotation rates, stellar winds) in the context of developing stellar evolutionary models, understanding binary evolution and using them as the closest analogues of the first stars that re-ionised the Universe. 

\subsection{UV Signatures of Multiple Stellar Populations in MC Clusters: Insights from photometric and spectroscopic studies} \label{MPs in MCs}
Even after decades of research, the origin of MPs in GCs remains an enigma. The existence of MPs is characterised by split or broadened RGBs in CMDs of Galactic GCs exhibiting variations in He and light-element abundances. Rich (massive) GCs in the MCs provided an opportunity to explore MPs in clusters aged $<$10--12 Gyr and with cluster masses one order of magnitude smaller than those of Galactic GCs. In recent years, several studies have been dedicated to the study of MPs in GCs in the MCs by combining UV filters (e.g., F275W, F336W, F343N) with {\sl HST} optical and near-IR filters, and distinguishing different populations using CMDs and/or pseudo-colour--colour diagrams, and using spectroscopic data to reveal chemically distinct subpopulations (C--N--O variations and measurable He variations) in MC GCs, with ages ranging from 2 to 11~Gyr.

As an example, \citet{Dalessandro+2016} found that NGC~121, the only GC in the SMC similar in age to the Galactic GCs (11~Gyr), showed a clear broadening and split of its RGB. The authors used F336W versus differential colour distribution ($\Delta_{C(\mathrm{F336W, F438W, F814W})}$) CMDs and Very Large Telescope (VLT)/UVES--FLAMES spectroscopic data to detect light-element abundances and found their variation to be similar to that of Galactic GCs. \citet{Niederhofer+2017} carried out a survey using the {\sl HST} UV (F336W, F343N), optical (F438W, F555W) and near-IR (F814W) bands for MC clusters with masses on the order of $10^5 M_\odot$ and ages between 100~Myr (e.g., NGC~1850) and 11~Gyr (e.g., NGC~121). Their results for NGC~121 confirmed the findings of \citet{Dalessandro+2016}. The authors used CMDs and differential colour moduli to detect a split in the RGB of NGC~121, suggesting two populations with different abundances: one nitrogen-rich but carbon-poorer than the other. The colour spread along the HB of NGC~121 suggested a He abundance spread ($\Delta\,Y = 0.025$). Studies by \citet{Lagioia+2019a} explored He variations in four SMC GCs aged 6--11~Gyr (NGC~121, NGC~339, NGC~416, and Lindsay~1), determining slight ($\Delta\,Y =0.01$) He variations between their pairs of different RGB populations. They also found that the 2P stars in all clusters are, on average, enhanced in nitrogen and depleted in carbon and oxygen, similar to those in Galactic GCs. Thus, studies have well established that cluster mass plays a role in the onset of MPs in GC-like objects, with star-to-star variation in light-element abundances becoming more prominent in higher-mass clusters \citep[see][]{Milone2017MNRAS.464.3636M}. Similar indications reflected in studies of MPs in rich MC clusters, where clusters with ages similar to GCs were found to exhibit light-element variations and multiple RGBs in CMDs \citep[e.g.,][]{Dalessandro+2016, Niederhofer+2017, Lagioia2019b}.

However, some studies in rich MC clusters aged $\sim$1.7--2 Gyr, i.e., significantly younger than in Galactic GCs, indicated the presence of MPs \citep[for example, see][]{Martocchia+2018,Kapse2022ApJ}. These works highlighted that age, too, could play a role in defining the onset and properties of chemical anomalies. The variations in light-element abundances appeared to vary as a function of age. For example, from the analysis of RGBs width and comparison with spread due to photometric errors for massive MC clusters, \citet{Martocchia+2019} found indications that older clusters (ages similar to Galactic GCs) show larger nitrogen spread as compared with younger counterparts ($\sim$2~Gyr). Thus, both age and mass appear to correlate with the occurrence of MPs; however, the physical processes involved in their formation are not well understood. Studies suggest that the possible astrophysical processes could be an unknown effect of stellar evolution for low-mass stars with RGB mass of $\sim1.5\,M_\odot$ at $\sim$2~Gyr or due to environmental effects within a galaxy.

Results from {\sl AstroSat}/UVIT have opened up a window for using UV photometric bands beyond the {\sl HST} to identify signatures of MPs in intermediate-age star clusters. By combining near-UV (N242W; 200--280 nm) data from UVIT and optical data from the {\sl HST}, \citet{Nayak2021MNRAS.503.5291N} reported a spread in the distribution of red clump stars in colour and magnitude in the near-UV versus (near-UV$-$optical) CMD of Kron~3, a cluster aged 6--8~Gyr with highly debated metallicity estimates. The authors argue that the red clump spread is not caused by contamination by field stars or differential reddening, but possibly by the presence of MPs that correspond to a small range in age (6.5--7.5~Gyr), metallicity (0.20~dex) and a variation in initial He abundance ($Y_\mathrm{ini}=0.23$--0.28). 

\subsection{Rotation, Extended MSTOs and UV-Dim Stars}
Extended MSTOs (eMSTOs) in young and intermediate-age ($\leq2$ Gyr-old) clusters are now widely interpreted as reflections of a distribution of stellar rotation rates rather than large age spreads \citep{BastianDeMink2009, Li2014, Niederhofer2015, Georgy2019}. Rotation modifies internal mixing and stellar lifetimes, while gravity darkening and inclination effects reshape observed colours and magnitudes, producing broadened or bifurcated MSTOs \citep{Cordoni2018}. UV passbands accentuate temperature- and line blanketing-driven colour shifts, enhancing the diagnostic contrast between slow and fast rotators \citep{Georgy2014, Milone2018MNRAS.481.5098M, Kamann2020}. Joint UV--optical SED fitting, CMD morphologies and spectroscopic $v\sin i$ measurements constrain the rotation distribution and its correlation with eMSTO structure \citep{Dupree2017, Marino2018}. In several clusters, the inferred fraction of fast rotators correlates with the extent of UV colour spreads, reinforcing rotation as the primary driver of eMSTOs \citep{Milone2018MNRAS.481.5098M, Sun2019, Kamann2020}.

While rotation successfully reproduces the broad eMSTO morphologies observed in most young clusters, high-precision UV photometry has revealed that the upper main sequence and turn-off region host additional stellar populations that do not fit standard single-star or standard rotating models. These so-called UV-dim stars, characterised by distinctly red UV--optical colours, were first identified within the eMSTO regions of young MC clusters. Whether they represent a distinct formation channel---such as binary evolution or dust-enshrouded fast rotators---that contributes directly to the extended main sequence, or merely a contaminating population whose presence complicates the eMSTO diagnosis, remains actively debated. We therefore examine their observed properties and possible origins before returning to their connection with cluster age and environment.

\begin{figure}
    \centering
    \includegraphics[width=0.6\textwidth]{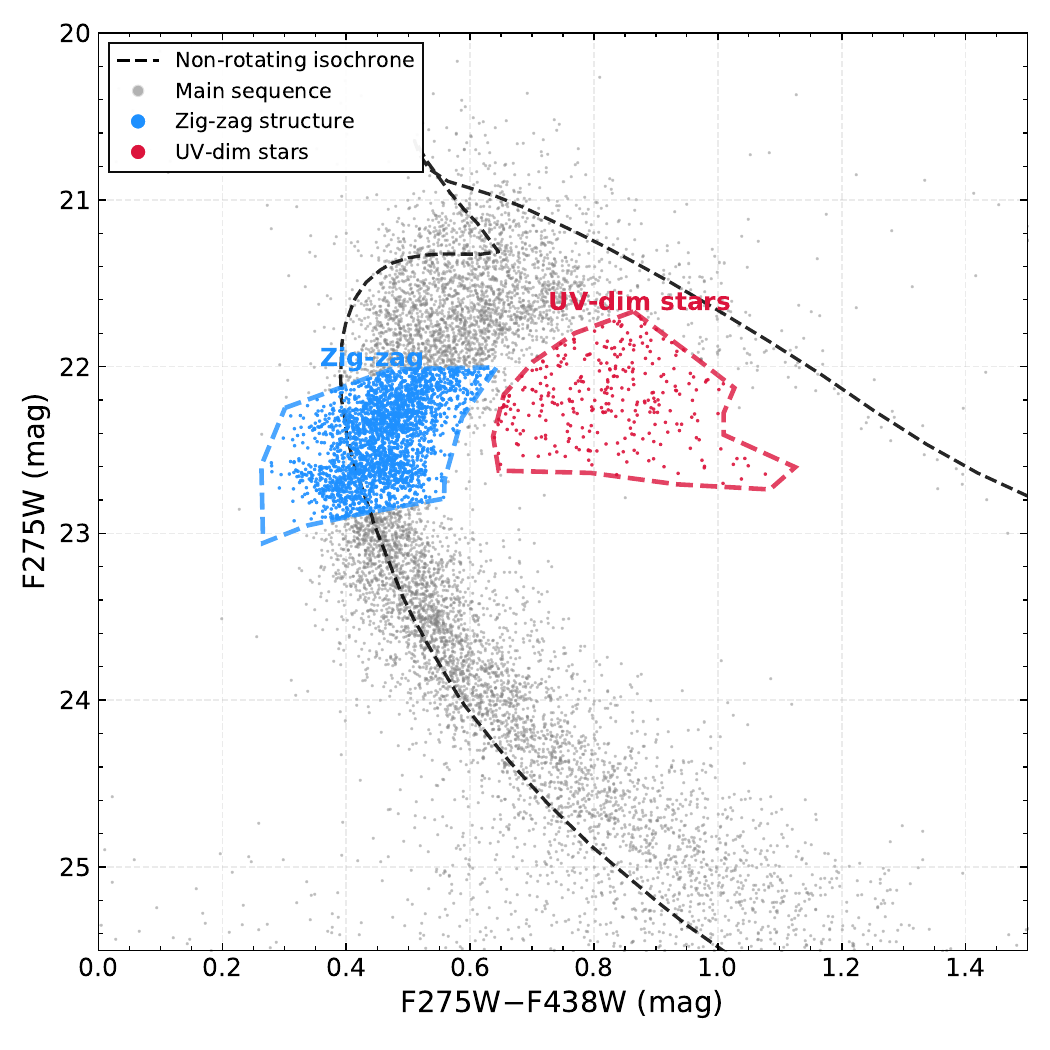}
    \caption{F275W versus F275W$-$F438W colour--magnitude diagram of the LMC cluster NGC~1783. 
    \textit{Grey points}: main-sequence and turn-off stars. 
    \textit{Blue points}: stars belonging to the zig-zag structure on the blue side of the main sequence. 
    \textit{Red points}: UV-dim stars redward of the bulk sequence. 
    The black dashed line shows a non-rotating isochrone for an age of $\sim$1.7~Gyr, ${\rm [Fe/H]}=-0.5$,  distance modulus $(m-M)_0=18.5$, and $A_V=0.18$~mag. 
    The blue and red dashed polygons highlight the approximate locations of the zig-zag and UV-dim populations, respectively. 
    (Data from A.~P.~Milone.)}
    \label{fig:uv_dim}
\end{figure}

UV-dim stars first appeared in {\sl HST} WFC3/UVIS F275W imaging \citep{Milone2023A&A...672A.161M} of the intermediate-age ($\sim$1.8 Gyr-old) LMC cluster NGC~1783, where they were detected as a discrete cloud of A-type stars lying redwards of the bulk main sequence in (F275W$-$F438W) versus (F336W$-$F814W) space \citep{Milone2022Univ....8..359M,Milone2023A&A...672A.161M} (see Figure~\ref{fig:uv_dim}). Their location within the eMSTO region immediately raised the question of whether such stars contribute to, or even partially drive, the extended main sequences seen in clusters younger than 2 Gyr. Next, \citet{Milone2023MNRAS.524.6149M} analysed UV observations of a few young MC star clusters (NGC~1805, NGC~1818, NGC~1850 and NGC~2164 in the LMC; NGC~330 in the Small Magellanic Cloud/SMC), and suggested that the presence of UV-dim stars is a property of young ($\sim$200~Myr-old) MC star clusters. UV-dim stars are much redder than the bulk of stars when the colour index includes the {\sl HST}'s F225W or F275W UV filters, whereas they are well-mixed with other stars in optical bands \citep{Milone2023A&A...672A.161M,Milone2023MNRAS.524.6149M}. They occupy the bluer part of the UV--optical CMD, but they are observed as a separate sequence in colour--colour diagrams (e.g., (F225W$-$F336W) versus (F336W$-$F814W) or (F275W$-$F336W) versus (F336W$-$F814W)). 

Studies suggest that they may be Be stars, in which the UV dimming is caused by dusty circumstellar material, and that the star's orientation with respect to the line of sight affects this phenomenon \citep{2023MNRAS.521.4462D}. In such cases, the stronger absorption of the UV photons by the dust is likely to cause the UV-dim phenomenon. Fast rotation was suggested to be responsible for the formation of the circumstellar discs, and associated mass loss may contribute to the `zig-zag' morphology found in the UV CMD of the upper main sequence of NGC 1783 \citep{2023MNRAS.521.4462D}. This dust-extinction scenario was supported by the redder colours of shell Be stars (Be stars viewed through discs) than those of Be stars viewed pole-on \citep{2023MNRAS.518.1505K}. 

A significant drop in the fraction of UV-dim stars was found in MC star clusters older than $\sim 2$ Gyr, indicating that UV-dim stars may be correlated with fast-rotating early-type stars whose rotation is not magnetically braked \citep{2023MNRAS.520.4080M}. In the Milky Way, a population of A--F-type stars with H$\alpha$ emission and fast rotation was detected in the young ($\sim 340$ Myr-old) star cluster NGC 3532 \citep{2025ApJ...979..246H}. They may be possible counterparts of the UV-dim stars in the LMC cluster NGC 1783, based on the dust-extinction scenario. However, that latter scenario, in particular the proposed fast-rotating feature, seems inconsistent with several observations. In young MC star clusters, most UV-dim stars have been found along the blue main sequence in optical CMDs; stars along the blue main sequence are believed to be slowly rotating \citep{Milone2023MNRAS.524.6149M}. A recent spectroscopic study of UV-dim stars in NGC 1783 indeed confirmed that they are mainly slowly rotating stars \citep{2025A&A...698A..27L}. Moreover, a photometric study of 35 Galactic OCs combining data from {\sl Swift}/UVOT, SkyMapper and {\sl Gaia} photometry found rare UV-dim stars in Galactic OCs, indicating that their appearance may be affected by the environment where they formed \citep{2025MNRAS.541.2790C}. How UV-dim stars are formed is still an open question. Whether the UV-dim phenomenon is correlated with circumstellar discs, stellar metallicity, or star cluster masses remains to be explored further through theoretical and observational work.

\subsection{Massive Stars at Low Metallicity: Insights from spectroscopic studies and UV surveys}

Metal-poor stars are likely progenitors of energetic transients, including superluminous supernovae, long-duration gamma-ray bursts and coalescing massive black holes. The MCs are the nearest laboratories that facilitate studies of such metal-poor ($Z=0.5$ Z$_\odot$ for the LMC, $Z=0.2$ Z$_\odot$ for the SMC, and $Z< 0.2$ Z$_\odot$ in the Bridge) massive stars, i.e., of OB-type ($M\gtrsim8$ M$_\odot$; e.g., \citealt{Poelarends2008}). These stars emit copious amounts of far-UV radiation from $\lambda = 91.2$ nm to $\lambda \approx 200$ nm. They influence young star clusters and star-forming regions through ionising UV radiation, and stellar winds which impart mass, momentum and heavy elements. As an example, the Wolf--Rayet (WR) stars are evolutionary descendants of massive O-type stars and have a significant impact on their environments owing to their strong ionising radiation fields, and the MCs have been a key laboratory for studying them. Their studies using UV and optical spectra have revealed that their population in the SMC are strikingly different from their counterparts in environments with higher metallicity, such as the MW, LMC, and M~31 \citep{Hainich2015A&A}, revealing fundamental differences in how massive stars evolve under metal-poor conditions.

A series of significant UV photometric surveys, spectroscopic surveys and their follow-up analysis over the past decade---primarily the VLT--FLAMES Tarantula Survey \citep[VFTS;][]{VFTS2011}) in the LMC, the Runaways and Isolated O-Type Star Spectroscopic Survey of the SMC \citep[RIOTS4;][]{RIOTS42016} and the Binarity at LOw Metallicity campaign \citep[BLOeM;][]{BLOeM2024}---have transformed our understanding of O-type stellar populations, which dominate the UV flux in the MCs. Also, the recently initiated Hubble Ultraviolet Legacy Library of Young Stars as Essential Standards \citep[ULLYSES;][]{Roman2025ApJ...985..109R} programme is a major and unique effort aimed at producing a UV spectroscopic library of massive (OB-type) stars in the MCs and a few more nearby, metal-poor galaxies (Sextans A: 0.10 Z$_{\odot}$; NGC~3109: 0.15 Z$_{\odot}$).

\subsubsection{Insights from spectroscopic studies}

The VFTS survey delivered a comprehensive multi-epoch spectroscopic (3960--6200 {\AA}) census of $\approx$800 massive stars in the 30~Doradus (30~Dor) region of the LMC, which included $\sim$300 O-type stars in and around the young (1--2 Myr-old) star cluster Radcliffe~136 (R136). The survey enabled the first uniform assessment of binaries in massive stars in our closest starburst environment. Further follow-up studies, for example \citet{Sana2013}, analysed 360 O-type stars and, after correcting for observational biases, they derived an intrinsic spectroscopic binary fraction of $\approx 50$ per cent. This value of the binary fraction was later revised to $\approx 60$ per cent \citep{Almeida2017A&A}. These results indicate that binary evolution (mass transfer/mergers) affects about half of all O-type stars, underscoring the importance of identifying binaries among massive stars in the MCs and beyond. With regards to rotational velocities of massive stars, \citet{RamirezAgudelo2013} investigated their distribution extending up to high-velocity tails that reach up to 600~km s$^{-1}$. These fast-rotating O-type stars are consistent with products of early binary interactions, particularly mass transfer and mergers. Additional signatures supporting such an origin include surface abundance anomalies (e.g., nitrogen enrichment), runaway velocities, and the presence of stripped companions detectable through UV spectroscopy. These systems are also relevant for investigating gamma-ray burst progenitor pathways. Meanwhile, the abundance of slowly rotating massive stars suggested the possibility of an early braking mechanism operating immediately after their formation. 

Further studies of the most massive VFTS stars \citep[$\gtrsim 100 M_{\odot}$;][]{2014A&A...570A..38B} have proven their role in powering 30~Dor’s ionisation. \citet{Brands2022A&A} simultaneously analysed optical and UV spectroscopy of massive stars in R136, and derived an initial stellar mass of $\gtrsim\,250 M_{\odot}$ for the most massive star (R136a1) in their sample. In fact, recent UV analysis combined with dedicated spectroscopic binary monitoring in the optical supports current masses of at least $\sim 150$--$200 M_{\odot}$ \citep{Shenar2023A&A} in R136, ruling out companions more massive than $\sim 50M_{\odot}$. Thus, UV spectroscopy and complementary datasets have helped set a record for the most massive stars at the LMC's metallicity.

Our observational understanding of low-metallicity (low-$Z$) massive stars has been strengthened by moving beyond the LMC and conducting spectroscopic studies in the SMC \citep[for example, see][]{Bouret2013A&A, Dufton2019A&A, Ramachandran2019A&A}. Only a few massive star-forming regions at sub-SMC metallicities have been observed in distant ($\gtrsim1$~Mpc) dwarf galaxies, such as Leo~P \citep[see references in][]{Telford2023ApJ}. Similarly, a catalogue of OB-type stars (over 150 stars) was presented for Sextans~A, which shows a dearth of objects with masses $\gtrsim 30M_{\odot}$ \citep{Lorenzo2022MNRAS}.

The focus of spectroscopic surveys of massive stars in the SMC has been on creating a spectral library for massive stars, their spectral classification, detecting binaries, estimating radial velocities and deciphering the origin of runaway OB stars. In this context, the RIOTS4 survey carried out a spatially complete survey of $\sim 400$ uniformly selected field OB stars (including a small fraction located in the SMC's bar) with the help of the IMACS and MIKE spectrographs on the 6.5~m Magellan Telescope. The survey resulted in spectral classifications and radial velocity estimates. Their multi-epoch spectroscopic data encompassed 29 stars that had 17 normal OB stars (non-Oe/Be). Out of these seventeen massive stars, $\approx$60 per cent were found to be members of spectroscopic binaries, consistent with results in the Galaxy. These studies confirmed the existence of a steep upper IMF in the SMC field, apparently caused by the inability of the most massive stars to form in the smallest clusters \citep{Schootemeijer2021}.

The most recent BLOeM campaign, a European Southern Observatory large programme, is designed to obtain multiple-epoch spectroscopy of over 900 massive stars in the SMC using the VLT/FLAMES. The survey aims to provide the binary fraction and orbital configurations of systems with periods $<$3~yr, identify dormant black-hole binary candidates and create a legacy database of physical parameters of massive stars at low metallicity ($Z=0.2$ Z$_\odot$). Recent results from this campaign by \citet{Sana2025NatAs} suggest that at least $\sim\,70$ per cent of their carefully selected sample of O-type stars from BLOeM campaign are in close binaries, and that $\sim\,68$ per cent of all O-type stars interact with their companion during their lifetime. Furthermore, the authors found no statistically significant trend in multiplicity properties with metallicity, suggesting that multiplicity and binary interactions govern the evolution of massive stars and determine their cosmic feedback and explosive fates.

There are other subcategories of massive stars that are UV-bright, and UV investigations in the MCs aim to play a key role towards our understanding of their formation and evolution. For example, OBe-type stars are extremely fast-rotating OB-type stars that have circumstellar gaseous decretion discs around them, which cause characteristic emission lines in their spectra. The formation pathways for such fast rotators remain an open question and could result from binary interactions or from the evolution of a single star. In the past, \citet{Schootemeijer2022A&A} used UV photometry with optical datasets to create a census of OBe type stars in the MCs and nearby metal-poor star-forming dwarf galaxies. They found that their fraction increases with decreasing metallicity, with the highest in Sextans~A (~$30$ per cent). The recent spectroscopic investigations by BLOeM of SMC OBe stars have attempted to constrain understanding of their formation channel and suggest that these stars are mostly post-interaction products \citep{Bodensteiner2025A&A}. However, to address the open question of their formation channel, it is necessary to conduct UV observations beyond the Galaxy and MCs, and compare them with stellar evolutionary models to decipher the formation channels of such stars.

\subsubsection {Insights from UV photometry and spectroscopic studies in the Magellanic Bridge}

Photometric sky surveys in the far- and near-UV bands with {\sl GALEX} \citep{Martin2005ApJ...619L...1M} have played a critical role in identifying massive stars in inter-Cloud regions and in estimating the UV-ionising photon flux. Based on near-UV photometric data from {\sl GALEX}, $V$- (APASS) and $J$-band (2MASS) data, and proper-motion catalogues, \citet{Casetti-Dinescu2012} used colour--colour diagrams to select OB-type candidates and create their spatial distribution in a 7900$^\circ$ area of the MCs to map recently formed stars in the MCs' periphery and the Magellanic Bridge, and to explore the presence of such stars in the Leading Arm and Magellanic Stream. The Bridge region, located closer to the SMC's western area, was found to be densely populated with such candidates compared with other locations. The spatial distribution of OB-type candidates in the Bridge and the \citet{Bica_2008} catalogue of extended objects suggested a stronger correlation between OB-type candidates and associations, rather than with star clusters and emission nebulae. 

Recently, three O-type stars were discovered ($M\,\approx17$--$25\,M_\odot$, $T_\mathrm{eff}\,\approx 33,000~K,\, \mathrm{log}g\,\approx\,3.5--4.1,$ $[Fe/H]=-0.60$ to $-1.35~dex$) in the Bridge \citep{Ramachandran2021}, where one star had a mean metallicity similar to that of the LMC ($-0.25$ dex), whereas the other two had SMC-like mean metallicity ($-1.10$ and $-1.36$ dex). In fact, \citet{Schosser2025A&A} studied these two SMC-like metallicity stars by analysing their UV forest of iron lines using high-resolution HST-COS far-UV spectra, to precisely measure their metallicities, and found that they are the most metal-poor O-type stars that have been discovered to date. This indicates the possible lack of LMC--SMC interstellar medium and the presence of a low-density dynamic gas that is still forming stars. More recently, \citet{Choudhury2026MNRAS} examined the first high-resolution far-UV images of seven star clusters in the Magellanic Bridge region using {\sl AstroSat}/UVIT and found their ages to be $\lesssim$100~Myr, indicative of \textit{in situ} star formation in the Bridge due to recent tidal interactions between the MCs, in agreement with previous works. The authors modelled the SEDs of MS stars in these clusters by combining UVIT with complementary datasets, which revealed five hot MS stars with $T_{\mathrm{eff}}$ ranging between 32,500 ($\pm$1250)~K to 42,500 ($\pm$1250)~K and $\log g$ from 3.5 ($\pm$0.1)~dex to 4.0 ($\pm$0.1)~dex, consistent with late O-type stars. Figure~\ref{fig:sed_mc} shows one such star in cluster BS~245. These studies emphasise that O-type can form in extreme environments like the Bridge, which are apparently not conducive for massive star formation given their low-gas density, metal-poor and tidally disrupted gas plus interstellar medium.

\subsubsection {The quest for UV studies: ULLYSES, X-ULLYSES and UV stellar catalogues of the Magellanic Clouds}

The objectives of the ULLYSES programme as regards massive stars are to characterise their winds and photospheres at low metallicity and produce a complete spectroscopic library of massive stars that can be used to test stellar evolutionary theories and enable population synthesis in the metallicity range typical of cosmic noon. To achieve this, the programme obtained medium-resolution {\sl HST}/COS and {\sl HST}/STIS UV spectra of 240 massive stars in the MCs, and six stars in the more distant, even lower-metallicity Local Group galaxies NGC~3109 and Sextans~A (10--20 per cent and 10 per cent solar metallicity, respectively). The MC target stars were strategically selected to uniformly sample spectral types and luminosity classes (including WR stars) while leveraging archival observations.

As a complementary programme to the ULLYSES, the X-ULLYSES \citep{Vink2023A&A} programme has demonstrated that the combination of UV spectra plus optical spectra from the VLT/X-shooter spectrograph is a more powerful framework to estimate properties of massive stars in the MCs, compared with either UV or optical spectra alone. These properties include key parameters such as mass-loss rate, wind structure, and surface abundances, all of which play a crucial role in the evolution of massive stars.

The above programmes have been built on decades of preliminary work on massive stars, which includes characterising stellar winds and mass loss in OB-type stars \citep[see references in][]{Vink2022ARA&A}, surface abundance studies \citep[for example, see][]{Bouret2013A&A}, and attempts to advance our understanding of stellar feedback processes and the ionising output of massive stars \citep[for example, see][]{Doran2013A&A}.

\begin{figure}
    \centering
    \includegraphics[width=0.95\textwidth]{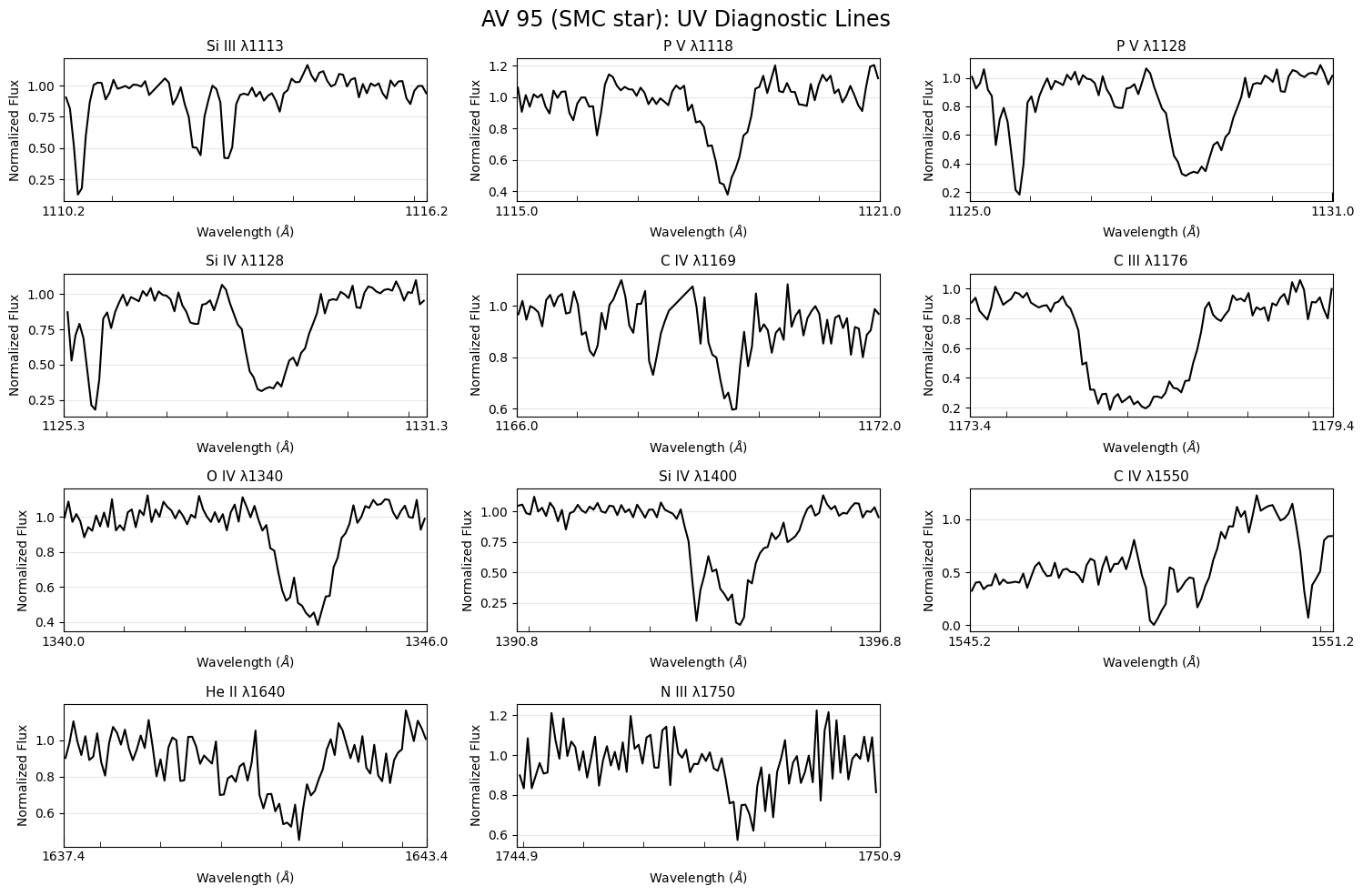}
    \caption{Normalised ULLYSES spectra of an SMC O-type (AV~95) star, with a $T_\mathrm{eff}\,\approx\,38,250~K$ and $\mathrm{log}g\approx3.64$, studied by \citet{Backs2024A&A}. Individual panels show the major lines used for UV diagnosis of the spectra.}
    \label{fig:star_ullyses}
\end{figure}

\begin{figure}
    \centering
    \includegraphics[width=0.7\textwidth]{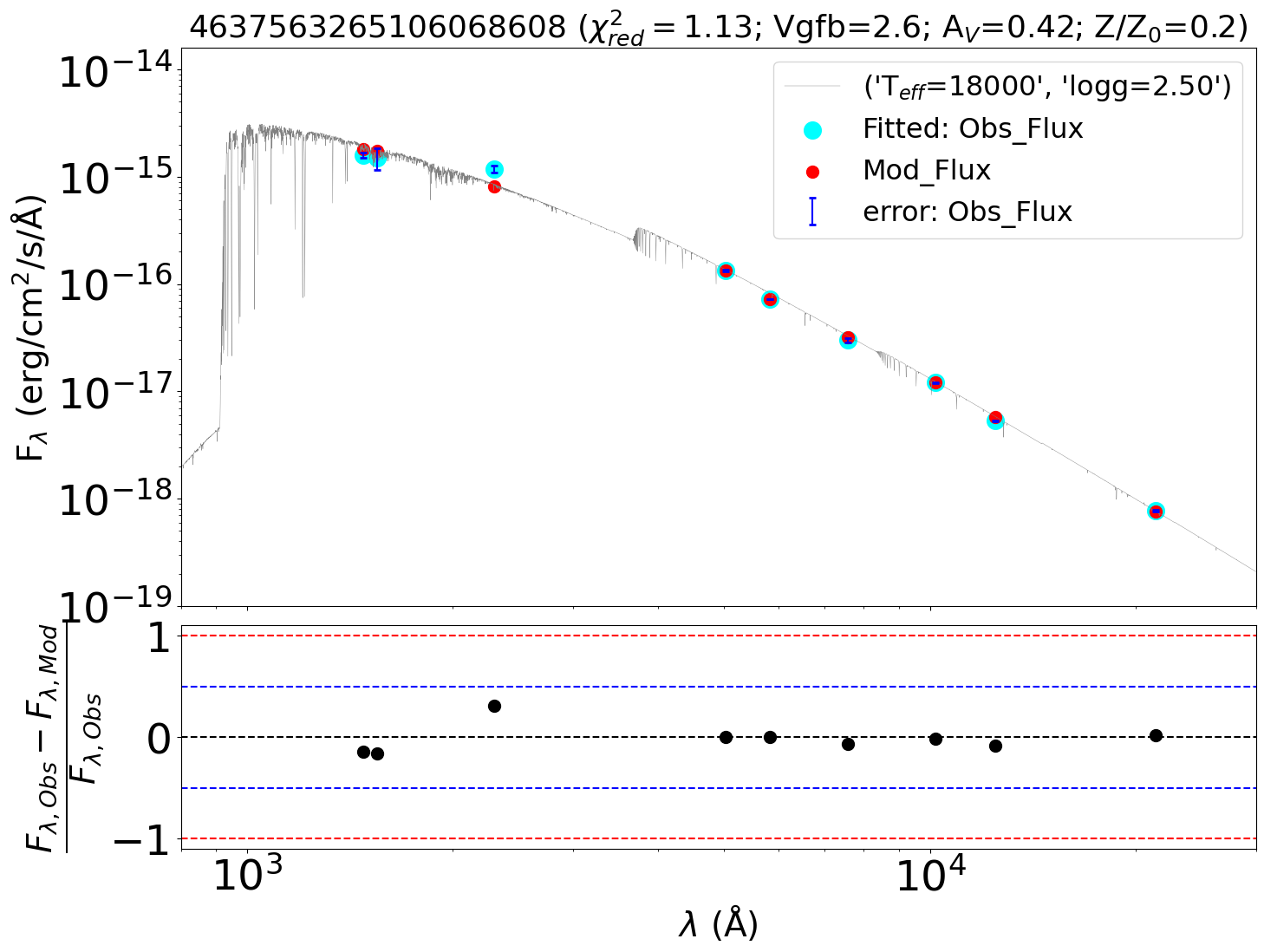}
    \caption{Well-fitted SED constructed from multi-wavelength flux (UV to NIR) of a hot-MS star with $T_{\mathrm{eff}} =42,500\,\pm\,1,250\,K,$ and $\log g =4.0\,\pm\,0.12$~dex, in cluster BS~245, located in the Bridge. {\sc tlusty} atmospheric models (grey line) were fitted to the observed fluxes, assuming a single star. For details, see \citet{Choudhury2026MNRAS}.}
    \label{fig:sed_mc}
\end{figure}

We briefly touch upon some of the key X-ULLYSES results concerning massive stars, published recently. From studies of 13 O-type giants and supergiant stars in SMC, \citet{Backs2024A&A} found that the scaling of the mass-loss rate ($\dot{M}$) with metallicity varies strongly as a function of luminosity, contrary to theoretical predictions. Some of the UV spectroscopic diagnostic line profiles for the star AvZ~95 in the SMC are shown in Figure~\ref{fig:star_ullyses} \citep{Backs2024A&A}. In contrast, studies by \citet{Hawcroft2024A&A} in the LMC show agreement between empirical and theoretical estimations of the mass-loss rate. Together, these results demonstrate that good agreement between empirical and predicted mass loss in one part of the parameter space cannot be extrapolated across the full luminosity--metallicity range. With regard to dependence of terminal wind speed on metallicity ($Z$), the analysis of 150 massive stars in the MCs by \citet{Hawcroft2024A&A} revealed a steeper dependence on metallicity as compared with previous works involving a Galactic sample. Their work further suggested that the effective surface temperature of the star ($T_\mathrm{eff}$), and not escape velocity, may be a better predictor of terminal wind speed. These studies motivate extending wind analyses to stars spanning wider metallicity ranges, across sub-SMC environments, and broader luminosity regimes to better constrain the dependence of wind-structure parameters on stellar properties \citep{Brands2025A&A}. In abundance studies, \citet{Martins2024A&A} found that the surface nitrogen abundance of massive stars is higher in MCs than in the Galaxy. Also, the SMC stars with higher $\log\mathrm{g}$ showed nitrogen enrichment, a trend supported by stellar evolutionary models that incorporate chemical transport by stellar rotation. The authors emphasised the need for an analysis of a larger, unbiased stellar sample to better constrain chemical mixing and angular momentum transport in massive single and binary stars.

Thus, UV spectroscopy combined with optical spectroscopic studies has paved the way for understanding the physics of massive stars. However, such efforts are still small in numbers. As alternate initiatives, high-angular-resolution UV photometric survey data, when combined with optical and NIR photometric surveys, have proven extremely beneficial for identifying massive stars near crowded regions, such as star clusters, and for determining their first-hand stellar parameters, such as, e.g., $T_\mathrm{eff}$ or $\mathrm{log}g$ \citep{Choudhury2026MNRAS}), and populations with UV-excess, using SEDs. As an example, in recent advances on the studies of UV-bright stellar populations in the MCs, \citet{Hota2025} released a far-UV catalogue of young stellar populations in the SMC based on {\sl AstroSat}/UVIT observations. This catalogue will support hierarchical star-formation studies across the MCs, deciphering the most recent star-formation history (100--300~Myr), studying the kinematics of young stellar populations, and estimating stellar parameters using SED diagrams (to identify targets for future spectroscopic studies of OB- and A-type stars). Similarly, \citet{Choudhury2026MNRAS} released a catalogue of FUV-bright MS stars identified using {\sl AstroSat}/UVIT, in seven star clusters in the Bridge, to facilitate future spectroscopic investigations. In addition to massive stars, even intermediate-mass (2--$8 M_\odot$) and He-rich stars are producers of UV flux. Such populations result from binary interactions between massive stars and their binary companions and are producers of excess UV emission. In the past, only a few stripped stars had been confirmed earlier owing to the scarcity of wide-field, high-angular-resolution UV surveys of stellar populations with reliable distances and extinction estimates. This has motivated the recent Stripped-Star Ultraviolet Magellanic Clouds Survey \citep[SUMS;][]{Ludwig2025arXiv250518632L}, which used {\sl Swift}/UVOT. The resulting catalogue from SUMS contains several hundred candidate intermediate-mass stripped stars, identified by the UV light they contribute to the SEDs of their host systems. 

\subsection{Reflections}

The MCs are the nearest resolved galaxies hosting low-metallicity stellar populations. Hence, they represent an ideal test bed for various star-formation and stellar evolutionary models across a wider metallicity range and in shallower gravitational potentials than in the Milky Way. So far, only a modest collection of UV data is available. UV studies of star clusters in the MCs push the boundaries of MP analyses to clusters beyond the Milky Way. Although the origin of MPs remains an open question, studies in the MCs have established that the onset of MPs is not only dependent on cluster mass, but also strongly dependent on cluster age. These insights have been driven primarily by observations using {\sl HST} UV passbands and by {\sl AstroSat}/UVIT, combined with complementary spectroscopic studies. Moreover, {\sl HST} UV photometric data, combined with observations in optical filters, have proven a powerful diagnostic tool for identifying UV-dim stars along the main sequence. Such stars may contribute to the eMSTOs seen in star clusters aged $\sim$2~Gyr and younger. However, the origin of these UV-dim stars remains an open question and requires further theoretical and observational work. 

Spectroscopic surveys such as VFTS, RIOTS4 and BLOeM, which focus on massive (O-type) stars in MC clusters, have shown that binary evolution (mergers or mass transfer) plays a significant role in their formation. Such studies have placed constraints on their mass limits, placed constraints on their binary fraction, highlighted their significance in ionising the surrounding interstellar medium and revealed a possibly steep upper IMF in the SMC. Studies from X-ULLYSES project investigated relations between stellar-wind structures and luminosity class, and estimated empirical relation between stellar wind properties with stellar parameters and metallicities, that will propel future studies. UV photometric data from {\sl GALEX}, combined with multi-epoch spectroscopic data, have enabled the discovery of O-type stars in the Magellanic Bridge. These results raise fundamental questions about the formation of massive stars in low-metallicity, low-gas-density and tidally perturbed environments. While many aspects of massive-star formation and evolution remain unresolved, the ULLYSES project is paving the way for future investigations by assembling a UV spectral library for massive stars in MC-like metal-poor galaxies. This library will enable characterisation of stellar winds (velocity law, density structure, mass-loss rate) and photospheric properties (temperature, gravity, chemical abundances, rotational velocity) as functions of metallicity, spectral type and luminosity class. When combined with theoretical models, these data will help improve our understanding of ionising sources during cosmic noon.

In addition, exploitation of the publicly available far-UV catalogues of stellar populations from {\sl AstroSat}/UVIT will have a broad impact, enabling studies that range from stellar-scale analyses (e.g., spectroscopic observations of OB and A-type stars) to galactic-scale properties of the MCs (e.g., recent star-formation and interaction histories). Similarly, UV catalogues from the SUMS survey will motivate both observational and theoretical efforts to better understand the formation and origin of intermediate-mass and He-rich stellar populations.

In summary, testing MPs in star clusters across wide ranges in mass and age, mapping the high-mass stars, estimating stellar properties of massive stars and investigating stellar mergers and post-mass-transfer systems all require high-resolution UV imaging and spectroscopic data in the MCs. We are hopeful that upcoming UV missions will be equipped with instruments capable of probing the full range of stellar populations across the MCs.

\section{Extragalactic star clusters: Windows into star formation beyond the Local Group}

Young, extragalactic massive star clusters, the building blocks of galaxies, are particularly amenable to UV studies: their light is dominated by short-lived, massive stars whose strong UV continua and resonance lines encode information about stellar evolution, feedback and the interaction between stars and their host galaxies.

Whereas decades of work on the Milky Way and the MCs have laid the groundwork for our understanding of star clusters in general, it is only through observations of systems beyond the Local Group that we can fully explore how the environment shapes cluster populations. However, UV observations of extragalactic star clusters have long been hampered by the limited spatial resolution of the prevailing instrumentation, the relatively faint UV fluxes of most except for the very youngest stellar populations and the inhomogeneity of UV emission mechanisms: in addition to actively star-forming populations, UV emission is also associated with the `UV upturn' phenomenon in the oldest stellar populations and with specific line fluxes at a range of ages. This apparently haphazard origin of UV emission mechanisms has long limited dedicated studies to targeting individual objects such as the young extragalactic H{\sc ii} regions powering natal star clusters. Recent large-scale imaging surveys and spectroscopic programmes have dramatically advanced this effort, now placing the UV firmly at the heart of extragalactic star cluster astrophysics.

\subsection{The UV advantage for extragalactic clusters}

Massive stars emit the bulk of their light in the far-UV (912--2000 {\AA}) and the near-UV (2000--3000 {\AA}). Because their lifetimes are short---typically a few million years---far- and near-UV emission provides a sensitive clock for dating star-formation episodes. In integrated star cluster SEDs, UV colours and spectral features reveal both ages and details of the underlying stellar IMFs, particularly the presence of the most massive stars ($\ge 100 M_\odot$). Crucially, UV lines such as C\,{\sc iv} $\lambda$1550 {\AA}, Si\,{\sc iv} $\lambda$1400 {\AA} and He\,{\sc ii} $\lambda$1640 {\AA} trace stellar winds and ionising continua, thus offering access to direct probes of feedback processes. In contrast, optical and IR diagnostics are more strongly affected by dust and evolved stars. For extragalactic work---where the spatial resolution of current instrumentation is limited---the UV offers both sensitivity to the youngest stars and discriminatory power between star cluster populations and diffuse galactic-field light.

The first UV observations of extragalactic star clusters were obtained with {\sl IUE}. Observations of giant H{\sc ii} regions---such as NGC 604 in M33 and NGC 5471 in M101---revealed `P Cygni' profiles in resonance lines, evidence of powerful O-star winds, superposed on faint nebular emission \citep{1980A&A....85L..21R}. By the early 1980s, evolutionary models \citep{1981A&A...103..305L} had made clear that far-UV luminosities could not be explained by continuous star formation alone; rather, they implied short starbursts, thus setting the stage for the burst-dominated paradigm of extragalactic clusters.

{\sl IUE} studies of amorphous star-forming galaxies such as NGC 1705 demonstrated that their stellar populations included normal OB stars, suggesting no anomalies in their IMFs \citep{1984STIN...8519904L}. Meanwhile, discoveries of WR features in giant extragalactic H{\sc ii} regions like Tololo 89 \citep[Tol 89;][]{1985A&A...143..347D} revealed young bursts dominated by massive stars with strong winds, ages of just a few Myr and metallicities significantly below solar. These early works firmly established UV spectroscopy as a critical tool for identifying the ages, feedback and chemical context of extragalactic clusters.

By the 1990s, the {\sl HST}'s Faint Object Spectrograph (FOS) enabled higher-resolution UV spectroscopy of star clusters in galaxies such as M101. FOS-based studies showed that short-duration bursts with conventional IMFs sufficed to explain the UV continua, although dust attenuation laws appeared to differ from the dominant trends in the Milky Way \citep{1994A&A...291....1R}. This was a key step in demonstrating that star cluster UV properties could be understood within population synthesis frameworks, but it also highlighted the sensitivity of UV spectra to variations in extinction laws.

\subsection{From targeted studies to systematic surveys}

In the first decade of the current century, systematic efforts broadened the sample available for UV studies. Studies of M31 and M87 GCs with {\sl GALEX} and {\sl HST} (with either STIS or WFPC2) revealed unexpectedly strong far-UV fluxes, often requiring hot HB stars or He-enriched 2P populations to reconcile observations with canonical models \citep{2007MNRAS.377..987K, 2017MNRAS.464..713P}. This represented early recognition that UV emission from star clusters could diagnose MPs \citep[e.g.,][]{2018PhyS...93b4001D}, thus extending our physical understanding beyond the young-cluster regime.

Photometric studies combining {\sl GALEX} with optical and IR data \citep[e.g.,][]{2007ApJ...659..359M, 2009AJ....137.4884M} demonstrated the power of UV--optical SEDs to constrain ages of well-populated extragalactic star clusters, although uncertainties increased significantly for ages in excess of $\sim$10 Gyr. Near-UV synthesis models \citep{1998AAS...193.6002P, 2002AAS...200.4016P} enabled quantitative age dating of GCs in M31 and beyond. In addition, a systematic analysis of SED coverage in starburst galaxies \citep{2003MNRAS.342..259D} highlighted the necessity of UV data to avoid degeneracies in derived cluster ages \citep[see also][]{2016MNRAS.457.4296W}.

More recently, the Panchromatic Hubble Andromeda Treasury (PHAT) survey \citep{2011sca..conf..129J} expanded UV--optical cluster catalogues in M31 by an order of magnitude, for the first time providing unprecedented statistics on cluster demographics in a large spiral galaxy outside the Milky Way. Meanwhile, the Legacy ExtraGalactic UV Survey \citep[LEGUS;][]{2015AJ....149...51C} extended integrated UV photometry of reasonable spatial resolution to 50 galaxies within 12 Mpc, allowing the authors to create comprehensive catalogues of thousands of star clusters across diverse galactic environments.

Some of the most important LEGUS results include:

\begin{itemize}
\item Cluster disruption timescales: young, low-mass star clusters dissolve rapidly ($<$100 Myr), while massive clusters survive significantly longer \citep{2019MNRAS.490.4648H, 2022MNRAS.512.1294H}. UV observations enabled robust identification and characterisation of the youngest ($<$10 Myr) populations.

\item Environmental dependence: higher star-formation-rate surface densities correlate with higher cluster-formation efficiency, sometimes in excess of 30 per cent \citep{2020SSRv..216...69A}.

\item Morphological evolution: H$\alpha$ and UV imaging of LEGUS clusters shows that gas clearing occurs on 1--3 Myr timescales \citep{2019MNRAS.490.4648H, 2022MNRAS.512.1294H}, in turn implying that feedback operates on short timescales.
\end{itemize}

Jointly, PHAT and LEGUS demonstrated that UV-selected star cluster samples are indispensable for constraining both the demographics and the underlying physics, with thousands of star clusters now appropriately placed into coherent evolutionary frameworks.

{\sl AstroSat}/UVIT has significantly advanced space-based UV imaging, simultaneously in far-UV ($\sim$130--180 nm), near-UV ($\sim$200--300 nm) and visible ($\sim$320--550 nm) bands with a large $\approx$28 arcmin diameter field and a typical point-source FWHM of $\approx$1.0--1.5 arcsec after careful calibration \citep{2017AJ....154..128T, 2020AJ....159..158T, 2020ApJS..247...47L}. UVIT's combination of relatively high spatial resolution and wide sky coverage has enabled resolved UV studies of entire nearby galaxies and their cluster populations. The M31 UVIT survey produced extensive far-/near-UV point-source catalogues and improved astrometry that permit secure matches to {\sl HST}/PHAT sources and X-ray counterparts \citep{2020ApJS..247...47L}. UVIT mapping of nearby spirals and dwarf galaxies (e.g., NGC 7793, NGC 628, WLM, IC 2574 and NGC 7252) has been used to identify hundreds of young star-forming clumps and compact clusters, derive luminosity and mass functions down to $\lesssim 10^3 M_\odot$ in favourable cases and characterise radial trends in star formation at sub-hundred-pc scales \citep{Subramaniam2016ApJ...833L..27S,2018A&A...613L...9G, 2018AJ....156..109M, 2019AJ....158..229M,2021ApJ...909..203M, 
2022MNRAS.516.2171U}.

Beyond young populations, UVIT has been productive in evolved cluster science and in feedback studies. High-resolution far-UV imaging and matched photometry have exposed UV-bright HB and post-AGB populations in external GC systems, strengthening evidence for multiple/He-enhanced populations outside the Milky Way (e.g., targeted studies of M31 clusters) \citep[see][]{2020ApJS..247...47L}. In actively star-forming or post-merger systems, combined UVIT--H$\alpha$ and multi-wavelength analyses have mapped the clearing of natal gas and the morphology of feedback-driven cavities on sub-kpc scales \citep{2018A&A...613L...9G,2024MNRAS.528.4432R}. Collectively, UVIT data have created an observational niche---wide field, moderate-resolution UV imaging---that complements {\sl HST} and {\sl GALEX} while opening up UV cluster science to new targets and larger samples, particularly in the southern sky.

\subsection{Spectroscopic insights}

Meanwhile, spectroscopy has managed to keep pace with imaging. {\sl HST}/COS surveys such as the COS Legacy Archive Spectroscopy Survey \citep[CLASSY;][]{2022ApJS..261...31B, 2022ApJS..262...37J} have provided the most comprehensive atlas of cluster-dominated starburst spectra to date, covering 45 galaxies and a range of metallicities. CLASSY spectra have revealed systematic underpredictions of He\,{\sc ii} $\lambda$1640 {\AA} emission by most current population synthesis models, suggesting that the presence of binary stellar systems, stellar rotation or very massive stars must be invoked \citep{2020Galax...8...13L}. Wind lines such as C\,{\sc iv} and Si\,{\sc iv} varied with metallicity, as expected, but scatter exceeded theoretical expectations---underscoring the need for better stellar-wind prescriptions.

The CLusters in the UV as EngineS (CLUES) survey \citep{2022AJ....164..208S} focused specifically on the young, $<$20 Myr-old massive star clusters identified in LEGUS galaxies, combining far-UV spectroscopy with photometry. This programme demonstrated excellent consistency between spectroscopic and photometric ages for clusters younger than $\sim$5 Myr but systematic discrepancies at older ages. Importantly, CLUES showed that UV spectroscopy better constrains metallicities than optical colours alone, thus reaffirming the UV as a key tool for probing both chemistry and stellar content.

Large UV surveys have revealed that clusters form hierarchically. LEGUS analyses of spatial clustering \citep{2017ApJ...840..113G, 2018PhDT........36G} found that star clusters and stellar associations are distributed in turbulent, hierarchical patterns which tend to disperse within $\times$40 Myr. This supports a picture of turbulence-driven star formation where star clusters trace unbound stellar associations that dissolve quickly \citep[e.g.,][and references therein]{2022MNRAS.512.1196M, 2025arXiv250608951H}.

UV and H$\alpha$-based star cluster H{\sc ii} morphologies further confirm that natal gas clearing is rapid, often occurring on 1--2 Myr timescales \citep{2022MNRAS.512.1294H}. This highlights how UV-bright star clusters are short-lived signposts of the most recent star-formation episode in a galaxy, tightly linked to feedback-driven evolution of the interstellar medium.

UV studies are not limited to young clusters, however. Extragalactic GCs frequently display UV upturns driven by hot HB and post-AGB stars. Observations of M87 and M31 clusters \citep{2007MNRAS.377..987K, 2017MNRAS.464..713P} showed that some metal-rich star clusters are unexpectedly far-UV-bright, implying the presence of He-enriched 2P populations. More recent far-UV imaging of NGC 1399 clusters \citep{2023MNRAS.518...87D} confirmed widespread hot HB populations, reinforcing the conclusion that MPs are a ubiquitous feature of GCs.

\subsection{Towards cosmic analogues}

Perhaps the most exciting role of extragalactic UV star cluster studies is their relevance to galaxy formation in the early Universe. The rest-frame far-UV corresponds to the optical/near-IR wavelength range observed for galaxies at redshifts $z > 4$. CLASSY spectra, with their comprehensive libraries of wind lines and nebular continua, now serve as empirical templates for interpreting James Webb Space Telescope ({\sl JWST}) spectra of reionisation-era galaxies. Comparisons already suggest that young starbursts at $z > 6$ may be powered by star cluster populations similar to local young, very massive clusters \citep[sometimes erroneously referred to as `super' star clusters;][]{2005HiA....13..363D} but with lower metallicities and harder ionising spectra.

Despite remarkable progress over the past four decades, major challenges remain, however. Dust attenuation strongly affects UV fluxes, and extinction curves seem to vary among galaxies, in turn complicating corrections. Stochastic IMF sampling introduces large scatter in UV properties of low-mass, often poorly sampled star clusters, making it difficult to infer universal IMF parameters \citep[but see][]{2023ApJ...954..136J}. Spatial resolution remains a limiting factor: beyond $\sim$30 Mpc, even {\sl HST} cannot reliably resolve clusters (owing to its longer-wavelength operation, {\sl JWST} offers similar resolution beyond 1 $\mu$m, despite its larger mirror diameter). Finally, stellar population models still lag behind observations, particularly in predicting ionising fluxes and He\,{\sc ii} emission.

Future missions promise relief. ULLYSES is delivering deep UV stellar libraries across a range of metallicities, aimed at improving stellar population synthesis inputs. The Ultraviolet Transient Astronomical Satellite ({\sl ULTRASAT}) will open the time-domain UV sky \citep{2024ApJ...964...74S}, whereas forthcoming survey missions such as the ({\sl UVEX} will provide wide-field UV imaging of the Galactic plane and nearby galaxies, enabling the identification of low-mass, low-metallicity star-forming systems and facilitating targeted spectroscopic follow-up \citep{Kulkarni2021UVEX}. At the same time, concepts for large UV--optical space telescopes continue to evolve: earlier mission studies such as the Large UV/Optical/IR Surveyor ({\sl LUVOIR}) and the Habitable Exoplanet ({\sl HabEx}) Imaging Mission have now been consolidated into the NASA-led Habitable Worlds Observatory ({\sl HWO}), which aims to deliver transformative gains in sensitivity, spatial resolution and spectroscopic capability across the UV and optical regimes \citep{2026arXiv260106386P}. Such a facility would revolutionise extragalactic star cluster studies, enabling resolved UV spectroscopy of cluster populations well beyond the Local Volume. In the mean time, synergy between {\sl HST} UV and {\sl JWST} IR observations continues to offer powerful insights into the embedded-to-exposed star cluster transition.

\subsection{Reflections}

UV studies of extragalactic star clusters have evolved from pioneering {\sl IUE} spectra in the 1980s to today's systematic surveys and spectral atlases. From identifying OB stars in NGC 1705, one of the earliest UV-bright cluster laboratories, to cataloguing thousands of LEGUS star clusters, to dissecting stellar winds in CLASSY, UV observations have reshaped our understanding of star cluster demographics, feedback and stellar evolution. They have also illuminated the role of star clusters as analogues to high-redshift galaxies, making them essential to understanding cosmic history at large.

As future UV-capable observatories expand our reach, extragalactic star clusters will remain vital laboratories for stellar astrophysics, galactic ecology and the evolution of the Universe on the largest spatial scales.

The trajectory of UV star cluster studies suggests that the field is entering a golden era of synthesis, marrying theory, models and multi-wavelength observations. The combination of UV data with {\sl JWST}'s near-IR spectra will allow us to cover the full star cluster lifecycle, from embedded to gas-cleared phases, with direct comparisons to reionisation-era galaxies. Planned missions like {\sl HWO} are expected to provide the spectral sensitivity and angular resolution to resolve star clusters at distances far beyond the current $\sim$30 Mpc limit, thus finally extending demographic studies across a representative slice of the Universe.

At the same time, theoretical models of stellar populations must continue to evolve. The inclusion of stellar binaries, rotation and very massive stars in population synthesis codes has already started to improve the match to observed UV line strengths. As more comprehensive stellar libraries are compiled through programmes like ULLYSES, and as time-domain UV observations become available with {\sl ULTRASAT}, the next generation of models will be more predictive and better physically grounded.

Extragalactic UV star cluster astrophysics research has clearly evolved to become more than an observational niche. It is a keystone in bridging the physics of individual stars to the formation and evolution of galaxies across cosmic time. The questions sharpened by four decades of UV work---from IMF variations to the mechanisms of feedback---will continue to drive discovery for decades to come.

\section{Outstanding Science Questions}

\begin{itemize}
    \item Can we use star clusters to calibrate stellar atmospheric models in the UV? 
    Star clusters provide pseudo-homogeneous populations for building averaged UV isochrones and spectral libraries, enabling robust calibration of stellar atmospheric models despite limited individual stellar data and non-photospheric contamination. However, identifying non-standard evolution is also crucial for such calibration.

    \item How do UV evolutionary pathways vary with metallicity, particularly among Galactic GCs, MC clusters and extragalactic star cluster systems? More fundamentally, how does metallicity regulate key processes in massive-star evolution, including mass loss, rotation, binarity, and the production of ionising radiation?

    \item What is the nature and prevalence of stripped stars produced through binary interactions?
    What are their UV spectral signatures, lifetimes, and contribution to the ionising photon budget of star clusters, particularly at low metallicities?

    \item How do MPs manifest in the UV? Can UV photometry and spectroscopy uniquely constrain abundance variations (e.g., He, CNO) and their distribution within star clusters, and what does this reveal about the formation and evolutionary pathways of MPs?

    \item How do UV colours quantitatively map to abundance variations within (e.g., MPs) and across clusters (e.g., metallicity, density, and dynamical differences)? Can UV diagnostics disentangle the effects of helium enrichment, light-element variations, and stellar rotation?

    \item What is the physical origin of MPs, and how can future UV observations constrain the interplay between helium enrichment, stellar dynamics, and chemical evolution?

    \item How do the undersampled observations of Galactic OCs outside the Galactic disc bias our understanding of cluster stellar populations? Better coverage of the Galactic disc will enable us to identify such biases.

    \item Can we distinguish among various classes of compact objects such as WDs, sdOB stars, stripped stars and HB stars with multi-band SEDs in the absence of UV spectroscopy? Multi-epoch data might provide supplementary information for verification; however, a UV spectroscopic mission would be the most powerful tool to resolve any degeneracy. Such spectroscopy will also help distinguish between chromospheric activity and emission from a compact companion.

    \item How many Blue Lurkers are there? UV analysis of a handful of clusters has shown that the number of Blue Lurkers is comparable to the number of BSS. Deeper UV observations of more clusters are necessary to build a statistically significant sample and to understand stellar interactions in lower-mass stars.

    \item How do UV colours quantitatively map to abundance variations within (e.g., MPs) and across (e.g., metallicity, density and dynamical differences) clusters?

    \item What physical mechanism primarily drives the UV-dim phenomenon?
    Is it circumstellar dust extinction, gravity-darkening effects, metallicity or something as yet unidentified?

    \item How does massive-star formation proceed in extremely low-density environments such as the Magellanic Bridge?

    \item How accurately can UV spectral features trace the most massive stars in unresolved or partially resolved extragalactic clusters? Can we use these diagnostics to identify variations in the stellar IMF?
\end{itemize}

\section*{Acknowledgements}
This article is part of a collection of reviews resulting from the April 2025 Workshop on `Space-based stellar astrophysics in the UV' organised at the International Space Science Institute--Beijing. We are grateful to the institute's staff for excellent organisational and generous financial support. Richard de Grijs acknowledges financial support in the form of a Distinguished Senior Fellowship from the President’s International Fellowship Initiative (PIFI), Chinese Academy of Sciences (Grant No. 2025PVA0007). Snehalata Sahu received funding from the European Research Council under the European Union’s Horizon 2020 research and innovation programme number 101020057 (WDPLANETS).

\par\noindent
{\bf Conflicts of interests:} The authors declare no competing conflicts of interest of any kind.

\bibliography{starclusters}

@ARTICLE{2021ApJ...909..203M,
       author = {{Mondal}, Chayan and {Subramaniam}, Annapurni and {George}, Koshy and {Postma}, Joseph E. and {Subramanian}, Smitha and {Barway}, Sudhanshu},
        title = "{Tracing Young Star-forming Clumps in the Nearby Flocculent Spiral Galaxy NGC 7793 with UVIT Imaging}",
      journal = {\apj},
         year = 2021,
        month = mar,
       volume = {909},
       number = {2},
          eid = {203},
        pages = {203},
          doi = {10.3847/1538-4357/abe0b4},
archivePrefix = {arXiv},
       eprint = {2101.11314},
 primaryClass = {astro-ph.GA},
       adsurl = {https://ui.adsabs.harvard.edu/abs/2021ApJ...909..203M}
}

@ARTICLE{2019AJ....158..229M,
       author = {{Mondal}, Chayan and {Subramaniam}, Annapurni and {George}, Koshy},
        title = "{Ultraviolet Imaging Telescope View of Dwarf Irregular Galaxy IC 2574: Is the Star Formation Triggered Due to Expanding H I Shells?}",
      journal = {\aj},
         year = 2019,
        month = dec,
       volume = {158},
       number = {6},
          eid = {229},
        pages = {229},
          doi = {10.3847/1538-3881/ab4ea1},
archivePrefix = {arXiv},
       eprint = {1906.10660},
 primaryClass = {astro-ph.GA},
       adsurl = {https://ui.adsabs.harvard.edu/abs/2019AJ....158..229M}
}

@ARTICLE{2018AJ....156..109M,
       author = {{Mondal}, Chayan and {Subramaniam}, Annapurni and {George}, Koshy},
        title = "{UVIT Imaging of WLM: Demographics of Star-forming Regions in the Nearby Dwarf Irregular Galaxy}",
      journal = {\aj},
         year = 2018,
        month = sep,
       volume = {156},
       number = {3},
          eid = {109},
        pages = {109},
          doi = {10.3847/1538-3881/aad4f6},
archivePrefix = {arXiv},
       eprint = {1807.07359},
 primaryClass = {astro-ph.GA},
       adsurl = {https://ui.adsabs.harvard.edu/abs/2018AJ....156..109M}
}

@ARTICLE{2023MNRAS.523L..58D,
       author = {{Dattatrey}, Arvind K. and {Yadav}, R.~K.~S. and {Kumawat}, Gourav and {Rani}, Sharmila and {Singh}, Gaurav and {Subramaniam}, Annapurni and {Singh}, Ravi S.},
        title = "{GlobULeS - V. UVIT/AstroSat studies of stellar populations in NGC 362: detection of blue lurkers in a globular cluster}",
      journal = {\mnras},
         year = 2023,
        month = jul,
       volume = {523},
       number = {1},
        pages = {L58-L63},
          doi = {10.1093/mnrasl/slad059},
archivePrefix = {arXiv},
       eprint = {2305.09723},
 primaryClass = {astro-ph.SR},
       adsurl = {https://ui.adsabs.harvard.edu/abs/2023MNRAS.523L..58D}
}

@ARTICLE{2025ApJ...990L..62Y,
       author = {{Yadav}, R.~K.~S. and {Dattatrey}, Arvind K. and {Subramaniam}, Annapurni and {Rangwal}, Geeta and {Singh}, Ravi S.},
        title = "{Detection of Young Massive White Dwarfs in Core-collapsed Globular Cluster NGC 362}",
      journal = {\apjl},
         year = 2025,
        month = sep,
       volume = {990},
       number = {2},
          eid = {L62},
        pages = {L62},
          doi = {10.3847/2041-8213/adf9d9},
       adsurl = {https://ui.adsabs.harvard.edu/abs/2025ApJ...990L..62Y}
}

@ARTICLE{1996AJ....112.1487H,
       author = {{Harris}, William E.},
        title = "{A Catalog of Parameters for Globular Clusters in the Milky Way}",
      journal = {\aj},
         year = 1996,
        month = oct,
       volume = {112},
        pages = {1487},
          doi = {10.1086/118116},
       adsurl = {https://ui.adsabs.harvard.edu/abs/1996AJ....112.1487H}
}

@ARTICLE{2010arXiv1012.3224H,
       author = {{Harris}, William E.},
        title = "{A New Catalog of Globular Clusters in the Milky Way}",
      journal = {arXiv e-prints},
         year = 2010,
        month = dec,
          eid = {arXiv:1012.3224},
        pages = {arXiv:1012.3224},
          doi = {10.48550/arXiv.1012.3224},
archivePrefix = {arXiv},
       eprint = {1012.3224},
 primaryClass = {astro-ph.GA},
       adsurl = {https://ui.adsabs.harvard.edu/abs/2010arXiv1012.3224H}
}

@ARTICLE{1993A&AS...98..477M,
       author = {{Meynet}, G. and {Mermilliod}, J.-C. and {Maeder}, A.},
        title = "{New dating of galactic open clusters.}",
      journal = {\aaps},
         year = 1993,
        month = may,
       volume = {98},
        pages = {477-504},
       adsurl = {https://ui.adsabs.harvard.edu/abs/1993A&AS...98..477M}
}

@ARTICLE{1990ApJ...364...35G,
       author = {{Greggio}, Laura and {Renzini}, Alvio},
        title = "{Clues on the Hot Star Content and the Ultraviolet Output of Elliptical Galaxies}",
      journal = {\apj},
         year = 1990,
        month = nov,
       volume = {364},
        pages = {35},
          doi = {10.1086/169384},
       adsurl = {https://ui.adsabs.harvard.edu/abs/1990ApJ...364...35G}
}

@ARTICLE{2011Ap&SS.335...51B,
       author = {{Bianchi}, Luciana},
        title = "{GALEX and star formation}",
      journal = {\apss},
         year = 2011,
        month = sep,
       volume = {335},
       number = {1},
        pages = {51-60},
          doi = {10.1007/s10509-011-0612-2},
       adsurl = {https://ui.adsabs.harvard.edu/abs/2011Ap&SS.335...51B}
}

@ARTICLE{2018ApJ...864...33D,
       author = {{Dalessandro}, E. and {Mucciarelli}, A. and {Bellazzini}, M. and {Sollima}, A. and {Vesperini}, E. and {Hong}, J. and {H{\'e}nault-Brunet}, Vincent and {Ferraro}, F.~R. and {Ibata}, R. and {Lanzoni}, B. and {Massari}, D. and {Salaris}, M.},
        title = "{The Unexpected Kinematics of Multiple Populations in NGC 6362: Do Binaries Play a Role?}",
      journal = {\apj},
         year = 2018,
        month = sep,
       volume = {864},
       number = {1},
          eid = {33},
        pages = {33},
          doi = {10.3847/1538-4357/aad4b3},
archivePrefix = {arXiv},
       eprint = {1807.07918},
 primaryClass = {astro-ph.GA},
       adsurl = {https://ui.adsabs.harvard.edu/abs/2018ApJ...864...33D}
}

@ARTICLE{Ferraro2012Natur.492..393F,
       author = {{Ferraro}, F.~R. and {Lanzoni}, B. and {Dalessandro}, E. and {Beccari}, G. and {Pasquato}, M. and {Miocchi}, P. and {Rood}, R.~T. and {Sigurdsson}, S. and {Sills}, A. and {Vesperini}, E. and {Mapelli}, M. and {Contreras}, R. and {Sanna}, N. and {Mucciarelli}, A.},
        title = "{Dynamical age differences among coeval star clusters as revealed by blue stragglers}",
      journal = {\nat},
         year = 2012,
        month = dec,
       volume = {492},
       number = {7429},
        pages = {393-395},
          doi = {10.1038/nature11686},
archivePrefix = {arXiv},
       eprint = {1212.5071},
 primaryClass = {astro-ph.SR},
       adsurl = {https://ui.adsabs.harvard.edu/abs/2012Natur.492..393F}
}

@ARTICLE{Schiavon2012AJ....143..121S,
       author = {{Schiavon}, Ricardo P. and {Dalessandro}, Emanuele and {Sohn}, Sangmo T. and {Rood}, Robert T. and {O'Connell}, Robert W. and {Ferraro}, Francesco R. and {Lanzoni}, Barbara and {Beccari}, Giacomo and {Rey}, Soo-Chang and {Rhee}, Jaehyon and {Rich}, R. Michael and {Yoon}, Suk-Jin and {Lee}, Young-Wook},
        title = "{Ultraviolet Properties of Galactic Globular Clusters with GALEX. I. The Color-Magnitude Diagrams}",
      journal = {\aj},
         year = 2012,
        month = may,
       volume = {143},
       number = {5},
          eid = {121},
        pages = {121},
          doi = {10.1088/0004-6256/143/5/121},
archivePrefix = {arXiv},
       eprint = {1201.5377},
 primaryClass = {astro-ph.GA},
       adsurl = {https://ui.adsabs.harvard.edu/abs/2012AJ....143..121S}
}

@ARTICLE{2005ARA&A..43..387G,
       author = {{Gallart}, C. and {Zoccali}, M. and {Aparicio}, A.},
        title = "{The Adequacy of Stellar Evolution Models for the Interpretation of the Color-Magnitude Diagrams of Resolved Stellar Populations}",
      journal = {\araa},
         year = 2005,
        month = sep,
       volume = {43},
       number = {1},
        pages = {387-434},
          doi = {10.1146/annurev.astro.43.072103.150608},
       adsurl = {https://ui.adsabs.harvard.edu/abs/2005ARA&A..43..387G}
}

@ARTICLE{2002ARA&A..40..487F,
       author = {{Freeman}, Ken and {Bland-Hawthorn}, Joss},
        title = "{The New Galaxy: Signatures of Its Formation}",
      journal = {\araa},
         year = 2002,
        month = jan,
       volume = {40},
        pages = {487-537},
          doi = {10.1146/annurev.astro.40.060401.093840},
archivePrefix = {arXiv},
       eprint = {astro-ph/0208106},
 primaryClass = {astro-ph},
       adsurl = {https://ui.adsabs.harvard.edu/abs/2002ARA&A..40..487F}
}

@ARTICLE{2016A&A...586A.148N,
       author = {{Niederhofer}, F. and {Bastian}, N. and {Kozhurina-Platais}, V. and {Hilker}, M. and {de Mink}, S.~E. and {Cabrera-Ziri}, I. and {Li}, C. and {Ercolano}, B.},
        title = "{Controversial age spreads from the main sequence turn-off and red clump in intermediate-age clusters in the LMC}",
      journal = {\aap},
         year = 2016,
        month = feb,
       volume = {586},
          eid = {A148},
        pages = {A148},
          doi = {10.1051/0004-6361/201526484},
archivePrefix = {arXiv},
       eprint = {1510.08476},
 primaryClass = {astro-ph.SR},
       adsurl = {https://ui.adsabs.harvard.edu/abs/2016A&A...586A.148N}
}

@ARTICLE{2011AJ....142...36G,
       author = {{Glatt}, Katharina and {Grebel}, Eva K. and {Jordi}, Katrin and {Gallagher}, III, John S. and {Da Costa}, Gary and {Clementini}, Gisella and {Tosi}, Monica and {Harbeck}, Daniel and {Nota}, Antonella and {Sabbi}, Elena and {Sirianni}, Marco},
        title = "{Present-day Mass Function of Six Small Magellanic Cloud Intermediate-age and Old Star Clusters}",
      journal = {\aj},
         year = 2011,
        month = aug,
       volume = {142},
       number = {2},
          eid = {36},
        pages = {36},
          doi = {10.1088/0004-6256/142/2/36},
       adsurl = {https://ui.adsabs.harvard.edu/abs/2011AJ....142...36G}
}

@ARTICLE{2013ApJ...775..134V,
       author = {{VandenBerg}, Don A. and {Brogaard}, K. and {Leaman}, R. and {Casagrande}, L.},
        title = "{The Ages of 55 Globular Clusters as Determined Using an Improved $\Delta V^{HB}_{TO}$ Method along with Color-Magnitude Diagram Constraints, and Their Implications for Broader Issues}",
      journal = {\apj},
         year = 2013,
        month = oct,
       volume = {775},
       number = {2},
          eid = {134},
        pages = {134},
          doi = {10.1088/0004-637X/775/2/134},
archivePrefix = {arXiv},
       eprint = {1308.2257},
 primaryClass = {astro-ph.GA},
       adsurl = {https://ui.adsabs.harvard.edu/abs/2013ApJ...775..134V}
}

@ARTICLE{2014A&A...570A..38B,
       author = {{Bestenlehner}, J.~M. and {Gr{\"a}fener}, G. and {Vink}, J.~S. and {Najarro}, F. and {de Koter}, A. and {Sana}, H. and {Evans}, C.~J. and {Crowther}, P.~A. and {H{\'e}nault-Brunet}, V. and {Herrero}, A. and {Langer}, N. and {Schneider}, F.~R.~N. and {Sim{\'o}n-D{\'\i}az}, S. and {Taylor}, W.~D. and {Walborn}, N.~R.},
        title = "{The VLT-FLAMES Tarantula Survey. XVII. Physical and wind properties of massive stars at the top of the main sequence}",
      journal = {\aap},
         year = 2014,
        month = oct,
       volume = {570},
          eid = {A38},
        pages = {A38},
          doi = {10.1051/0004-6361/201423643},
archivePrefix = {arXiv},
       eprint = {1407.1837},
 primaryClass = {astro-ph.SR},
       adsurl = {https://ui.adsabs.harvard.edu/abs/2014A&A...570A..38B}
}

@ARTICLE{Moehler+2011A&A,
       author = {{Moehler}, S. and {Dreizler}, S. and {Lanz}, T. and {Bono}, G. and {Sweigart}, A.~V. and {Calamida}, A. and {Nonino}, M.},
        title = "{The hot horizontal-branch stars in {\ensuremath{\omega}} Centauri}",
      journal = {\aap},
         year = 2011,
        month = feb,
       volume = {526},
          eid = {A136},
        pages = {A136},
          doi = {10.1051/0004-6361/201015020},
archivePrefix = {arXiv},
       eprint = {1009.3191},
 primaryClass = {astro-ph.SR},
       adsurl = {https://ui.adsabs.harvard.edu/abs/2011A&A...526A.136M}
}

@ARTICLE{Ferraro+1998ApJ,
       author = {{Ferraro}, Francesco R. and {Paltrinieri}, Barbara and {Fusi Pecci}, Flavio and {Rood}, Robert T. and {Dorman}, Ben},
        title = "{Multimodal Distributions along the Horizontal Branch}",
      journal = {\apj},
         year = 1998,
        month = jun,
       volume = {500},
       number = {1},
        pages = {311-319},
          doi = {10.1086/305712},
archivePrefix = {arXiv},
       eprint = {astro-ph/9708210},
 primaryClass = {astro-ph},
       adsurl = {https://ui.adsabs.harvard.edu/abs/1998ApJ...500..311F}
}

@ARTICLE{Brown+2010ApJ,
       author = {{Brown}, Thomas M. and {Sweigart}, Allen V. and {Lanz}, Thierry and {Smith}, Ed and {Landsman}, Wayne B. and {Hubeny}, Ivan},
        title = "{The Blue Hook Populations of Massive Globular Clusters}",
      journal = {\apj},
         year = 2010,
        month = aug,
       volume = {718},
       number = {2},
        pages = {1332-1344},
          doi = {10.1088/0004-637X/718/2/1332},
archivePrefix = {arXiv},
       eprint = {1006.1591},
 primaryClass = {astro-ph.SR},
       adsurl = {https://ui.adsabs.harvard.edu/abs/2010ApJ...718.1332B}
}

@ARTICLE{Schosser2025A&A,
       author = {{Sch{\"o}sser}, E.~C. and {Ramachandran}, V. and {Sander}, A.~A.~C. and {Gallagher}, J.~S. and {Bernini-Peron}, M. and {Gonz{\'a}lez-Tor{\`a}}, G. and {Josiek}, J. and {Lefever}, R.~R. and {Hamann}, W.-R. and {Oskinova}, L.~M.},
        title = "{Extremely iron-poor O-type stars in the Magellanic Bridge}",
      journal = {\aap},
         year = 2025,
        month = apr,
       volume = {696},
          eid = {L3},
        pages = {L3},
          doi = {10.1051/0004-6361/202554027},
archivePrefix = {arXiv},
       eprint = {2503.10769},
 primaryClass = {astro-ph.GA},
       adsurl = {https://ui.adsabs.harvard.edu/abs/2025A&A...696L...3S}
}

@ARTICLE{1995ApJ...442..105D,
       author = {{Dorman}, Ben and {O'Connell}, Robert W. and {Rood}, Robert T.},
        title = "{Ultraviolet Radiation from Evolved Stellar Populations. II. The Ultraviolet Upturn Phenomenon in Elliptical Galaxies}",
      journal = {\apj},
         year = 1995,
        month = mar,
       volume = {442},
        pages = {105},
          doi = {10.1086/175428},
archivePrefix = {arXiv},
       eprint = {astro-ph/9405030},
 primaryClass = {astro-ph},
       adsurl = {https://ui.adsabs.harvard.edu/abs/1995ApJ...442..105D}
}

@ARTICLE{Choudhury2026MNRAS,
       author = {{Choudhury}, Samyaday and {Nayak}, Prasanta K. and {Dhanush}, S.~R. and {Sahu}, Snehalata and {de Grijs}, Richard},
        title = "{UVIT Magellanic BRidge Analysis (UMBRA) - I. Far-UV-Gaia study of seven star clusters}",
      journal = {\mnras},
         year = 2026,
        month = may,
       volume = {548},
       number = {3},
          eid = {stag594},
        pages = {stag594},
          doi = {10.1093/mnras/stag594},
       adsurl = {https://ui.adsabs.harvard.edu/abs/2026MNRAS.548ag594C}
}

@ARTICLE{2026SSRv..222...42F,
       author = {{Fu}, Xiaoting and {Del Zanna}, Giulio and {Ji}, Li and {Geier}, Stephan and {Dorsch}, Matti and {Jadhav}, Vikrant and {Sutaria}, Firoza and {Li}, Chengyuan and {Parker}, Quentin and {Fang}, Xuan},
        title = "{Stellar Astrophysics in the Ultraviolet: Setting the Scene}",
      journal = {\ssr},
         year = 2026,
        month = apr,
       volume = {222},
       number = {4},
          eid = {42},
        pages = {42},
          doi = {10.1007/s11214-026-01301-x},
       adsurl = {https://ui.adsabs.harvard.edu/abs/2026SSRv..222...42F}
}

@ARTICLE{Doran2013A&A,
       author = {{Doran}, E.~I. and {Crowther}, P.~A. and {de Koter}, A. and {Evans}, C.~J. and {McEvoy}, C. and {Walborn}, N.~R. and {Bastian}, N. and {Bestenlehner}, J.~M. and {Gr{\"a}fener}, G. and {Herrero}, A. and {K{\"o}hler}, K. and {Ma{\'\i}z Apell{\'a}niz}, J. and {Najarro}, F. and {Puls}, J. and {Sana}, H. and {Schneider}, F.~R.~N. and {Taylor}, W.~D. and {van Loon}, J. Th. and {Vink}, J.~S.},
        title = "{The VLT-FLAMES Tarantula Survey. XI. A census of the hot luminous stars and their feedback in 30 Doradus}",
      journal = {\aap},
         year = 2013,
        month = oct,
       volume = {558},
          eid = {A134},
        pages = {A134},
          doi = {10.1051/0004-6361/201321824},
archivePrefix = {arXiv},
       eprint = {1308.3412},
 primaryClass = {astro-ph.SR},
       adsurl = {https://ui.adsabs.harvard.edu/abs/2013A&A...558A.134D}
}

@ARTICLE{Vink2022ARA&A,
       author = {{Vink}, Jorick S.},
        title = "{Theory and Diagnostics of Hot Star Mass Loss}",
      journal = {\araa},
         year = 2022,
        month = aug,
       volume = {60},
        pages = {203-246},
          doi = {10.1146/annurev-astro-052920-094949},
archivePrefix = {arXiv},
       eprint = {2109.08164},
 primaryClass = {astro-ph.SR},
       adsurl = {https://ui.adsabs.harvard.edu/abs/2022ARA&A..60..203V}
}

@ARTICLE{Bouret2013A&A,
       author = {{Bouret}, J.-C. and {Lanz}, T. and {Martins}, F. and {Marcolino}, W.~L.~F. and {Hillier}, D.~J. and {Depagne}, E. and {Hubeny}, I.},
        title = "{Massive stars at low metallicity. Evolution and surface abundances of O dwarfs in the SMC}",
      journal = {\aap},
         year = 2013,
        month = jul,
       volume = {555},
          eid = {A1},
        pages = {A1},
          doi = {10.1051/0004-6361/201220798},
archivePrefix = {arXiv},
       eprint = {1304.6923},
 primaryClass = {astro-ph.SR},
       adsurl = {https://ui.adsabs.harvard.edu/abs/2013A&A...555A...1B}
}

@ARTICLE{Telford2023ApJ,
       author = {{Telford}, O. Grace and {McQuinn}, Kristen B.~W. and {Chisholm}, John and {Berg}, Danielle A.},
        title = "{The Ionizing Spectra of Extremely Metal-poor O Stars: Constraints from the Only H II Region in Leo P}",
      journal = {\apj},
         year = 2023,
        month = jan,
       volume = {943},
       number = {1},
          eid = {65},
        pages = {65},
          doi = {10.3847/1538-4357/aca896},
archivePrefix = {arXiv},
       eprint = {2210.17535},
 primaryClass = {astro-ph.SR},
       adsurl = {https://ui.adsabs.harvard.edu/abs/2023ApJ...943...65T}
}

@ARTICLE{Lorenzo2022MNRAS,
       author = {{Lorenzo}, M. and {Garcia}, M. and {Najarro}, F. and {Herrero}, A. and {Cervi{\~n}o}, M. and {Castro}, N.},
        title = "{A new reference catalogue for the very metal-poor Universe: +150 OB stars in Sextans A}",
      journal = {\mnras},
         year = 2022,
        month = nov,
       volume = {516},
       number = {3},
        pages = {4164-4179},
          doi = {10.1093/mnras/stac2050},
archivePrefix = {arXiv},
       eprint = {2207.09700},
 primaryClass = {astro-ph.GA},
       adsurl = {https://ui.adsabs.harvard.edu/abs/2022MNRAS.516.4164L}
}

@ARTICLE{Brands2025A&A,
       author = {{Brands}, Sarah A. and {Backs}, Frank and {de Koter}, Alex and {Puls}, Joachim and {Crowther}, Paul A. and {Sana}, Hugues and {Tramper}, Frank and {Kaper}, Lex and {Sundqvist}, Jon O. and {Bestenlehner}, Joachim M. and {Driessen}, Florian A. and {Erba}, Christiana and {Hawcroft}, Calum and {Herrero}, Artemio and {John Hillier}, D. and {Ignace}, Richard and {Lefever}, Roel R. and {Dylan Kee}, N. and {Kub{\'a}tov{\'a}}, Brankica and {Mahy}, Laurent and {Moffat}, Anthony F.~J. and {Najarro}, Francisco and {Prinja}, Raman K. and {Ramachandran}, Varsha and {Sander}, Andreas A.~C. and {Vink}, Jorick S. and {XShootU Collaboration}},
        title = "{X-Shooting ULLYSES: Massive stars at low metallicity: XII. Clumped winds of O-type (super)giants in the Large Magellanic Cloud}",
      journal = {\aap},
         year = 2025,
        month = may,
       volume = {697},
          eid = {A54},
        pages = {A54},
          doi = {10.1051/0004-6361/202452784},
archivePrefix = {arXiv},
       eprint = {2503.14687},
 primaryClass = {astro-ph.SR},
       adsurl = {https://ui.adsabs.harvard.edu/abs/2025A&A...697A..54B}
}

@ARTICLE{Backs2024A&A,
       author = {{Backs}, F. and {Brands}, S.~A. and {de Koter}, A. and {Kaper}, L. and {Vink}, J.~S. and {Puls}, J. and {Sundqvist}, J. and {Tramper}, F. and {Sana}, H. and {Bernini-Peron}, M. and {Bestenlehner}, J.~M. and {Crowther}, P.~A. and {Hawcroft}, C. and {Ignace}, R. and {Kuiper}, R. and {van Loon}, J. Th. and {Mahy}, L. and {Marcolino}, W. and {Najarro}, F. and {Oskinova}, L.~M. and {Pauli}, D. and {Ramachandran}, V. and {Sander}, A.~A.~C. and {Verhamme}, O.},
        title = "{X-Shooting ULLYSES: Massive stars at low metallicity: VI. Atmosphere and mass-loss properties of O-type giants in the Small Magellanic Cloud}",
      journal = {\aap},
         year = 2024,
        month = dec,
       volume = {692},
          eid = {A88},
        pages = {A88},
          doi = {10.1051/0004-6361/202451893},
archivePrefix = {arXiv},
       eprint = {2411.06884},
 primaryClass = {astro-ph.SR},
       adsurl = {https://ui.adsabs.harvard.edu/abs/2024A&A...692A..88B}
}

@ARTICLE{Martins2024A&A,
       author = {{Martins}, F. and {Bouret}, J.-C. and {Hillier}, D.~J. and {Brands}, S.~A. and {Crowther}, P.~A. and {Herrero}, A. and {Najarro}, F. and {Pauli}, D. and {Puls}, J. and {Ramachandran}, V. and {Sander}, A.~A.~C. and {Vink}, J.~S. and {XShootU Collaboration}},
        title = "{X-Shooting ULLYSES: Massive stars at low metallicity: V. Effect of metallicity on surface abundances of O stars}",
      journal = {\aap},
         year = 2024,
        month = sep,
       volume = {689},
          eid = {A31},
        pages = {A31},
          doi = {10.1051/0004-6361/202449457},
archivePrefix = {arXiv},
       eprint = {2405.01267},
 primaryClass = {astro-ph.SR},
       adsurl = {https://ui.adsabs.harvard.edu/abs/2024A&A...689A..31M}
}

@ARTICLE{Brands2022A&A,
       author = {{Brands}, Sarah A. and {de Koter}, Alex and {Bestenlehner}, Joachim M. and {Crowther}, Paul A. and {Sundqvist}, Jon O. and {Puls}, Joachim and {Caballero-Nieves}, Saida M. and {Abdul-Masih}, Michael and {Driessen}, Florian A. and {Garc{\'\i}a}, Miriam and {Geen}, Sam and {Gr{\"a}fener}, G{\"o}tz and {Hawcroft}, Calum and {Kaper}, Lex and {Keszthelyi}, Zsolt and {Langer}, Norbert and {Sana}, Hugues and {Schneider}, Fabian R.~N. and {Shenar}, Tomer and {Vink}, Jorick S.},
        title = "{The R136 star cluster dissected with Hubble Space Telescope/STIS. III. The most massive stars and their clumped winds}",
      journal = {\aap},
         year = 2022,
        month = jul,
       volume = {663},
          eid = {A36},
        pages = {A36},
          doi = {10.1051/0004-6361/202142742},
archivePrefix = {arXiv},
       eprint = {2202.11080},
 primaryClass = {astro-ph.SR},
       adsurl = {https://ui.adsabs.harvard.edu/abs/2022A&A...663A..36B}
}

@ARTICLE{Hawcroft2024A&A,
       author = {{Hawcroft}, C. and {Sana}, H. and {Mahy}, L. and {Sundqvist}, J.~O. and {de Koter}, A. and {Crowther}, P.~A. and {Bestenlehner}, J.~M. and {Brands}, S.~A. and {David-Uraz}, A. and {Decin}, L. and {Erba}, C. and {Garcia}, M. and {Hamann}, W.-R. and {Herrero}, A. and {Ignace}, R. and {Kee}, N.~D. and {Kub{\'a}tov{\'a}}, B. and {Lefever}, R. and {Moffat}, A. and {Najarro}, F. and {Oskinova}, L. and {Pauli}, D. and {Prinja}, R. and {Puls}, J. and {Sander}, A.~A.~C. and {Shenar}, T. and {St-Louis}, N. and {ud-Doula}, A. and {Vink}, J.~S.},
        title = "{X-Shooting ULLYSES: Massive stars at low metallicity. III. Terminal wind speeds of ULLYSES massive stars}",
      journal = {\aap},
         year = 2024,
        month = aug,
       volume = {688},
          eid = {A105},
        pages = {A105},
          doi = {10.1051/0004-6361/202245588},
archivePrefix = {arXiv},
       eprint = {2303.12165},
 primaryClass = {astro-ph.SR},
       adsurl = {https://ui.adsabs.harvard.edu/abs/2024A&A...688A.105H}
}

@ARTICLE{Vink2023A&A,
       author = {{Vink}, Jorick S. and {Mehner}, A. and {Crowther}, P.~A. and {Fullerton}, A. and {Garcia}, M. and {Martins}, F. and {Morrell}, N. and {Oskinova}, L.~M. and {St-Louis}, N. and {ud-Doula}, A. and {Sander}, A.~A.~C. and {Sana}, H. and {Bouret}, J.-C. and {Kub{\'a}tov{\'a}}, B. and {Marchant}, P. and {Martins}, L.~P. and {Wofford}, A. and {van Loon}, J. Th. and {Grace Telford}, O. and {G{\"o}tberg}, Y. and {Bowman}, D.~M. and {Erba}, C. and {Kalari}, V.~M. and {Abdul-Masih}, M. and {Alkousa}, T. and {Backs}, F. and {Barbosa}, C.~L. and {Berlanas}, S.~R. and {Bernini-Peron}, M. and {Bestenlehner}, J.~M. and {Blomme}, R. and {Bodensteiner}, J. and {Brands}, S.~A. and {Evans}, C.~J. and {David-Uraz}, A. and {Driessen}, F.~A. and {Dsilva}, K. and {Geen}, S. and {G{\'o}mez-Gonz{\'a}lez}, V.~M.~A. and {Grassitelli}, L. and {Hamann}, W.-R. and {Hawcroft}, C. and {Herrero}, A. and {Higgins}, E.~R. and {John Hillier}, D. and {Ignace}, R. and {Istrate}, A.~G. and {Kaper}, L. and {Kee}, N.~D. and {Kehrig}, C. and {Keszthelyi}, Z. and {Klencki}, J. and {de Koter}, A. and {Kuiper}, R. and {Laplace}, E. and {Larkin}, C.~J.~K. and {Lefever}, R.~R. and {Leitherer}, C. and {Lennon}, D.~J. and {Mahy}, L. and {Ma{\'\i}z Apell{\'a}niz}, J. and {Maravelias}, G. and {Marcolino}, W. and {McLeod}, A.~F. and {de Mink}, S.~E. and {Najarro}, F. and {Oey}, M.~S. and {Parsons}, T.~N. and {Pauli}, D. and {Pedersen}, M.~G. and {Prinja}, R.~K. and {Ramachandran}, V. and {Ram{\'\i}rez-Tannus}, M.~C. and {Sabhahit}, G.~N. and {Schootemeijer}, A. and {Reyero Serantes}, S. and {Shenar}, T. and {Stringfellow}, G.~S. and {Sudnik}, N. and {Tramper}, F. and {Wang}, L.},
        title = "{X-Shooting ULLYSES: Massive stars at low metallicity. I. Project description}",
      journal = {\aap},
         year = 2023,
        month = jul,
       volume = {675},
          eid = {A154},
        pages = {A154},
          doi = {10.1051/0004-6361/202245650},
archivePrefix = {arXiv},
       eprint = {2305.06376},
 primaryClass = {astro-ph.SR},
       adsurl = {https://ui.adsabs.harvard.edu/abs/2023A&A...675A.154V}
}

@ARTICLE{Almeida2017A&A,
       author = {{Almeida}, L.~A. and {Sana}, H. and {Taylor}, W. and {Barb{\'a}}, R. and {Bonanos}, A.~Z. and {Crowther}, P. and {Damineli}, A. and {de Koter}, A. and {de Mink}, S.~E. and {Evans}, C.~J. and {Gieles}, M. and {Grin}, N.~J. and {H{\'e}nault-Brunet}, V. and {Langer}, N. and {Lennon}, D. and {Lockwood}, S. and {Ma{\'\i}z Apell{\'a}niz}, J. and {Moffat}, A.~F.~J. and {Neijssel}, C. and {Norman}, C. and {Ram{\'\i}rez-Agudelo}, O.~H. and {Richardson}, N.~D. and {Schootemeijer}, A. and {Shenar}, T. and {Soszy{\'n}ski}, I. and {Tramper}, F. and {Vink}, J.~S.},
        title = "{The Tarantula Massive Binary Monitoring. I. Observational campaign and OB-type spectroscopic binaries}",
      journal = {\aap},
         year = 2017,
        month = feb,
       volume = {598},
          eid = {A84},
        pages = {A84},
          doi = {10.1051/0004-6361/201629844},
archivePrefix = {arXiv},
       eprint = {1610.03500},
 primaryClass = {astro-ph.SR},
       adsurl = {https://ui.adsabs.harvard.edu/abs/2017A&A...598A..84A}
}

@ARTICLE{Schootemeijer2021,
       author = {{Schootemeijer}, A. and {Langer}, N. and {Lennon}, D. and {Evans}, C.~J. and {Crowther}, P.~A. and {Geen}, S. and {Howarth}, I. and {de Koter}, A. and {Menten}, K.~M. and {Vink}, J.~S.},
        title = "{A dearth of young and bright massive stars in the Small Magellanic Cloud}",
      journal = {\aap},
         year = 2021,
        month = feb,
       volume = {646},
          eid = {A106},
        pages = {A106},
          doi = {10.1051/0004-6361/202038789},
archivePrefix = {arXiv},
       eprint = {2012.05913},
 primaryClass = {astro-ph.GA},
       adsurl = {https://ui.adsabs.harvard.edu/abs/2021A&A...646A.106S}
}

@ARTICLE{Poelarends2008,
       author = {{Poelarends}, A.~J.~T. and {Herwig}, F. and {Langer}, N. and {Heger}, A.},
        title = "{The Supernova Channel of Super-AGB Stars}",
      journal = {\apj},
         year = 2008,
        month = mar,
       volume = {675},
       number = {1},
        pages = {614-625},
          doi = {10.1086/520872},
archivePrefix = {arXiv},
       eprint = {0705.4643},
 primaryClass = {astro-ph},
       adsurl = {https://ui.adsabs.harvard.edu/abs/2008ApJ...675..614P}
}

@ARTICLE{Calamida2008ApJ,
       author = {{Calamida}, A. and {Corsi}, C.~E. and {Bono}, G. and {Stetson}, P.~B. and {Prada Moroni}, P. and {Degl'Innocenti}, S. and {Ferraro}, I. and {Iannicola}, G. and {Koester}, D. and {Pulone}, L. and {Monelli}, M. and {Amico}, P. and {Buonanno}, R. and {Caputo}, F. and {D'Odorico}, S. and {Freyhammer}, L.~M. and {Marchetti}, E. and {Nonino}, M. and {Romaniello}, M.},
        title = "{On the White Dwarf Cooling Sequence of the Globular Cluster {\ensuremath{\omega}} Centauri}",
      journal = {\apjl},
         year = 2008,
        month = jan,
       volume = {673},
       number = {1},
        pages = {L29},
          doi = {10.1086/527436},
archivePrefix = {arXiv},
       eprint = {0712.0603},
 primaryClass = {astro-ph},
       adsurl = {https://ui.adsabs.harvard.edu/abs/2008ApJ...673L..29C}
}

@ARTICLE{Chen2022ApJ,
       author = {{Chen}, Jianxing and {Ferraro}, Francesco R. and {Cadelano}, Mario and {Salaris}, Maurizio and {Lanzoni}, Barbara and {Pallanca}, Cristina and {Althaus}, Leandro G. and {Cassisi}, Santi and {Dalessandro}, Emanuele},
        title = "{Slowly Cooling White Dwarfs in NGC 6752}",
      journal = {\apj},
         year = 2022,
        month = aug,
       volume = {934},
       number = {2},
          eid = {93},
        pages = {93},
          doi = {10.3847/1538-4357/ac7a45},
archivePrefix = {arXiv},
       eprint = {2206.10039},
 primaryClass = {astro-ph.SR},
       adsurl = {https://ui.adsabs.harvard.edu/abs/2022ApJ...934...93C}
}

@ARTICLE{Chen2021NatAs,
       author = {{Chen}, Jianxing and {Ferraro}, Francesco R. and {Cadelano}, Mario and {Salaris}, Maurizio and {Lanzoni}, Barbara and {Pallanca}, Cristina and {Althaus}, Leandro G. and {Dalessandro}, Emanuele},
        title = "{Slowly cooling white dwarfs in M13 from stable hydrogen burning}",
      journal = {Nature Astronomy},
         year = 2021,
        month = nov,
       volume = {5},
        pages = {1170-1177},
          doi = {10.1038/s41550-021-01445-6},
archivePrefix = {arXiv},
       eprint = {2109.02306},
 primaryClass = {astro-ph.GA},
       adsurl = {https://ui.adsabs.harvard.edu/abs/2021NatAs...5.1170C}
}

@ARTICLE{Gupta2025ApJ,
       author = {{Gupta}, Laksh and {Choudhury}, Samyaday and {Calamida}, Annalisa and {Johnson}, Christian I. and {Nardiello}, Domenico},
        title = "{An Excess of Luminous White Dwarfs in the Peculiar Galactic Globular Cluster NGC 2808}",
      journal = {\apj},
         year = 2025,
        month = nov,
       volume = {994},
       number = {1},
          eid = {97},
        pages = {97},
          doi = {10.3847/1538-4357/ae0ca6},
archivePrefix = {arXiv},
       eprint = {2509.26190},
 primaryClass = {astro-ph.SR},
       adsurl = {https://ui.adsabs.harvard.edu/abs/2025ApJ...994...97G}
}

@ARTICLE{Piatti+2018MNRAS,
       author = {{Piatti}, Andr{\'e}s E. and {Cole}, Andrew A. and {Emptage}, Bryn},
        title = "{Star cluster formation history along the minor axis of the Large Magellanic Cloud}",
      journal = {\mnras},
         year = 2018,
        month = jan,
       volume = {473},
       number = {1},
        pages = {105-115},
          doi = {10.1093/mnras/stx2418},
archivePrefix = {arXiv},
       eprint = {1709.05244},
 primaryClass = {astro-ph.GA},
       adsurl = {https://ui.adsabs.harvard.edu/abs/2018MNRAS.473..105P}
}

@ARTICLE{2024ApJ...971...71J,
       author = {{Jiang}, Yueyue and {Zhong}, Jing and {Qin}, Songmei and {Tang}, Tong and {Chen}, Li and {Hou}, Jinliang},
        title = "{On the Determination of Stellar Mass and Binary Fraction of Open Clusters within 500 pc from the Sun}",
      journal = {\apj},
         year = 2024,
        month = aug,
       volume = {971},
       number = {1},
          eid = {71},
        pages = {71},
          doi = {10.3847/1538-4357/ad5344},
archivePrefix = {arXiv},
       eprint = {2405.11853},
 primaryClass = {astro-ph.SR},
       adsurl = {https://ui.adsabs.harvard.edu/abs/2024ApJ...971...71J}
}

@ARTICLE{2021AJ....162..264J,
       author = {{Jadhav}, Vikrant V. and {Roy}, Kaustubh and {Joshi}, Naman and {Subramaniam}, Annapurni},
        title = "{High Mass-Ratio Binary Population in Open Clusters: Segregation of Early Type Binaries and an Increasing Binary Fraction with Mass}",
      journal = {\aj},
         year = 2021,
        month = dec,
       volume = {162},
       number = {6},
          eid = {264},
        pages = {264},
          doi = {10.3847/1538-3881/ac2571},
archivePrefix = {arXiv},
       eprint = {2109.03782},
 primaryClass = {astro-ph.SR},
       adsurl = {https://ui.adsabs.harvard.edu/abs/2021AJ....162..264J}
}

@ARTICLE{PU2000,
       author = {{Pietrzynski}, G. and {Udalski}, A.},
        title = "{The Optical Gravitational Lensing Experiment. Multiple Cluster Candidates in the Large Magellanic Cloud}",
      journal = {\actaa},
         year = 2000,
        month = sep,
       volume = {50},
        pages = {355-367},
          doi = {10.48550/arXiv.astro-ph/0010294},
archivePrefix = {arXiv},
       eprint = {astro-ph/0010294},
 primaryClass = {astro-ph},
       adsurl = {https://ui.adsabs.harvard.edu/abs/2000AcA....50..355P}
}

@ARTICLE{Hota2025,
       author = {{Hota}, S. and {Subramaniam}, A. and {Nayak}, P.~K. and {Subramanian}, S.},
        title = "{VizieR Online Data Catalog: U-SMAC II. FUV catalog of sources detected in SMC (Hota+, 2024)}",
 howpublished = {VizieR On-line Data Catalog: J/AJ/168/255. Originally published in: 2024AJ....168..255H},
         year = 2025,
      journal = {\aj},
        month = aug,
          eid = {J/AJ/168/255},
       adsurl = {https://ui.adsabs.harvard.edu/abs/2025yCat..51680255H}
}

@ARTICLE{BLOeM2024,
       author = {{Shenar}, T. and {Bodensteiner}, J. and {Sana}, H. and {Crowther}, P.~A. and {Lennon}, D.~J. and {Abdul-Masih}, M. and {Almeida}, L.~A. and {Backs}, F. and {Berlanas}, S.~R. and {Bernini-Peron}, M. and {Bestenlehner}, J.~M. and {Bowman}, D.~M. and {Bronner}, V.~A. and {Britavskiy}, N. and {de Koter}, A. and {de Mink}, S.~E. and {Deshmukh}, K. and {Evans}, C.~J. and {Fabry}, M. and {Gieles}, M. and {Gilkis}, A. and {Gonz{\'a}lez-Tor{\`a}}, G. and {Gr{\"a}fener}, G. and {G{\"o}tberg}, Y. and {Hawcroft}, C. and {H{\'e}nault-Brunet}, V. and {Herrero}, A. and {Holgado}, G. and {Janssens}, S. and {Johnston}, C. and {Josiek}, J. and {Justham}, S. and {Kalari}, V.~M. and {Katabi}, Z.~Z. and {Keszthelyi}, Z. and {Klencki}, J. and {Kub{\'a}t}, J. and {Kub{\'a}tov{\'a}}, B. and {Langer}, N. and {Lefever}, R.~R. and {Ludwig}, B. and {Mackey}, J. and {Mahy}, L. and {Ma{\'\i}z Apell{\'a}niz}, J. and {Mandel}, I. and {Maravelias}, G. and {Marchant}, P. and {Menon}, A. and {Najarro}, F. and {Oskinova}, L.~M. and {O'Grady}, A.~J.~G. and {Ovadia}, R. and {Patrick}, L.~R. and {Pauli}, D. and {Pawlak}, M. and {Ramachandran}, V. and {Renzo}, M. and {Rocha}, D.~F. and {Sander}, A.~A.~C. and {Sayada}, T. and {Schneider}, F.~R.~N. and {Schootemeijer}, A. and {Sch{\"o}sser}, E.~C. and {Sch{\"u}rmann}, C. and {Sen}, K. and {Shahaf}, S. and {Sim{\'o}n-D{\'\i}az}, S. and {Stoop}, M. and {Toonen}, S. and {Tramper}, F. and {van Loon}, J. Th. and {Valli}, R. and {van Son}, L.~A.~C. and {Vigna-G{\'o}mez}, A. and {Villase{\~n}or}, J.~I. and {Vink}, J.~S. and {Wang}, C. and {Willcox}, R.},
        title = "{Binarity at LOw Metallicity (BLOeM): A spectroscopic VLT monitoring survey of massive stars in the SMC}",
      journal = {\aap},
         year = 2024,
        month = oct,
       volume = {690},
          eid = {A289},
        pages = {A289},
          doi = {10.1051/0004-6361/202451586},
archivePrefix = {arXiv},
       eprint = {2407.14593},
 primaryClass = {astro-ph.SR},
       adsurl = {https://ui.adsabs.harvard.edu/abs/2024A&A...690A.289S}
}

@ARTICLE{RIOTS42016,
       author = {{Lamb}, J.~B. and {Oey}, M.~S. and {Segura-Cox}, D.~M. and {Graus}, A.~S. and {Kiminki}, D.~C. and {Golden-Marx}, J.~B. and {Parker}, J. Wm.},
        title = "{The Runaways and Isolated O-Type Star Spectroscopic Survey of the SMC (RIOTS4)}",
      journal = {\apj},
         year = 2016,
        month = feb,
       volume = {817},
       number = {2},
          eid = {113},
        pages = {113},
          doi = {10.3847/0004-637X/817/2/113},
archivePrefix = {arXiv},
       eprint = {1512.01233},
 primaryClass = {astro-ph.GA},
       adsurl = {https://ui.adsabs.harvard.edu/abs/2016ApJ...817..113L}
}

@ARTICLE{Shenar2023A&A,
       author = {{Shenar}, T. and {Sana}, H. and {Crowther}, P.~A. and {Bostroem}, K.~A. and {Mahy}, L. and {Najarro}, F. and {Oskinova}, L. and {Sander}, A.~A.~C.},
        title = "{Constraints on the multiplicity of the most massive stars known: R136 a1, a2, a3, and c}",
      journal = {\aap},
         year = 2023,
        month = nov,
       volume = {679},
          eid = {A36},
        pages = {A36},
          doi = {10.1051/0004-6361/202346930},
archivePrefix = {arXiv},
       eprint = {2309.13113},
 primaryClass = {astro-ph.SR},
       adsurl = {https://ui.adsabs.harvard.edu/abs/2023A&A...679A..36S}
}

@ARTICLE{VFTS2011,
       author = {{Evans}, C.~J. and {Taylor}, W.~D. and {H{\'e}nault-Brunet}, V. and {Sana}, H. and {de Koter}, A. and {Sim{\'o}n-D{\'\i}az}, S. and {Carraro}, G. and {Bagnoli}, T. and {Bastian}, N. and {Bestenlehner}, J.~M. and {Bonanos}, A.~Z. and {Bressert}, E. and {Brott}, I. and {Campbell}, M.~A. and {Cantiello}, M. and {Clark}, J.~S. and {Costa}, E. and {Crowther}, P.~A. and {de Mink}, S.~E. and {Doran}, E. and {Dufton}, P.~L. and {Dunstall}, P.~R. and {Friedrich}, K. and {Garcia}, M. and {Gieles}, M. and {Gr{\"a}fener}, G. and {Herrero}, A. and {Howarth}, I.~D. and {Izzard}, R.~G. and {Langer}, N. and {Lennon}, D.~J. and {Ma{\'\i}z Apell{\'a}niz}, J. and {Markova}, N. and {Najarro}, F. and {Puls}, J. and {Ramirez}, O.~H. and {Sab{\'\i}n-Sanjuli{\'a}n}, C. and {Smartt}, S.~J. and {Stroud}, V.~E. and {van Loon}, J. Th. and {Vink}, J.~S. and {Walborn}, N.~R.},
        title = "{The VLT-FLAMES Tarantula Survey. I. Introduction and observational overview}",
      journal = {\aap},
         year = 2011,
        month = jun,
       volume = {530},
          eid = {A108},
        pages = {A108},
          doi = {10.1051/0004-6361/201116782},
archivePrefix = {arXiv},
       eprint = {1103.5386},
 primaryClass = {astro-ph.SR},
       adsurl = {https://ui.adsabs.harvard.edu/abs/2011A&A...530A.108E}
}

@ARTICLE{Sana2013,
       author = {{Sana}, H. and {de Koter}, A. and {de Mink}, S.~E. and {Dunstall}, P.~R. and {Evans}, C.~J. and {H{\'e}nault-Brunet}, V. and {Ma{\'\i}z Apell{\'a}niz}, J. and {Ram{\'\i}rez-Agudelo}, O.~H. and {Taylor}, W.~D. and {Walborn}, N.~R. and {Clark}, J.~S. and {Crowther}, P.~A. and {Herrero}, A. and {Gieles}, M. and {Langer}, N. and {Lennon}, D.~J. and {Vink}, J.~S.},
        title = "{The VLT-FLAMES Tarantula Survey. VIII. Multiplicity properties of the O-type star population}",
      journal = {\aap},
         year = 2013,
        month = feb,
       volume = {550},
          eid = {A107},
        pages = {A107},
          doi = {10.1051/0004-6361/201219621},
archivePrefix = {arXiv},
       eprint = {1209.4638},
 primaryClass = {astro-ph.SR},
       adsurl = {https://ui.adsabs.harvard.edu/abs/2013A&A...550A.107S}
}

@article{RamirezAgudelo2013,
       author = {{Ramirez-Agudelo}, O.~H. and {Simon-Diaz}, S. and {Sana}, H. and {de Koter}, A. and {Sabin-Sanjulian}, C. and {de Mink}, S.~E. and {Dufton}, P.~L. and {Grafener}, G. and {Evans}, C.~J. and {Herrero}, A. and {Langer}, N. and {Lennon}, D.~J. and {Maiz Apellaniz}, J. and {Markova}, N. and {Najarro}, F. and {Puls}, J. and {Vink}, J.~S.},
        title = "{VizieR Online Data Catalog: O-stars in VLT-FLAMES Tarantula Survey (Ramirez-Agudelo+ 2013)}",
 howpublished = {VizieR On-line Data Catalog: J/A+A/560/A29. Originally published in: 2013A\&A...560A..29R},
         year = 2013,
      journal = {\aap},
        month = sep,
          eid = {J/A+A/560/A29},
          doi = {10.26093/cds/vizier.35600029},
       adsurl = {https://ui.adsabs.harvard.edu/abs/2013yCat..35600029R}
}

@ARTICLE{Martocchia+2018,
       author = {{Martocchia}, S. and {Niederhofer}, F. and {Dalessandro}, E. and {Bastian}, N. and {Kacharov}, N. and {Usher}, C. and {Cabrera-Ziri}, I. and {Lardo}, C. and {Cassisi}, S. and {Geisler}, D. and {Hilker}, M. and {Hollyhead}, K. and {Kozhurina-Platais}, V. and {Larsen}, S. and {Mackey}, D. and {Mucciarelli}, A. and {Platais}, I. and {Salaris}, M.},
        title = "{The search for multiple populations in Magellanic Cloud clusters - IV. Coeval multiple stellar populations in the young star cluster NGC 1978}",
      journal = {\mnras},
         year = 2018,
        month = jul,
       volume = {477},
       number = {4},
        pages = {4696-4705},
          doi = {10.1093/mnras/sty916},
archivePrefix = {arXiv},
       eprint = {1804.04141},
 primaryClass = {astro-ph.SR},
       adsurl = {https://ui.adsabs.harvard.edu/abs/2018MNRAS.477.4696M}
}

@ARTICLE{Martocchia+2019,
       author = {{Martocchia}, S. and {Dalessandro}, E. and {Lardo}, C. and {Cabrera-Ziri}, I. and {Bastian}, N. and {Kozhurina-Platais}, V. and {Salaris}, M. and {Chantereau}, W. and {Geisler}, D. and {Hilker}, M. and {Kacharov}, N. and {Larsen}, S. and {Mucciarelli}, A. and {Niederhofer}, F. and {Platais}, I. and {Usher}, C.},
        title = "{The search for multiple populations in Magellanic Clouds clusters - V. Correlation between cluster age and abundance spreads}",
      journal = {\mnras},
         year = 2019,
        month = aug,
       volume = {487},
       number = {4},
        pages = {5324-5334},
          doi = {10.1093/mnras/stz1596},
archivePrefix = {arXiv},
       eprint = {1906.03273},
 primaryClass = {astro-ph.SR},
       adsurl = {https://ui.adsabs.harvard.edu/abs/2019MNRAS.487.5324M}
}

@ARTICLE{Niederhofer+2017,
       author = {{Niederhofer}, F. and {Bastian}, N. and {Kozhurina-Platais}, V. and {Larsen}, S. and {Salaris}, M. and {Dalessandro}, E. and {Mucciarelli}, A. and {Cabrera-Ziri}, I. and {Cordero}, M. and {Geisler}, D. and {Hilker}, M. and {Hollyhead}, K. and {Kacharov}, N. and {Lardo}, C. and {Li}, C. and {Mackey}, D. and {Platais}, I.},
        title = "{The search for multiple populations in Magellanic Cloud clusters - I. Two stellar populations in the Small Magellanic Cloud globular cluster NGC 121}",
      journal = {\mnras},
         year = 2017,
        month = jan,
       volume = {464},
       number = {1},
        pages = {94-103},
          doi = {10.1093/mnras/stw2269},
archivePrefix = {arXiv},
       eprint = {1609.01595},
 primaryClass = {astro-ph.SR},
       adsurl = {https://ui.adsabs.harvard.edu/abs/2017MNRAS.464...94N}
}

@ARTICLE{Dalessandro+2016,
       author = {{Dalessandro}, E. and {Lapenna}, E. and {Mucciarelli}, A. and {Origlia}, L. and {Ferraro}, F.~R. and {Lanzoni}, B.},
        title = "{Multiple Populations in the Old and Massive Small Magellanic Cloud Globular Cluster NGC 121}",
      journal = {\apj},
         year = 2016,
        month = oct,
       volume = {829},
       number = {2},
          eid = {77},
        pages = {77},
          doi = {10.3847/0004-637X/829/2/77},
archivePrefix = {arXiv},
       eprint = {1607.05736},
 primaryClass = {astro-ph.SR},
       adsurl = {https://ui.adsabs.harvard.edu/abs/2016ApJ...829...77D}
}

@ARTICLE{Lagioia+2019a,
       author = {{Lagioia}, Edoardo P. and {Milone}, Antonino P. and {Marino}, Anna F. and {Dotter}, Aaron},
        title = "{Helium Variation in Four Small Magellanic Cloud Globular Clusters}",
      journal = {\apj},
         year = 2019,
        month = feb,
       volume = {871},
       number = {2},
          eid = {140},
        pages = {140},
          doi = {10.3847/1538-4357/aaf729},
       adsurl = {https://ui.adsabs.harvard.edu/abs/2019ApJ...871..140L}
}

@ARTICLE{Lagioia2019b,
       author = {{Lagioia}, Edoardo P. and {Milone}, Antonino P. and {Marino}, Anna F. and {Cordoni}, Giacomo and {Tailo}, Marco},
        title = "{The Role of Cluster Mass in the Multiple Populations of Galactic and Extragalactic Globular Clusters}",
      journal = {\aj},
         year = 2019,
        month = nov,
       volume = {158},
       number = {5},
          eid = {202},
        pages = {202},
          doi = {10.3847/1538-3881/ab45f2},
archivePrefix = {arXiv},
       eprint = {1909.08439},
 primaryClass = {astro-ph.SR},
       adsurl = {https://ui.adsabs.harvard.edu/abs/2019AJ....158..202L}
}

@ARTICLE{Choudhury2015,
       author = {{Choudhury}, Samyaday and {Subramaniam}, Annapurni and {Piatti}, Andr{\'e}s E.},
        title = "{Deep Washington Photometry of Inconspicuous Star Cluster Candidates in the Large Magellanic Cloud}",
      journal = {\aj},
         year = 2015,
        month = feb,
       volume = {149},
       number = {2},
          eid = {52},
        pages = {52},
          doi = {10.1088/0004-6256/149/2/52},
archivePrefix = {arXiv},
       eprint = {1410.7198},
 primaryClass = {astro-ph.GA},
       adsurl = {https://ui.adsabs.harvard.edu/abs/2015AJ....149...52C}
}

@ARTICLE{VISCACHA+2019,
       author = {{Maia}, Francisco F.~S. and {Dias}, Bruno and {Santos}, Jo{\~a}o F.~C. and {Kerber}, Leandro de O. and {Bica}, Eduardo and {Piatti}, Andr{\'e}s E. and {Barbuy}, Beatriz and {Quint}, Bruno and {Fraga}, Luciano and {Sanmartim}, David and {Angelo}, Mateus S. and {Hernandez-Jimenez}, Jose A. and {Katime Santrich}, Orlando J. and {Oliveira}, Raphael A.~P. and {P{\'e}rez-Villegas}, Angeles and {Souza}, Stefano O. and {Vieira}, Rodrigo G. and {Westera}, Pieter},
        title = "{The VISCACHA survey - I. Overview and first results}",
      journal = {\mnras},
         year = 2019,
        month = apr,
       volume = {484},
       number = {4},
        pages = {5702-5722},
          doi = {10.1093/mnras/stz369},
archivePrefix = {arXiv},
       eprint = {1902.01959},
 primaryClass = {astro-ph.GA},
       adsurl = {https://ui.adsabs.harvard.edu/abs/2019MNRAS.484.5702M}
}

@ARTICLE{Cioni2011,
       author = {{Cioni}, M. -R.~L. and {Clementini}, G. and {Girardi}, L. and {Guandalini}, R. and {Gullieuszik}, M. and {Miszalski}, B. and {Moretti}, M. -I. and {Ripepi}, V. and {Rubele}, S. and {Bagheri}, G. and {Bekki}, K. and {Cross}, N. and {de Blok}, W.~J.~G. and {de Grijs}, R. and {Emerson}, J.~P. and {Evans}, C.~J. and {Gibson}, B. and {Gonzales-Solares}, E. and {Groenewegen}, M.~A.~T. and {Irwin}, M. and {Ivanov}, V.~D. and {Lewis}, J. and {Marconi}, M. and {Marquette}, J. -B. and {Mastropietro}, C. and {Moore}, B. and {Napiwotzki}, R. and {Naylor}, T. and {Oliveira}, J.~M. and {Read}, M. and {Sutorius}, E. and {van Loon}, J. Th. and {Wilkinson}, M.~I. and {Wood}, P.~R.},
        title = "{The VMC survey. I. Strategy and first data}",
      journal = {\aap},
         year = 2011,
        month = mar,
       volume = {527},
          eid = {A116},
        pages = {A116},
          doi = {10.1051/0004-6361/201016137},
archivePrefix = {arXiv},
       eprint = {1012.5193},
 primaryClass = {astro-ph.CO},
       adsurl = {https://ui.adsabs.harvard.edu/abs/2011A&A...527A.116C}
}

@ARTICLE{SMASH2017,
       author = {{Nidever}, David L. and {Olsen}, Knut and {Walker}, Alistair R. and {Vivas}, A. Katherina and {Blum}, Robert D. and {Kaleida}, Catherine and {Choi}, Yumi and {Conn}, Blair C. and {Gruendl}, Robert A. and {Bell}, Eric F. and {Besla}, Gurtina and {Mu{\~n}oz}, Ricardo R. and {Gallart}, Carme and {Martin}, Nicolas F. and {Olszewski}, Edward W. and {Saha}, Abhijit and {Monachesi}, Antonela and {Monelli}, Matteo and {de Boer}, Thomas J.~L. and {Johnson}, L. Clifton and {Zaritsky}, Dennis and {Stringfellow}, Guy S. and {van der Marel}, Roeland P. and {Cioni}, Maria-Rosa L. and {Jin}, Shoko and {Majewski}, Steven R. and {Martinez-Delgado}, David and {Monteagudo}, Lara and {No{\"e}l}, Noelia E.~D. and {Bernard}, Edouard J. and {Kunder}, Andrea and {Chu}, You-Hua and {Bell}, Cameron P.~M. and {Santana}, Felipe and {Frechem}, Joshua and {Medina}, Gustavo E. and {Parkash}, Vaishali and {Navarrete}, J.~C. Ser{\'o}n and {Hayes}, Christian},
        title = "{SMASH: Survey of the MAgellanic Stellar History}",
      journal = {\aj},
         year = 2017,
        month = nov,
       volume = {154},
       number = {5},
          eid = {199},
        pages = {199},
          doi = {10.3847/1538-3881/aa8d1c},
archivePrefix = {arXiv},
       eprint = {1701.00502},
 primaryClass = {astro-ph.GA},
       adsurl = {https://ui.adsabs.harvard.edu/abs/2017AJ....154..199N}
}

@ARTICLE{Bica+1999AJ,
       author = {{Bica}, Eduardo L.~D. and {Schmitt}, Henrique R. and {Dutra}, Carlos M. and {Oliveira}, Humberto L.},
        title = "{A Revised and Extended Catalog of Magellanic System Clusters, Associations, and Emission Nebulae. II. The Large Magellanic Cloud}",
      journal = {\aj},
         year = 1999,
        month = jan,
       volume = {117},
       number = {1},
        pages = {238-246},
          doi = {10.1086/300687},
archivePrefix = {arXiv},
       eprint = {astro-ph/9810266},
 primaryClass = {astro-ph},
       adsurl = {https://ui.adsabs.harvard.edu/abs/1999AJ....117..238B}
}

@ARTICLE{Bica2020,
       author = {{Bica}, Eduardo and {Westera}, Pieter and {Kerber}, Leandro de O. and {Dias}, Bruno and {Maia}, Francisco and {Santos}, Jr., Jo{\~a}o F.~C. and {Barbuy}, Beatriz and {Oliveira}, Raphael A.~P.},
        title = "{An Updated Small Magellanic Cloud and Magellanic Bridge Catalog of Star Clusters, Associations, and Related Objects}",
      journal = {\aj},
         year = 2020,
        month = mar,
       volume = {159},
       number = {3},
          eid = {82},
        pages = {82},
          doi = {10.3847/1538-3881/ab6595},
archivePrefix = {arXiv},
       eprint = {1907.08642},
 primaryClass = {astro-ph.GA},
       adsurl = {https://ui.adsabs.harvard.edu/abs/2020AJ....159...82B}
}

@ARTICLE{Ramachandran2021,
       author = {{Ramachandran}, V. and {Oskinova}, L.~M. and {Hamann}, W. -R.},
        title = "{Discovery of O stars in the tidal Magellanic Bridge. Stellar parameters, abundances, and feedback of the nearest metal-poor massive stars and their implication for the Magellanic System ecology}",
      journal = {\aap},
         year = 2021,
        month = feb,
       volume = {646},
          eid = {A16},
        pages = {A16},
          doi = {10.1051/0004-6361/202039486},
archivePrefix = {arXiv},
       eprint = {2011.08006},
 primaryClass = {astro-ph.GA},
       adsurl = {https://ui.adsabs.harvard.edu/abs/2021A&A...646A..16R}
}

@ARTICLE{Bica_2008,
       author = {{Bica}, E. and {Bonatto}, C. and {Dutra}, C.~M. and {Santos}, J.~F.~C.},
        title = "{A general catalogue of extended objects in the Magellanic System}",
      journal = {\mnras},
         year = 2008,
        month = sep,
       volume = {389},
       number = {2},
        pages = {678-690},
          doi = {10.1111/j.1365-2966.2008.13612.x},
archivePrefix = {arXiv},
       eprint = {0806.3049},
 primaryClass = {astro-ph},
       adsurl = {https://ui.adsabs.harvard.edu/abs/2008MNRAS.389..678B}
}

@ARTICLE{Casetti-Dinescu2012,
       author = {{Casetti-Dinescu}, Dana I. and {Vieira}, Katherine and {Girard}, Terrence M. and {van Altena}, William F.},
        title = "{Constraints on the Magellanic Clouds' Interaction from the Distribution of OB Stars and the Kinematics of Giants}",
      journal = {\apj},
         year = 2012,
        month = jul,
       volume = {753},
       number = {2},
          eid = {123},
        pages = {123},
          doi = {10.1088/0004-637X/753/2/123},
archivePrefix = {arXiv},
       eprint = {1205.0989},
 primaryClass = {astro-ph.GA},
       adsurl = {https://ui.adsabs.harvard.edu/abs/2012ApJ...753..123C}
}

@ARTICLE{dhanushp1,
       author = {{Dhanush}, S.~R. and {Subramaniam}, A. and {Nayak}, Prasanta K. and {Subramanian}, S.},
        title = "{Spatiotemporal map of star clusters in the Magellanic Clouds using Gaia: synchronized peaks and radial shrinkage of cluster formation}",
      journal = {\mnras},
         year = 2024,
        month = feb,
       volume = {528},
       number = {2},
        pages = {2274-2298},
          doi = {10.1093/mnras/stae096},
archivePrefix = {arXiv},
       eprint = {2401.05307},
 primaryClass = {astro-ph.GA},
       adsurl = {https://ui.adsabs.harvard.edu/abs/2024MNRAS.528.2274D}
}

@ARTICLE{Nayak2016,
       author = {{Nayak}, P.~K. and {Subramaniam}, A. and {Choudhury}, S. and {Indu}, G. and {Sagar}, Ram},
        title = "{Star clusters in the Magellanic Clouds - I. Parametrization and classification of 1072 clusters in the LMC}",
      journal = {\mnras},
         year = 2016,
        month = dec,
       volume = {463},
       number = {2},
        pages = {1446-1461},
          doi = {10.1093/mnras/stw2043},
archivePrefix = {arXiv},
       eprint = {1608.06389},
 primaryClass = {astro-ph.GA},
       adsurl = {https://ui.adsabs.harvard.edu/abs/2016MNRAS.463.1446N}
}

@ARTICLE{Nayak2018,
       author = {{Nayak}, P.~K. and {Subramaniam}, A. and {Choudhury}, S. and {Sagar}, Ram},
        title = "{Star clusters in the Magellanic Clouds. II. Age-dating, classification, and spatio-temporal distribution of the SMC clusters}",
      journal = {\aap},
         year = 2018,
        month = sep,
       volume = {616},
          eid = {A187},
        pages = {A187},
          doi = {10.1051/0004-6361/201732227},
archivePrefix = {arXiv},
       eprint = {1804.00635},
 primaryClass = {astro-ph.GA},
       adsurl = {https://ui.adsabs.harvard.edu/abs/2018A&A...616A.187N}
}

@ARTICLE{Udalski+1997,
       author = {{Udalski}, A. and {Kubiak}, M. and {Szymanski}, M.},
        title = "{Optical Gravitational Lensing Experiment. OGLE-2 -- the Second Phase of the OGLE Project}",
      journal = {\actaa},
         year = 1997,
        month = jul,
       volume = {47},
        pages = {319-344},
          doi = {10.48550/arXiv.astro-ph/9710091},
archivePrefix = {arXiv},
       eprint = {astro-ph/9710091},
 primaryClass = {astro-ph},
       adsurl = {https://ui.adsabs.harvard.edu/abs/1997AcA....47..319U}
}

@ARTICLE{Zaritsky+2002AJ,
       author = {{Zaritsky}, Dennis and {Harris}, Jason and {Thompson}, Ian B. and {Grebel}, Eva K. and {Massey}, Philip},
        title = "{The Magellanic Clouds Photometric Survey: The Small Magellanic Cloud Stellar Catalog and Extinction Map}",
      journal = {\aj},
         year = 2002,
        month = feb,
       volume = {123},
       number = {2},
        pages = {855-872},
          doi = {10.1086/338437},
archivePrefix = {arXiv},
       eprint = {astro-ph/0110665},
 primaryClass = {astro-ph},
       adsurl = {https://ui.adsabs.harvard.edu/abs/2002AJ....123..855Z}
}

@ARTICLE{Zaritsky+2004,
       author = {{Zaritsky}, Dennis and {Harris}, Jason and {Thompson}, Ian B. and {Grebel}, Eva K.},
        title = "{The Magellanic Clouds Photometric Survey: The Large Magellanic Cloud Stellar Catalog and Extinction Map}",
      journal = {\aj},
         year = 2004,
        month = oct,
       volume = {128},
       number = {4},
        pages = {1606-1614},
          doi = {10.1086/423910},
archivePrefix = {arXiv},
       eprint = {astro-ph/0407006},
 primaryClass = {astro-ph},
       adsurl = {https://ui.adsabs.harvard.edu/abs/2004AJ....128.1606Z}
}

@ARTICLE{Sana2025NatAs,
       author = {{Sana}, H. and {Shenar}, T. and {Bodensteiner}, J. and {Britavskiy}, N. and {Langer}, N. and {Lennon}, D.~J. and {Mahy}, L. and {Mandel}, I. and {de Mink}, S.~E. and {Patrick}, L.~R. and {Villase{\~n}or}, J.~I. and {Dirickx}, M. and {Abdul-Masih}, M. and {Almeida}, L.~A. and {Backs}, F. and {Berlanas}, S.~R. and {Bernini-Peron}, M. and {Bowman}, D.~M. and {Bronner}, V.~A. and {Crowther}, P.~A. and {Deshmukh}, K. and {Evans}, C.~J. and {Fabry}, M. and {Gieles}, M. and {Gilkis}, A. and {Gonz{\'a}lez-Tor{\`a}}, G. and {Gr{\"a}fener}, G. and {G{\"o}tberg}, Y. and {Hawcroft}, C. and {H{\'e}nault-Brunet}, V. and {Herrero}, A. and {Holgado}, G. and {Izzard}, R.~G. and {de Koter}, A. and {Janssens}, S. and {Johnston}, C. and {Josiek}, J. and {Justham}, S. and {Kalari}, V.~M. and {Klencki}, J. and {Kub{\'a}t}, J. and {Kub{\'a}tov{\'a}}, B. and {Lefever}, R.~R. and {van Loon}, J. Th. and {Ludwig}, B. and {Mackey}, J. and {Ma{\'\i}z Apell{\'a}niz}, J. and {Maravelias}, G. and {Marchant}, P. and {Mazeh}, T. and {Menon}, A. and {Moe}, M. and {Najarro}, F. and {Oskinova}, L.~M. and {Ovadia}, R. and {Pauli}, D. and {Pawlak}, M. and {Ramachandran}, V. and {Renzo}, M. and {Rocha}, D.~F. and {Sander}, A.~A.~C. and {Schneider}, F.~R.~N. and {Schootemeijer}, A. and {Sch{\"o}sser}, E.~C. and {Sch{\"u}rmann}, C. and {Sen}, K. and {Shahaf}, S. and {Sim{\'o}n-D{\'\i}az}, S. and {van Son}, L.~A.~C. and {Stoop}, M. and {Toonen}, S. and {Tramper}, F. and {Valli}, R. and {Vigna-G{\'o}mez}, A. and {Vink}, J.~S. and {Wang}, C. and {Willcox}, R.},
        title = "{A high fraction of close massive binary stars at low metallicity}",
      journal = {Nature Astronomy},
         year = 2025,
        month = sep,
       volume = {9},
        pages = {1337-1346},
          doi = {10.1038/s41550-025-02610-x},
archivePrefix = {arXiv},
       eprint = {2509.12488},
 primaryClass = {astro-ph.SR},
       adsurl = {https://ui.adsabs.harvard.edu/abs/2025NatAs...9.1337S}
}

@ARTICLE{Hainich2015A&A,
       author = {{Hainich}, R. and {Pasemann}, D. and {Todt}, H. and {Shenar}, T. and {Sander}, A. and {Hamann}, W.-R.},
        title = "{Wolf-Rayet stars in the Small Magellanic Cloud. I. Analysis of the single WN stars}",
      journal = {\aap},
         year = 2015,
        month = sep,
       volume = {581},
          eid = {A21},
        pages = {A21},
          doi = {10.1051/0004-6361/201526241},
archivePrefix = {arXiv},
       eprint = {1507.04000},
 primaryClass = {astro-ph.SR},
       adsurl = {https://ui.adsabs.harvard.edu/abs/2015A&A...581A..21H}
}

@ARTICLE{Bodensteiner2025A&A,
       author = {{Bodensteiner}, J. and {Shenar}, T. and {Sana}, H. and {Britavskiy}, N. and {Crowther}, P.~A. and {Langer}, N. and {Lennon}, D.~J. and {Mahy}, L. and {Patrick}, L.~R. and {Villase{\~n}or}, J.~I. and {Abdul-Masih}, M. and {Bowman}, D.~M. and {de Koter}, A. and {de Mink}, S.~E. and {Deshmukh}, K. and {Fabry}, M. and {Gilkis}, A. and {G{\"o}tberg}, Y. and {Holgado}, G. and {Izzard}, R.~G. and {Janssens}, S. and {Kalari}, V.~M. and {Keszthelyi}, Z. and {Kub{\'a}t}, J. and {Mandel}, I. and {Maravelias}, G. and {Oskinova}, L.~M. and {Pauli}, D. and {Ramachandran}, V. and {Rocha}, D.~F. and {Renzo}, M. and {Sander}, A.~A.~C. and {Schneider}, F.~R.~N. and {Schootemeijer}, A. and {Sen}, K. and {Stoop}, M. and {Toonen}, S. and {van Loon}, J. Th. and {Valli}, R. and {Vigna-G{\'o}mez}, A. and {Vink}, J.~S. and {Wang}, C. and {Xu}, X.-T.},
        title = "{Binarity at LOw Metallicity (BLOeM): Multiplicity properties of Oe and Be stars}",
      journal = {\aap},
         year = 2025,
        month = jun,
       volume = {698},
          eid = {A38},
        pages = {A38},
          doi = {10.1051/0004-6361/202452623},
archivePrefix = {arXiv},
       eprint = {2502.02641},
 primaryClass = {astro-ph.SR},
       adsurl = {https://ui.adsabs.harvard.edu/abs/2025A&A...698A..38B}
}

@ARTICLE{Schootemeijer2022A&A,
       author = {{Schootemeijer}, A. and {Lennon}, D.~J. and {Garcia}, M. and {Langer}, N. and {Hastings}, B. and {Sch{\"u}rmann}, C.},
        title = "{A census of OBe stars in nearby metal-poor dwarf galaxies reveals a high fraction of extreme rotators}",
      journal = {\aap},
         year = 2022,
        month = nov,
       volume = {667},
          eid = {A100},
        pages = {A100},
          doi = {10.1051/0004-6361/202244730},
archivePrefix = {arXiv},
       eprint = {2209.04943},
 primaryClass = {astro-ph.GA},
       adsurl = {https://ui.adsabs.harvard.edu/abs/2022A&A...667A.100S}
}

@ARTICLE{1980A&A....85L..21R,
       author = {{Rosa}, M.},
        title = "{IUE UV spectra of giant extragalactic H II regions.}",
      journal = {\aap},
         year = 1980,
        month = may,
       volume = {85},
        pages = {L21-L24},
       adsurl = {https://ui.adsabs.harvard.edu/abs/1980A&A....85L..21R}
}

@ARTICLE{Subramaniam2016ApJ...833L..27S,
       author = {{Subramaniam}, Annapurni and {Sindhu}, N. and {Tandon}, S.~N. and {Kameswara Rao}, N. and {Postma}, J. and {C{\^o}t{\'e}}, Patrick and {Hutchings}, J.~B. and {Ghosh}, S.~K. and {George}, K. and {Girish}, V. and {Mohan}, R. and {Murthy}, J. and {Sankarasubramanian}, K. and {Stalin}, C.~S. and {Sutaria}, F. and {Mondal}, C. and {Sahu}, S.},
        title = "{A Hot Companion to a Blue Straggler in NGC 188 as Revealed by the Ultra-Violet Imaging Telescope (UVIT) on ASTROSAT}",
      journal = {\apjl},
         year = 2016,
        month = dec,
       volume = {833},
       number = {2},
          eid = {L27},
        pages = {L27},
          doi = {10.3847/2041-8213/833/2/L27},
       adsurl = {https://ui.adsabs.harvard.edu/abs/2016ApJ...833L..27S}
}

@ARTICLE{Jadhav2023A&A...676A..47J,
       author = {{Jadhav}, Vikrant V. and {Subramaniam}, Annapurni and {Sagar}, Ram},
        title = "{UOCS. X. Rich collection of post-mass-transfer systems in NGC 6791}",
      journal = {\aap},
         year = 2023,
        month = aug,
       volume = {676},
          eid = {A47},
        pages = {A47},
          doi = {10.1051/0004-6361/202345907},
archivePrefix = {arXiv},
       eprint = {2306.15396},
 primaryClass = {astro-ph.SR},
       adsurl = {https://ui.adsabs.harvard.edu/abs/2023A&A...676A..47J}
}

@ARTICLE{2021MNRAS.507.1699J,
       author = {{Jadhav}, Vikrant V. and {Subramaniam}, Annapurni},
        title = "{Blue straggler stars in open clusters using Gaia: dependence on cluster parameters and possible formation pathways}",
      journal = {\mnras},
         year = 2021,
        month = oct,
       volume = {507},
       number = {2},
        pages = {1699-1709},
          doi = {10.1093/mnras/stab2264},
archivePrefix = {arXiv},
       eprint = {2108.01396},
 primaryClass = {astro-ph.SR},
       adsurl = {https://ui.adsabs.harvard.edu/abs/2021MNRAS.507.1699J}
}

@ARTICLE{Jadhav2021JApA...42...89J,
       author = {{Jadhav}, Vikrant V. and {Pandey}, Sindhu and {Subramaniam}, Annapurni and {Sagar}, Ram},
        title = "{UOCS. IV. Discovery of diverse hot companions to blue stragglers in the old open cluster King 2}",
      journal = {Journal of Astrophysics and Astronomy},
         year = 2021,
        month = oct,
       volume = {42},
       number = {2},
          eid = {89},
        pages = {89},
          doi = {10.1007/s12036-021-09746-y},
archivePrefix = {arXiv},
       eprint = {2102.13375},
 primaryClass = {astro-ph.SR},
       adsurl = {https://ui.adsabs.harvard.edu/abs/2021JApA...42...89J}
}

@ARTICLE{Sindhu2019ApJ...882...43S,
       author = {{Sindhu}, N. and {Subramaniam}, Annapurni and {Jadhav}, Vikrant V. and {Chatterjee}, Sourav and {Geller}, Aaron M. and {Knigge}, Christian and {Leigh}, Nathan and {Puzia}, Thomas H. and {Shara}, Michael and {Simunovic}, Mirko},
        title = "{UVIT Open Cluster Study. I. Detection of a White Dwarf Companion to a Blue Straggler in M67: Evidence of Formation through Mass Transfer}",
      journal = {\apj},
         year = 2019,
        month = sep,
       volume = {882},
       number = {1},
          eid = {43},
        pages = {43},
          doi = {10.3847/1538-4357/ab31a8},
archivePrefix = {arXiv},
       eprint = {1907.05556},
 primaryClass = {astro-ph.SR},
       adsurl = {https://ui.adsabs.harvard.edu/abs/2019ApJ...882...43S}
}

@ARTICLE{Gosnell2015ApJ...814..163G,
       author = {{Gosnell}, Natalie M. and {Mathieu}, Robert D. and {Geller}, Aaron M. and {Sills}, Alison and {Leigh}, Nathan and {Knigge}, Christian},
        title = "{Implications for the Formation of Blue Straggler Stars from HST Ultraviolet Observations of NGC 188}",
      journal = {\apj},
         year = 2015,
        month = dec,
       volume = {814},
       number = {2},
          eid = {163},
        pages = {163},
          doi = {10.1088/0004-637X/814/2/163},
archivePrefix = {arXiv},
       eprint = {1510.04290},
 primaryClass = {astro-ph.SR},
       adsurl = {https://ui.adsabs.harvard.edu/abs/2015ApJ...814..163G}
}

@ARTICLE{Jadhav2019ApJ...886...13J,
       author = {{Jadhav}, Vikrant V. and {Sindhu}, N. and {Subramaniam}, Annapurni},
        title = "{UVIT Open Cluster Study. II. Detection of Extremely Low Mass White Dwarfs and Post-Mass Transfer Binaries in M67}",
      journal = {\apj},
         year = 2019,
        month = nov,
       volume = {886},
       number = {1},
          eid = {13},
        pages = {13},
          doi = {10.3847/1538-4357/ab4b43},
archivePrefix = {arXiv},
       eprint = {1910.01819},
 primaryClass = {astro-ph.SR},
       adsurl = {https://ui.adsabs.harvard.edu/abs/2019ApJ...886...13J}
}

@ARTICLE{Leiner2019ApJ...881...47L,
       author = {{Leiner}, Emily and {Mathieu}, Robert D. and {Vanderburg}, Andrew and {Gosnell}, Natalie M. and {Smith}, Jeffrey C.},
        title = "{Blue Lurkers: Hidden Blue Stragglers on the M67 Main Sequence Identified from Their Kepler/K2 Rotation Periods}",
      journal = {\apj},
         year = 2019,
        month = aug,
       volume = {881},
       number = {1},
          eid = {47},
        pages = {47},
          doi = {10.3847/1538-4357/ab2bf8},
archivePrefix = {arXiv},
       eprint = {1904.02169},
 primaryClass = {astro-ph.SR},
       adsurl = {https://ui.adsabs.harvard.edu/abs/2019ApJ...881...47L}
}

@ARTICLE{Geier2020A&A...635A.193G,
       author = {{Geier}, S.},
        title = "{The population of hot subdwarf stars studied with Gaia. III. Catalogue of known hot subdwarf stars: Data Release 2}",
      journal = {\aap},
         year = 2020,
        month = mar,
       volume = {635},
          eid = {A193},
        pages = {A193},
          doi = {10.1051/0004-6361/202037526},
archivePrefix = {arXiv},
       eprint = {2002.10896},
 primaryClass = {astro-ph.SR},
       adsurl = {https://ui.adsabs.harvard.edu/abs/2020A&A...635A.193G}
}

@ARTICLE{Brown2010ApJ...723.1072B,
       author = {{Brown}, Warren R. and {Kilic}, Mukremin and {Allende Prieto}, Carlos and {Kenyon}, Scott J.},
        title = "{The ELM Survey. I. A Complete Sample of Extremely Low-mass White Dwarfs}",
      journal = {\apj},
         year = 2010,
        month = nov,
       volume = {723},
       number = {2},
        pages = {1072-1081},
          doi = {10.1088/0004-637X/723/2/1072},
archivePrefix = {arXiv},
       eprint = {1011.3050},
 primaryClass = {astro-ph.GA},
       adsurl = {https://ui.adsabs.harvard.edu/abs/2010ApJ...723.1072B}
}

@ARTICLE{Bressan2012MNRAS.427..127B,
       author = {{Bressan}, Alessandro and {Marigo}, Paola and {Girardi}, L{\'e}o. and {Salasnich}, Bernardo and {Dal Cero}, Claudia and {Rubele}, Stefano and {Nanni}, Ambra},
        title = "{PARSEC: stellar tracks and isochrones with the PAdova and TRieste Stellar Evolution Code}",
      journal = {\mnras},
         year = 2012,
        month = nov,
       volume = {427},
       number = {1},
        pages = {127-145},
          doi = {10.1111/j.1365-2966.2012.21948.x},
archivePrefix = {arXiv},
       eprint = {1208.4498},
 primaryClass = {astro-ph.SR},
       adsurl = {https://ui.adsabs.harvard.edu/abs/2012MNRAS.427..127B}
}

@ARTICLE{Bedard2020ApJ...901...93B,
       author = {{B{\'e}dard}, A. and {Bergeron}, P. and {Brassard}, P. and {Fontaine}, G.},
        title = "{On the Spectral Evolution of Hot White Dwarf Stars. I. A Detailed Model Atmosphere Analysis of Hot White Dwarfs from SDSS DR12}",
      journal = {\apj},
         year = 2020,
        month = oct,
       volume = {901},
       number = {2},
          eid = {93},
        pages = {93},
          doi = {10.3847/1538-4357/abafbe},
archivePrefix = {arXiv},
       eprint = {2008.07469},
 primaryClass = {astro-ph.SR},
       adsurl = {https://ui.adsabs.harvard.edu/abs/2020ApJ...901...93B}
}

@ARTICLE{Gotberg2018A&A...615A..78G,
       author = {{G{\"o}tberg}, Y. and {de Mink}, S.~E. and {Groh}, J.~H. and {Kupfer}, T. and {Crowther}, P.~A. and {Zapartas}, E. and {Renzo}, M.},
        title = "{Spectral models for binary products: Unifying subdwarfs and Wolf-Rayet stars as a sequence of stripped-envelope stars}",
      journal = {\aap},
         year = 2018,
        month = jul,
       volume = {615},
          eid = {A78},
        pages = {A78},
          doi = {10.1051/0004-6361/201732274},
archivePrefix = {arXiv},
       eprint = {1802.03018},
 primaryClass = {astro-ph.SR},
       adsurl = {https://ui.adsabs.harvard.edu/abs/2018A&A...615A..78G}
}

@ARTICLE{Naoz2014ApJ...793..137N,
       author = {{Naoz}, Smadar and {Fabrycky}, Daniel C.},
        title = "{Mergers and Obliquities in Stellar Triples}",
      journal = {\apj},
         year = 2014,
        month = oct,
       volume = {793},
       number = {2},
          eid = {137},
        pages = {137},
          doi = {10.1088/0004-637X/793/2/137},
archivePrefix = {arXiv},
       eprint = {1405.5223},
 primaryClass = {astro-ph.SR},
       adsurl = {https://ui.adsabs.harvard.edu/abs/2014ApJ...793..137N}
}

@ARTICLE{Hills1976ApL....17...87H,
       author = {{Hills}, J.~G. and {Day}, C.~A.},
        title = "{Stellar Collisions in Globular Clusters}",
      journal = {\aplett},
         year = 1976,
        month = feb,
       volume = {17},
        pages = {87},
       adsurl = {https://ui.adsabs.harvard.edu/abs/1976ApL....17...87H}
}

@ARTICLE{McCrea1964MNRAS.128..147M,
       author = {{McCrea}, W.~H.},
        title = "{Extended main-sequence of some stellar clusters}",
      journal = {\mnras},
         year = 1964,
        month = jan,
       volume = {128},
        pages = {147},
          doi = {10.1093/mnras/128.2.147},
       adsurl = {https://ui.adsabs.harvard.edu/abs/1964MNRAS.128..147M}
}

@ARTICLE{2017ApJS..230...24B,
       author = {{Bianchi}, Luciana and {Shiao}, Bernie and {Thilker}, David},
        title = "{Revised Catalog of GALEX Ultraviolet Sources. I. The All-Sky Survey: GUVcat\_AIS}",
      journal = {\apjs},
         year = 2017,
        month = jun,
       volume = {230},
       number = {2},
          eid = {24},
        pages = {24},
          doi = {10.3847/1538-4365/aa7053},
archivePrefix = {arXiv},
       eprint = {1704.05903},
 primaryClass = {astro-ph.GA},
       adsurl = {https://ui.adsabs.harvard.edu/abs/2017ApJS..230...24B}
}

@ARTICLE{2024A&A...686A..42H,
       author = {{Hunt}, Emily L. and {Reffert}, Sabine},
        title = "{Improving the open cluster census. III. Using cluster masses, radii, and dynamics to create a cleaned open cluster catalogue}",
      journal = {\aap},
         year = 2024,
        month = jun,
       volume = {686},
          eid = {A42},
        pages = {A42},
          doi = {10.1051/0004-6361/202348662},
archivePrefix = {arXiv},
       eprint = {2403.05143},
 primaryClass = {astro-ph.GA},
       adsurl = {https://ui.adsabs.harvard.edu/abs/2024A&A...686A..42H}
}

@ARTICLE{1981A&A...103..305L,
       author = {{Lequeux}, J. and {Maucherat-Joubert}, M. and {Deharveng}, J.~M. and {Kunth}, D.},
        title = "{Star formation and extinction in extragalactic H II regions.}",
      journal = {\aap},
         year = 1981,
        month = nov,
       volume = {103},
        pages = {305-318},
       adsurl = {https://ui.adsabs.harvard.edu/abs/1981A&A...103..305L}
}

@MISC{1984STIN...8519904L,
       author = {{Lamb}, S.~A. and {Gallacgher}, J.~S. and {Hjellming}, M. and {Hunter}, D.~A.},
        title = "{International Ultraviolet Explorer observations of amorphous hot galaxies}",
         year = 1984,
        month = aug,
        pages = {19904},
       adsurl = {https://ui.adsabs.harvard.edu/abs/1984STIN...8519904L}
}

@ARTICLE{1985A&A...143..347D,
       author = {{Durret}, F. and {Bergeron}, J. and {Boksenberg}, A.},
        title = "{Gas and star content and spatial distribution in the giant extragalactic H II region TOL 89.}",
      journal = {\aap},
         year = 1985,
        month = feb,
       volume = {143},
        pages = {347-354},
       adsurl = {https://ui.adsabs.harvard.edu/abs/1985A&A...143..347D}
}

@ARTICLE{1994A&A...291....1R,
       author = {{Rosa}, M.~R. and {Benvenuti}, P.},
        title = "{The IMF and the extinction law in M 101: HST FOS spectra of extragalactic HII regions.}",
      journal = {\aap},
         year = 1994,
        month = nov,
       volume = {291},
        pages = {1-17},
       adsurl = {https://ui.adsabs.harvard.edu/abs/1994A&A...291....1R}
}

@ARTICLE{2007MNRAS.377..987K,
       author = {{Kaviraj}, S. and {Sohn}, S.~T. and {O'Connell}, R.~W. and {Yoon}, S. -J. and {Lee}, Y.~W. and {Yi}, S.~K.},
        title = "{UV bright globular clusters in M87: more evidence for super-He-rich stellar populations?}",
      journal = {\mnras},
         year = 2007,
        month = may,
       volume = {377},
       number = {3},
        pages = {987-996},
          doi = {10.1111/j.1365-2966.2007.11712.x},
archivePrefix = {arXiv},
       eprint = {astro-ph/0703198},
 primaryClass = {astro-ph},
       adsurl = {https://ui.adsabs.harvard.edu/abs/2007MNRAS.377..987K}
}

@ARTICLE{2017MNRAS.464..713P,
       author = {{Peacock}, Mark B. and {Zepf}, Stephen E. and {Kundu}, Arunav and {Chael}, Julia},
        title = "{Globular clusters in the far-ultraviolet: evidence for He-enriched second populations in extragalactic globular clusters?}",
      journal = {\mnras},
         year = 2017,
        month = jan,
       volume = {464},
       number = {1},
        pages = {713-720},
          doi = {10.1093/mnras/stw2382},
archivePrefix = {arXiv},
       eprint = {1609.04819},
 primaryClass = {astro-ph.GA},
       adsurl = {https://ui.adsabs.harvard.edu/abs/2017MNRAS.464..713P}
}

@ARTICLE{2007ApJ...659..359M,
       author = {{Ma}, Jun and {Yang}, Yanbin and {Burstein}, David and {Fan}, Zhou and {Wu}, Zhenyu and {Zhou}, Xu and {Wu}, Jianghua and {Jiang}, Zhaoji and {Chen}, Jiansheng},
        title = "{Age Constraints for an M31 Globular Cluster from SEDs Fit}",
      journal = {\apj},
         year = 2007,
        month = apr,
       volume = {659},
       number = {1},
        pages = {359-364},
          doi = {10.1086/511850},
archivePrefix = {arXiv},
       eprint = {astro-ph/0612368},
 primaryClass = {astro-ph},
       adsurl = {https://ui.adsabs.harvard.edu/abs/2007ApJ...659..359M}
}

@ARTICLE{2009AJ....137.4884M,
       author = {{Ma}, Jun and {Fan}, Zhou and {de Grijs}, Richard and {Wu}, Zhenyu and {Zhou}, Xu and {Wu}, Jianghua and {Jiang}, Zhaoji and {Chen}, Jiansheng},
        title = "{Spectral Energy Distributions and Age Estimates of 39 Globular Clusters in M31}",
      journal = {\aj},
         year = 2009,
        month = jun,
       volume = {137},
       number = {6},
        pages = {4884-4896},
          doi = {10.1088/0004-6256/137/6/4884},
archivePrefix = {arXiv},
       eprint = {0904.0553},
 primaryClass = {astro-ph.CO},
       adsurl = {https://ui.adsabs.harvard.edu/abs/2009AJ....137.4884M}
}

@INPROCEEDINGS{1998AAS...193.6002P,
       author = {{Peterson}, R.~C.},
        title = "{Modeling the Near-UV Light of Extragalactic Globular Clusters}",
    booktitle = {American Astronomical Society Meeting Abstracts},
         year = 1998,
       series = {American Astronomical Society Meeting Abstracts},
       volume = {193},
        month = dec,
          eid = {60.02},
        pages = {60.02},
       adsurl = {https://ui.adsabs.harvard.edu/abs/1998AAS...193.6002P}
}

@INPROCEEDINGS{2002AAS...200.4016P,
       author = {{Peterson}, R.~C. and {Carney}, B.~W. and {Dorman}, B. and {Landsman}, W. and {Green}, E.~M. and {Liebert}, J. and {O'Connell}, R.~W. and {Rood}, R.~T.},
        title = "{Ages and Metallicities of Extragalactic Globular Clusters and Galaxies from Near-UV Spectra}",
    booktitle = {American Astronomical Society Meeting Abstracts \#200},
         year = 2002,
       series = {American Astronomical Society Meeting Abstracts},
       volume = {200},
        month = may,
          eid = {40.16},
        pages = {40.16},
       adsurl = {https://ui.adsabs.harvard.edu/abs/2002AAS...200.4016P}
}

@ARTICLE{2003MNRAS.342..259D,
       author = {{de Grijs}, R. and {Fritze-v. Alvensleben}, U. and {Anders}, P. and {Gallagher}, J.~S. and {Bastian}, N. and {Taylor}, V.~A. and {Windhorst}, R.~A.},
        title = "{Star cluster formation and evolution in nearby starburst galaxies - I. Systematic uncertainties}",
      journal = {\mnras},
         year = 2003,
        month = jun,
       volume = {342},
       number = {1},
        pages = {259-273},
          doi = {10.1046/j.1365-8711.2003.06536.x},
archivePrefix = {arXiv},
       eprint = {astro-ph/0302286},
 primaryClass = {astro-ph},
       adsurl = {https://ui.adsabs.harvard.edu/abs/2003MNRAS.342..259D}
}

@INPROCEEDINGS{2011sca..conf..129J,
       author = {{Johnson}, L.~C. and {Seth}, A.~C. and {Dalcanton}, J.~J. and {Caldwell}, N. and {Gouliermis}, D.~A. and {Hodge}, P.~W. and {Larsen}, S.~S. and {Olsen}, K.~A.~G. and {San Roman}, I. and {Sarajedini}, A. and {Weisz}, D.~R. and {Phat Collaboration}},
        title = "{Stellar Clusters in M 31 from PHAT: Survey Overview and First Results}",
    booktitle = {Stellar Clusters \& Associations: A RIA Workshop on Gaia},
         year = 2011,
        month = jan,
        pages = {129-132},
          doi = {10.48550/arXiv.1107.2668},
archivePrefix = {arXiv},
       eprint = {1107.2668},
 primaryClass = {astro-ph.CO},
       adsurl = {https://ui.adsabs.harvard.edu/abs/2011sca..conf..129J}
}

@ARTICLE{2015AJ....149...51C,
       author = {{Calzetti}, D. and {Lee}, J.~C. and {Sabbi}, E. and {Adamo}, A. and {Smith}, L.~J. and {Andrews}, J.~E. and {Ubeda}, L. and {Bright}, S.~N. and {Thilker}, D. and {Aloisi}, A. and {Brown}, T.~M. and {Chandar}, R. and {Christian}, C. and {Cignoni}, M. and {Clayton}, G.~C. and {da Silva}, R. and {de Mink}, S.~E. and {Dobbs}, C. and {Elmegreen}, B.~G. and {Elmegreen}, D.~M. and {Evans}, A.~S. and {Fumagalli}, M. and {Gallagher}, III, J.~S. and {Gouliermis}, D.~A. and {Grebel}, E.~K. and {Herrero}, A. and {Hunter}, D.~A. and {Johnson}, K.~E. and {Kennicutt}, R.~C. and {Kim}, H. and {Krumholz}, M.~R. and {Lennon}, D. and {Levay}, K. and {Martin}, C. and {Nair}, P. and {Nota}, A. and {{\"O}stlin}, G. and {Pellerin}, A. and {Prieto}, J. and {Regan}, M.~W. and {Ryon}, J.~E. and {Schaerer}, D. and {Schiminovich}, D. and {Tosi}, M. and {Van Dyk}, S.~D. and {Walterbos}, R. and {Whitmore}, B.~C. and {Wofford}, A.},
        title = "{Legacy Extragalactic UV Survey (LEGUS) With the Hubble Space Telescope. I. Survey Description}",
      journal = {\aj},
         year = 2015,
        month = feb,
       volume = {149},
       number = {2},
          eid = {51},
        pages = {51},
          doi = {10.1088/0004-6256/149/2/51},
archivePrefix = {arXiv},
       eprint = {1410.7456},
        primaryClass = {astro-ph.GA},
       adsurl = {https://ui.adsabs.harvard.edu/abs/2015AJ....149...51C}
}

@ARTICLE{2020SSRv..216...69A,
       author = {{Adamo}, Angela and {Zeidler}, Peter and {Kruijssen}, J.~M. Diederik and {Chevance}, M{\'e}lanie and {Gieles}, Mark and {Calzetti}, Daniela and {Charbonnel}, Corinne and {Zinnecker}, Hans and {Krause}, Martin G.~H.},
        title = "{Star Clusters Near and Far; Tracing Star Formation Across Cosmic Time}",
      journal = {\ssr},
         year = 2020,
        month = jun,
       volume = {216},
       number = {4},
          eid = {69},
        pages = {69},
          doi = {10.1007/s11214-020-00690-x},
archivePrefix = {arXiv},
       eprint = {2005.06188},
 primaryClass = {astro-ph.GA},
       adsurl = {https://ui.adsabs.harvard.edu/abs/2020SSRv..216...69A}
}

@ARTICLE{Carrasco2025A&A...699A.142C,
       author = {{Carrasco-Varela}, Francisco F. and {Nayak}, Prasanta K. and {Puzia}, Thomas H.},
        title = "{Beyond the main sequence: Binary evolution pathways to blue stragglers in the Gaia era: I. Galactic open and globular clusters}",
      journal = {\aap},
         year = 2025,
        month = jul,
       volume = {699},
          eid = {A142},
        pages = {A142},
          doi = {10.1051/0004-6361/202452195},
archivePrefix = {arXiv},
       eprint = {2503.21051},
 primaryClass = {astro-ph.SR},
       adsurl = {https://ui.adsabs.harvard.edu/abs/2025A&A...699A.142C}
}

@ARTICLE{Landsman1997ApJ...481L..93L,
       author = {{Landsman}, W. and {Aparicio}, J. and {Bergeron}, P. and {Di Stefano}, R. and {Stecher}, T.~P.},
        title = "{S1040 in M67: A Post--Mass Transfer Binary with a Helium Core White Dwarf}",
      journal = {\apjl},
         year = 1997,
        month = jun,
       volume = {481},
       number = {2},
        pages = {L93-L96},
          doi = {10.1086/310654},
archivePrefix = {arXiv},
       eprint = {astro-ph/9703053},
 primaryClass = {astro-ph},
       adsurl = {https://ui.adsabs.harvard.edu/abs/1997ApJ...481L..93L}
}

@ARTICLE{Wang2025ApJ...984...52W,
       author = {{Wang}, Li and {Jiang}, Dengkai and {Li}, Chengyuan and {Deng}, Licai and {Milone}, Antonino P. and {Wang}, Long},
        title = "{Unveiling Bifurcated Blue Straggler Sequences in NGC 2173: Insights from Binary Evolution}",
      journal = {\apj},
         year = 2025,
        month = may,
       volume = {984},
       number = {1},
          eid = {52},
        pages = {52},
          doi = {10.3847/1538-4357/adc575},
archivePrefix = {arXiv},
       eprint = {2503.19966},
 primaryClass = {astro-ph.SR},
       adsurl = {https://ui.adsabs.harvard.edu/abs/2025ApJ...984...52W}
}

@ARTICLE{Dalessandro2012,
       author = {{Dalessandro}, Emanuele and {Schiavon}, Ricardo P. and {Rood}, Robert T. and {Ferraro}, Francesco R. and {Sohn}, Sangmo T. and {Lanzoni}, Barbara and {O'Connell}, Robert W.},
        title = "{Ultraviolet Properties of Galactic Globular Clusters with GALEX. II. Integrated Colors}",
      journal = {\aj},
         year = 2012,
        month = nov,
       volume = {144},
       number = {5},
          eid = {126},
        pages = {126},
          doi = {10.1088/0004-6256/144/5/126},
archivePrefix = {arXiv},
       eprint = {1208.5698},
 primaryClass = {astro-ph.GA},
       adsurl = {https://ui.adsabs.harvard.edu/abs/2012AJ....144..126D}
}

@ARTICLE{Lardo2011,
       author = {{Lardo}, C. and {Bellazzini}, M. and {Pancino}, E. and {Carretta}, E. and {Bragaglia}, A. and {Dalessandro}, E.},
        title = "{Mining SDSS in search of multiple populations in globular clusters}",
      journal = {\aap},
         year = 2011,
        month = jan,
       volume = {525},
          eid = {A114},
        pages = {A114},
          doi = {10.1051/0004-6361/201015662},
archivePrefix = {arXiv},
       eprint = {1010.4697},
 primaryClass = {astro-ph.GA},
       adsurl = {https://ui.adsabs.harvard.edu/abs/2011A&A...525A.114L}
}

@ARTICLE{Milone2012,
       author = {{Milone}, A.~P. and {Piotto}, G. and {Bedin}, L.~R. and {King}, I.~R. and {Anderson}, J. and {Marino}, A.~F. and {Bellini}, A. and {Gratton}, R. and {Renzini}, A. and {Stetson}, P.~B. and {Cassisi}, S. and {Aparicio}, A. and {Bragaglia}, A. and {Carretta}, E. and {D'Antona}, F. and {Di Criscienzo}, M. and {Lucatello}, S. and {Monelli}, M. and {Pietrinferni}, A.},
        title = "{Multiple Stellar Populations in 47 Tucanae}",
      journal = {\apj},
         year = 2012,
        month = jan,
       volume = {744},
       number = {1},
          eid = {58},
        pages = {58},
          doi = {10.1088/0004-637X/744/1/5810.1086/141918},
archivePrefix = {arXiv},
       eprint = {1109.0900},
 primaryClass = {astro-ph.SR},
       adsurl = {https://ui.adsabs.harvard.edu/abs/2012ApJ...744...58M}
}

@ARTICLE{Cordoni2020,
       author = {{Cordoni}, G. and {Milone}, A.~P. and {Marino}, A.~F. and {Da Costa}, G.~S. and {Dondoglio}, E. and {Jerjen}, H. and {Lagioia}, E.~P. and {Mastrobuono-Battisti}, A. and {Norris}, J.~E. and {Tailo}, M. and {Yong}, D.},
        title = "{Gaia and Hubble Unveil the Kinematics of Stellar Populations in the Type II Globular Clusters {\ensuremath{\omega}} Centauri and M22}",
      journal = {\apj},
         year = 2020,
        month = aug,
       volume = {898},
       number = {2},
          eid = {147},
        pages = {147},
          doi = {10.3847/1538-4357/aba04b},
archivePrefix = {arXiv},
       eprint = {2006.16355},
 primaryClass = {astro-ph.SR},
       adsurl = {https://ui.adsabs.harvard.edu/abs/2020ApJ...898..147C}
}

@ARTICLE{Martocchia2018,
       author = {{Martocchia}, S. and {Cabrera-Ziri}, I. and {Lardo}, C. and {Dalessandro}, E. and {Bastian}, N. and {Kozhurina-Platais}, V. and {Usher}, C. and {Niederhofer}, F. and {Cordero}, M. and {Geisler}, D. and {Hollyhead}, K. and {Kacharov}, N. and {Larsen}, S. and {Li}, C. and {Mackey}, D. and {Hilker}, M. and {Mucciarelli}, A. and {Platais}, I. and {Salaris}, M.},
        title = "{Age as a major factor in the onset of multiple populations in stellar clusters}",
      journal = {\mnras},
         year = 2018,
        month = jan,
       volume = {473},
       number = {2},
        pages = {2688-2700},
          doi = {10.1093/mnras/stx2556},
archivePrefix = {arXiv},
       eprint = {1710.00831},
 primaryClass = {astro-ph.GA},
       adsurl = {https://ui.adsabs.harvard.edu/abs/2018MNRAS.473.2688M}
}

@ARTICLE{Li2019,
       author = {{Li}, Chengyuan and {de Grijs}, Richard},
        title = "{When Does the Onset of Multiple Stellar Populations in Star Clusters Occur? Detection of Enriched Stellar Populations in NGC 2121}",
      journal = {\apj},
         year = 2019,
        month = may,
       volume = {876},
       number = {2},
          eid = {94},
        pages = {94},
          doi = {10.3847/1538-4357/ab153b},
archivePrefix = {arXiv},
       eprint = {1904.00508},
 primaryClass = {astro-ph.SR},
       adsurl = {https://ui.adsabs.harvard.edu/abs/2019ApJ...876...94L}
}

@ARTICLE{Saracino2020,
       author = {{Saracino}, S. and {Martocchia}, S. and {Bastian}, N. and {Kozhurina-Platais}, V. and {Chantereau}, W. and {Salaris}, M. and {Cabrera-Ziri}, I. and {Dalessandro}, E. and {Kacharov}, N. and {Lardo}, C. and {Larsen}, S.~S. and {Platais}, I.},
        title = "{Chromosome maps of young LMC clusters: an additional case of coeval multiple populations}",
      journal = {\mnras},
         year = 2020,
        month = apr,
       volume = {493},
       number = {4},
        pages = {6060-6070},
          doi = {10.1093/mnras/staa644},
archivePrefix = {arXiv},
       eprint = {2003.01780},
 primaryClass = {astro-ph.SR},
       adsurl = {https://ui.adsabs.harvard.edu/abs/2020MNRAS.493.6060S}
}

@ARTICLE{BastianDeMink2009,
       author = {{Bastian}, N. and {de Mink}, S.~E.},
        title = "{The effect of stellar rotation on colour-magnitude diagrams: on the apparent presence of multiple populations in intermediate age stellar clusters}",
      journal = {\mnras},
         year = 2009,
        month = sep,
       volume = {398},
       number = {1},
        pages = {L11-L15},
          doi = {10.1111/j.1745-3933.2009.00696.x},
archivePrefix = {arXiv},
       eprint = {0906.1590},
 primaryClass = {astro-ph.GA},
       adsurl = {https://ui.adsabs.harvard.edu/abs/2009MNRAS.398L..11B}
}

@ARTICLE{Li2014,
       author = {{Li}, Chengyuan and {de Grijs}, Richard and {Deng}, Licai},
        title = "{The exclusion of a significant range of ages in a massive star cluster}",
      journal = {\nat},
         year = 2014,
        month = dec,
       volume = {516},
       number = {7531},
        pages = {367-369},
          doi = {10.1038/nature13969},
archivePrefix = {arXiv},
       eprint = {1412.5368},
 primaryClass = {astro-ph.SR},
       adsurl = {https://ui.adsabs.harvard.edu/abs/2014Natur.516..367L}
}

@ARTICLE{Niederhofer2015,
       author = {{Niederhofer}, F. and {Georgy}, C. and {Bastian}, N. and {Ekstr{\"o}m}, S.},
        title = "{Apparent age spreads in clusters and the role of stellar rotation}",
      journal = {\mnras},
         year = 2015,
        month = oct,
       volume = {453},
       number = {2},
        pages = {2070-2074},
          doi = {10.1093/mnras/stv1791},
archivePrefix = {arXiv},
       eprint = {1507.07561},
 primaryClass = {astro-ph.SR},
       adsurl = {https://ui.adsabs.harvard.edu/abs/2015MNRAS.453.2070N}
}

@ARTICLE{Georgy2019,
       author = {{Georgy}, C. and {Charbonnel}, C. and {Amard}, L. and {Bastian}, N. and {Ekstr{\"o}m}, S. and {Lardo}, C. and {Palacios}, A. and {Eggenberger}, P. and {Cabrera-Ziri}, I. and {Gallet}, F. and {Lagarde}, N.},
        title = "{Disappearance of the extended main sequence turn-off in intermediate age clusters as a consequence of magnetic braking}",
      journal = {\aap},
         year = 2019,
        month = feb,
       volume = {622},
          eid = {A66},
        pages = {A66},
          doi = {10.1051/0004-6361/201834505},
archivePrefix = {arXiv},
       eprint = {1812.05544},
 primaryClass = {astro-ph.SR},
       adsurl = {https://ui.adsabs.harvard.edu/abs/2019A&A...622A..66G}
}

@ARTICLE{Sun2019,
       author = {{Sun}, Weijia and {Li}, Chengyuan and {Deng}, Licai and {de Grijs}, Richard},
        title = "{Tidal-locking-induced Stellar Rotation Dichotomy in the Open Cluster NGC 2287?}",
      journal = {\apj},
         year = 2019,
        month = oct,
       volume = {883},
       number = {2},
          eid = {182},
        pages = {182},
          doi = {10.3847/1538-4357/ab3cd0},
archivePrefix = {arXiv},
       eprint = {1908.06530},
 primaryClass = {astro-ph.SR},
       adsurl = {https://ui.adsabs.harvard.edu/abs/2019ApJ...883..182S}
}

@ARTICLE{Cordoni2018,
       author = {{Cordoni}, G. and {Milone}, A.~P. and {Marino}, A.~F. and {Di Criscienzo}, M. and {D'Antona}, F. and {Dotter}, A. and {Lagioia}, E.~P. and {Tailo}, M.},
        title = "{Extended Main-sequence Turnoff as a Common Feature of Milky Way Open Clusters}",
      journal = {\apj},
         year = 2018,
        month = dec,
       volume = {869},
       number = {2},
          eid = {139},
        pages = {139},
          doi = {10.3847/1538-4357/aaedc1},
archivePrefix = {arXiv},
       eprint = {1811.01192},
 primaryClass = {astro-ph.SR},
       adsurl = {https://ui.adsabs.harvard.edu/abs/2018ApJ...869..139C}
}

@ARTICLE{Georgy2014,
       author = {{Georgy}, C. and {Granada}, A. and {Ekstr{\"o}m}, S. and {Meynet}, G. and {Anderson}, R.~I. and {Wyttenbach}, A. and {Eggenberger}, P. and {Maeder}, A.},
        title = "{Populations of rotating stars. III. SYCLIST, the new Geneva population synthesis code}",
      journal = {\aap},
         year = 2014,
        month = jun,
       volume = {566},
          eid = {A21},
        pages = {A21},
          doi = {10.1051/0004-6361/201423881},
archivePrefix = {arXiv},
       eprint = {1404.6952},
 primaryClass = {astro-ph.SR},
       adsurl = {https://ui.adsabs.harvard.edu/abs/2014A&A...566A..21G}
}

@ARTICLE{Kamann2020,
       author = {{Kamann}, S. and {Bastian}, N. and {Gossage}, S. and {Baade}, D. and {Cabrera-Ziri}, I. and {Da Costa}, G. and {de Mink}, S.~E. and {Georgy}, C. and {Giesers}, B. and {G{\"o}ttgens}, F. and {Hilker}, M. and {Husser}, T. -O. and {Lardo}, C. and {Larsen}, S.~S. and {Mackey}, D. and {Martocchia}, S. and {Mucciarelli}, A. and {Platais}, I. and {Roth}, M.~M. and {Salaris}, M. and {Usher}, C. and {Yong}, D.},
        title = "{How stellar rotation shapes the colour-magnitude diagram of the massive intermediate-age star cluster NGC 1846}",
      journal = {\mnras},
         year = 2020,
        month = feb,
       volume = {492},
       number = {2},
        pages = {2177-2192},
          doi = {10.1093/mnras/stz3583},
archivePrefix = {arXiv},
       eprint = {2001.01731},
 primaryClass = {astro-ph.SR},
       adsurl = {https://ui.adsabs.harvard.edu/abs/2020MNRAS.492.2177K}
}

@ARTICLE{Dupree2017,
       author = {{Dupree}, A.~K. and {Dotter}, A. and {Johnson}, C.~I. and {Marino}, A.~F. and {Milone}, A.~P. and {Bailey}, III, J.~I. and {Crane}, J.~D. and {Mateo}, M. and {Olszewski}, E.~W.},
        title = "{NGC 1866: First Spectroscopic Detection of Fast-rotating Stars in a Young LMC Cluster}",
      journal = {\apjl},
         year = 2017,
        month = sep,
       volume = {846},
       number = {1},
          eid = {L1},
        pages = {L1},
          doi = {10.3847/2041-8213/aa85dd},
archivePrefix = {arXiv},
       eprint = {1708.03386},
 primaryClass = {astro-ph.GA},
       adsurl = {https://ui.adsabs.harvard.edu/abs/2017ApJ...846L...1D}
}

@ARTICLE{Marino2018,
       author = {{Marino}, A.~F. and {Przybilla}, N. and {Milone}, A.~P. and {Da Costa}, G. and {D'Antona}, F. and {Dotter}, A. and {Dupree}, A.},
        title = "{Different Stellar Rotations in the Two Main Sequences of the Young Globular Cluster NGC 1818: The First Direct Spectroscopic Evidence}",
      journal = {\aj},
         year = 2018,
        month = sep,
       volume = {156},
       number = {3},
          eid = {116},
        pages = {116},
          doi = {10.3847/1538-3881/aad3cd},
archivePrefix = {arXiv},
       eprint = {1807.04493},
 primaryClass = {astro-ph.SR},
       adsurl = {https://ui.adsabs.harvard.edu/abs/2018AJ....156..116M}
}

@ARTICLE{1995A&AS..109..375A,
       author = {{Ahumada}, J. and {Lapasset}, E.},
        title = "{Catalogue of blue stragglers in open clusters.}",
      journal = {\aaps},
         year = 1995,
        month = feb,
       volume = {109},
        pages = {375-382},
       adsurl = {https://ui.adsabs.harvard.edu/abs/1995A&AS..109..375A}
}

@ARTICLE{2007A&A...463..789A,
       author = {{Ahumada}, J.~A. and {Lapasset}, E.},
        title = "{New catalogue of blue stragglers in open clusters}",
      journal = {\aap},
         year = 2007,
        month = feb,
       volume = {463},
       number = {2},
        pages = {789-797},
          doi = {10.1051/0004-6361:20054590},
       adsurl = {https://ui.adsabs.harvard.edu/abs/2007A&A...463..789A}
}

@ARTICLE{1997A&A...324..915F,
       author = {{Ferraro}, F.~R. and {Paltrinieri}, B. and {Fusi Pecci}, F. and {Cacciari}, C. and {Dorman}, B. and {Rood}, R.~T. and {Buonanno}, R. and {Corsi}, C.~E. and {Burgarella}, D. and {Laget}, M.},
        title = "{HST observations of blue Straggler stars in the core of the globular cluster M 3.}",
      journal = {\aap},
         year = 1997,
        month = aug,
       volume = {324},
        pages = {915-928},
          doi = {10.48550/arXiv.astro-ph/9703026},
archivePrefix = {arXiv},
       eprint = {astro-ph/9703026},
 primaryClass = {astro-ph},
       adsurl = {https://ui.adsabs.harvard.edu/abs/1997A&A...324..915F}
}

@ARTICLE{1974ApJS...28..157G,
       author = {{Greenstein}, Jesse L. and {Sargent}, Anneila I.},
        title = "{The Nature of Faint Blue Stars in the Halo. II}",
      journal = {\apjs},
         year = 1974,
        month = nov,
       volume = {28},
        pages = {157},
          doi = {10.1086/190315},
       adsurl = {https://ui.adsabs.harvard.edu/abs/1974ApJS...28..157G}
}

@ARTICLE{2016ApJ...833L..29L,
       author = {{Lanzoni}, B. and {Ferraro}, F.~R. and {Alessandrini}, E. and {Dalessandro}, E. and {Vesperini}, E. and {Raso}, S.},
        title = "{Refining the Dynamical Clock for Star Clusters}",
      journal = {\apjl},
         year = 2016,
        month = dec,
       volume = {833},
       number = {2},
          eid = {L29},
        pages = {L29},
          doi = {10.3847/2041-8213/833/2/L29},
       adsurl = {https://ui.adsabs.harvard.edu/abs/2016ApJ...833L..29L}
}

@ARTICLE{2009Natur.462.1028F,
       author = {{Ferraro}, F.~R. and {Beccari}, G. and {Dalessandro}, E. and {Lanzoni}, B. and {Sills}, A. and {Rood}, R.~T. and {Pecci}, F. Fusi and {Karakas}, A.~I. and {Miocchi}, P. and {Bovinelli}, S.},
        title = "{Two distinct sequences of blue straggler stars in the globular cluster M 30}",
      journal = {\nat},
         year = 2009,
        month = dec,
       volume = {462},
       number = {7276},
        pages = {1028-1031},
          doi = {10.1038/nature08607},
archivePrefix = {arXiv},
       eprint = {1001.1096},
 primaryClass = {astro-ph.GA},
       adsurl = {https://ui.adsabs.harvard.edu/abs/2009Natur.462.1028F}
}

@INPROCEEDINGS{2015ASSL..413...99F,
       author = {{Ferraro}, Francesco R. and {Lanzoni}, Barbara and {Dalessandro}, Emanuele and {Mucciarelli}, Alessio and {Lovisi}, Loredana},
        title = "{Blue Straggler Stars in Globular Clusters: A Powerful Tool to Probe the Internal Dynamical Evolution of Stellar Systems}",
    booktitle = {Astrophysics and Space Science Library},
         year = 2015,
       editor = {{Boffin}, Henri M.~J. and {Carraro}, Giovanni and {Beccari}, Giacomo},
       series = {Astrophysics and Space Science Library},
       volume = {413},
        month = jan,
        pages = {99},
          doi = {10.1007/978-3-662-44434-4_5},
archivePrefix = {arXiv},
       eprint = {1406.3471},
 primaryClass = {astro-ph.SR},
       adsurl = {https://ui.adsabs.harvard.edu/abs/2015ASSL..413...99F}
}

@ARTICLE{2018ApJ...860...36F,
       author = {{Ferraro}, F.~R. and {Lanzoni}, B. and {Raso}, S. and {Nardiello}, D. and {Dalessandro}, E. and {Vesperini}, E. and {Piotto}, G. and {Pallanca}, C. and {Beccari}, G. and {Bellini}, A. and {Libralato}, M. and {Anderson}, J. and {Aparicio}, A. and {Bedin}, L.~R. and {Cassisi}, S. and {Milone}, A.~P. and {Ortolani}, S. and {Renzini}, A. and {Salaris}, M. and {van der Marel}, R.~P.},
        title = "{The Hubble Space Telescope UV Legacy Survey of Galactic Globular Clusters. XV. The Dynamical Clock: Reading Cluster Dynamical Evolution from the Segregation Level of Blue Straggler Stars}",
      journal = {\apj},
         year = 2018,
        month = jun,
       volume = {860},
       number = {1},
          eid = {36},
        pages = {36},
          doi = {10.3847/1538-4357/aac01c},
archivePrefix = {arXiv},
       eprint = {1805.00968},
 primaryClass = {astro-ph.GA},
       adsurl = {https://ui.adsabs.harvard.edu/abs/2018ApJ...860...36F}
}

@ARTICLE{2011Natur.478..356G,
       author = {{Geller}, Aaron M. and {Mathieu}, Robert D.},
        title = "{A mass transfer origin for blue stragglers in NGC 188 as revealed by half-solar-mass companions}",
      journal = {\nat},
         year = 2011,
        month = oct,
       volume = {478},
       number = {7369},
        pages = {356-359},
          doi = {10.1038/nature10512},
archivePrefix = {arXiv},
       eprint = {1110.3793},
 primaryClass = {astro-ph.SR},
       adsurl = {https://ui.adsabs.harvard.edu/abs/2011Natur.478..356G}
}

@ARTICLE{2014ApJ...783L...8G,
       author = {{Gosnell}, Natalie M. and {Mathieu}, Robert D. and {Geller}, Aaron M. and {Sills}, Alison and {Leigh}, Nathan and {Knigge}, Christian},
        title = "{Detection of White Dwarf Companions to Blue Stragglers in the Open Cluster NGC 188: Direct Evidence for Recent Mass Transfer}",
      journal = {\apjl},
         year = 2014,
        month = mar,
       volume = {783},
       number = {1},
          eid = {L8},
        pages = {L8},
          doi = {10.1088/2041-8205/783/1/L8},
archivePrefix = {arXiv},
       eprint = {1401.7670},
 primaryClass = {astro-ph.SR},
       adsurl = {https://ui.adsabs.harvard.edu/abs/2014ApJ...783L...8G}
}

@ARTICLE{2019ApJ...885...45G,
       author = {{Gosnell}, Natalie M. and {Leiner}, Emily M. and {Mathieu}, Robert D. and {Geller}, Aaron M. and {Knigge}, Christian and {Sills}, Alison and {Leigh}, Nathan W.~C.},
        title = "{Constraining Mass-transfer Histories of Blue Straggler Stars with COS Spectroscopy of White Dwarf Companions}",
      journal = {\apj},
         year = 2019,
        month = nov,
       volume = {885},
       number = {1},
          eid = {45},
        pages = {45},
          doi = {10.3847/1538-4357/ab4273},
archivePrefix = {arXiv},
       eprint = {1904.02280},
 primaryClass = {astro-ph.SR},
       adsurl = {https://ui.adsabs.harvard.edu/abs/2019ApJ...885...45G}
}

@ARTICLE{2021MNRAS.508.4919R,
       author = {{Rao}, Khushboo K. and {Vaidya}, Kaushar and {Agarwal}, Manan and {Bhattacharya}, Souradeep},
        title = "{Determination of dynamical ages of open clusters through the A$^{+}$ parameter - I}",
      journal = {\mnras},
         year = 2021,
        month = dec,
       volume = {508},
       number = {4},
        pages = {4919-4937},
          doi = {10.1093/mnras/stab2894},
archivePrefix = {arXiv},
       eprint = {2102.07409},
 primaryClass = {astro-ph.GA},
       adsurl = {https://ui.adsabs.harvard.edu/abs/2021MNRAS.508.4919R}
}

@ARTICLE{2023MNRAS.526.1057R,
       author = {{Rao}, Khushboo K. and {Vaidya}, Kaushar and {Agarwal}, Manan and {Balan}, Shanmugha and {Bhattacharya}, Souradeep},
        title = "{Determination of dynamical ages of open clusters through the A$^{+}$ parameter - II}",
      journal = {\mnras},
         year = 2023,
        month = nov,
       volume = {526},
       number = {1},
        pages = {1057-1074},
          doi = {10.1093/mnras/stad2755},
archivePrefix = {arXiv},
       eprint = {2309.02746},
 primaryClass = {astro-ph.GA},
       adsurl = {https://ui.adsabs.harvard.edu/abs/2023MNRAS.526.1057R}
}

@ARTICLE{2021A&A...650A..67R,
       author = {{Rain}, M.~J. and {Ahumada}, J.~A. and {Carraro}, G.},
        title = "{A new, Gaia-based, catalogue of blue straggler stars in open clusters}",
      journal = {\aap},
         year = 2021,
        month = jun,
       volume = {650},
          eid = {A67},
        pages = {A67},
          doi = {10.1051/0004-6361/202040072},
archivePrefix = {arXiv},
       eprint = {2103.06004},
 primaryClass = {astro-ph.SR},
       adsurl = {https://ui.adsabs.harvard.edu/abs/2021A&A...650A..67R}
}

@ARTICLE{Rain2024A&A...685A..33R,
       author = {{Rain}, M.~J. and {Pera}, M.~S. and {Perren}, G.~I. and {Benvenuto}, O.~G. and {Panei}, J.~A. and {De Vito}, M.~A. and {Carraro}, G. and {Villanova}, S.},
        title = "{Binary origin of blue straggler stars in Galactic star clusters}",
      journal = {\aap},
         year = 2024,
        month = may,
       volume = {685},
          eid = {A33},
        pages = {A33},
          doi = {10.1051/0004-6361/202347499},
archivePrefix = {arXiv},
       eprint = {2402.14990},
 primaryClass = {astro-ph.SR},
       adsurl = {https://ui.adsabs.harvard.edu/abs/2024A&A...685A..33R}
}

@ARTICLE{2019MNRAS.490.4648H,
       author = {{Hannon}, Stephen and {Lee}, Janice C. and {Whitmore}, B.~C. and {Chandar}, R. and {Adamo}, A. and {Mobasher}, B. and {Aloisi}, A. and {Calzetti}, D. and {Cignoni}, M. and {Cook}, D.~O. and {Dale}, D. and {Deger}, S. and {Della Bruna}, L. and {Elmegreen}, D.~M. and {Gouliermis}, D.~A. and {Grasha}, K. and {Grebel}, E.~K. and {Herrero}, A. and {Hunter}, D.~A. and {Johnson}, K.~E. and {Kennicutt}, R. and {Kim}, H. and {Sacchi}, E. and {Smith}, L. and {Thilker}, D. and {Turner}, J. and {Walterbos}, R.~A.~M. and {Wofford}, A.},
        title = "{H {\ensuremath{\alpha}} morphologies of star clusters: a LEGUS study of H II region evolution time-scales and stochasticity in low-mass clusters}",
      journal = {\mnras},
         year = 2019,
        month = dec,
       volume = {490},
       number = {4},
        pages = {4648-4665},
          doi = {10.1093/mnras/stz2820},
archivePrefix = {arXiv},
       eprint = {1910.02983},
 primaryClass = {astro-ph.GA},
       adsurl = {https://ui.adsabs.harvard.edu/abs/2019MNRAS.490.4648H}
}

@ARTICLE{2022MNRAS.512.1294H,
       author = {{Hannon}, Stephen and {Lee}, Janice C. and {Whitmore}, B.~C. and {Mobasher}, B. and {Thilker}, D. and {Chandar}, R. and {Adamo}, A. and {Wofford}, A. and {Orozco-Duarte}, R. and {Calzetti}, D. and {Della Bruna}, L. and {Kreckel}, K. and {Groves}, B. and {Barnes}, A.~T. and {Boquien}, M. and {Belfiore}, F. and {Linden}, S.},
        title = "{H {\ensuremath{\alpha}} morphologies of star clusters in 16 LEGUS galaxies: Constraints on H II region evolution time-scales}",
      journal = {\mnras},
     year = 2022,
        month = may,
       volume = {512},
       number = {1},
        pages = {1294-1316},
          doi = {10.1093/mnras/stac550},
archivePrefix = {arXiv},
       eprint = {2203.01339},
 primaryClass = {astro-ph.GA},
       adsurl = {https://ui.adsabs.harvard.edu/abs/2022MNRAS.512.1294H}
}

@ARTICLE{2022ApJS..261...31B,
       author = {{Berg}, Danielle A. and {James}, Bethan L. and {King}, Teagan and {McDonald}, Meaghan and {Chen}, Zuyi and {Chisholm}, John and {Heckman}, Timothy and {Martin}, Crystal L. and {Stark}, Dan P. and {Aloisi}, Alessandra and {Amor{\'\i}n}, Ricardo O. and {Arellano-C{\'o}rdova}, Karla Z. and {Bayliss}, Matthew and {Bordoloi}, Rongmon and {Brinchmann}, Jarle and {Charlot}, St{\'e}phane and {Chevallard}, Jacopo and {Clark}, Ilyse and {Erb}, Dawn K. and {Feltre}, Anna and {Gronke}, Max and {Hayes}, Matthew and {Henry}, Alaina and {Hernandez}, Svea and {Jaskot}, Anne and {Jones}, Tucker and {Kewley}, Lisa J. and {Kumari}, Nimisha and {Leitherer}, Claus and {Llerena}, Mario and {Maseda}, Michael and {Mingozzi}, Matilde and {Nanayakkara}, Themiya and {Ouchi}, Masami and {Plat}, Adele and {Pogge}, Richard W. and {Ravindranath}, Swara and {Rigby}, Jane R. and {Sanders}, Ryan and {Scarlata}, Claudia and {Senchyna}, Peter and {Skillman}, Evan D. and {Steidel}, Charles C. and {Strom}, Allison L. and {Sugahara}, Yuma and {Wilkins}, Stephen M. and {Wofford}, Aida and {Xu}, Xinfeng and {Classy Team}},
        title = "{The COS Legacy Archive Spectroscopy Survey (CLASSY) Treasury Atlas}",
      journal = {\apjs},
         year = 2022,
        month = aug,
       volume = {261},
       number = {2},
          eid = {31},
        pages = {31},
          doi = {10.3847/1538-4365/ac6c03},
archivePrefix = {arXiv},
       eprint = {2203.07357},
 primaryClass = {astro-ph.GA},
       adsurl = {https://ui.adsabs.harvard.edu/abs/2022ApJS..261...31B}
}

@ARTICLE{2022ApJS..262...37J,
       author = {{James}, Bethan L. and {Berg}, Danielle A. and {King}, Teagan and {Sahnow}, David J. and {Mingozzi}, Matilde and {Chisholm}, John and {Heckman}, Timothy and {Martin}, Crystal L. and {Stark}, Dan P. and {Aloisi}, Alessandra and {Amor{\'\i}n}, Ricardo O. and {Arellano-C{\'o}rdova}, Karla Z. and {Bayliss}, Matthew and {Bordoloi}, Rongmon and {Brinchmann}, Jarle and {Charlot}, St{\'e}phane and {Chen}, Zuyi and {Chevallard}, Jacopo and {Clark}, Ilyse and {Erb}, Dawn K. and {Feltre}, Anna and {Hayes}, Matthew and {Henry}, Alaina and {Hernandez}, Svea and {Jaskot}, Anne and {Kewley}, Lisa J. and {Kumari}, Nimisha and {Leitherer}, Claus and {Llerena}, Mario and {Maseda}, Michael and {Nanayakkara}, Themiya and {Ouchi}, Masami and {Plat}, Adele and {Pogge}, Richard W. and {Ravindranath}, Swara and {Rigby}, Jane R. and {Scarlata}, Claudia and {Senchyna}, Peter and {Skillman}, Evan D. and {Steidel}, Charles C. and {Strom}, Allison L. and {Sugahara}, Yuma and {Wilkins}, Stephen M. and {Wofford}, Aida and {Xu}, Xinfeng and {Classy Team}},
        title = "{CLASSY. II. A Technical Overview of the COS Legacy Archive Spectroscopic Survey}",
      journal = {\apjs},
         year = 2022,
        month = oct,
       volume = {262},
       number = {2},
          eid = {37},
        pages = {37},
          doi = {10.3847/1538-4365/ac8008},
archivePrefix = {arXiv},
       eprint = {2206.01224},
 primaryClass = {astro-ph.GA},
       adsurl = {https://ui.adsabs.harvard.edu/abs/2022ApJS..262...37J}
}

@ARTICLE{2020Galax...8...13L,
       author = {{Leitherer}, Claus},
        title = "{Massive Star Formation in the Ultraviolet Observed with the Hubble Space Telescope}",
      journal = {Galaxies},
         year = 2020,
        month = feb,
       volume = {8},
       number = {1},
          eid = {13},
        pages = {13},
          doi = {10.3390/galaxies8010013},
archivePrefix = {arXiv},
       eprint = {2005.01761},
 primaryClass = {astro-ph.GA},
       adsurl = {https://ui.adsabs.harvard.edu/abs/2020Galax...8...13L}
}

@ARTICLE{2022AJ....164..208S,
       author = {{Sirressi}, Mattia and {Adamo}, Angela and {Hayes}, Matthew and {Osborne}, Shannon and {Hernandez}, Svea and {Chisholm}, John and {Messa}, Matteo and {Smith}, Linda J. and {Aloisi}, Alessandra and {Wofford}, Aida and {Fox}, Andrew J. and {Mizener}, Andrew and {Usher}, Christopher and {Bik}, Arjan and {Calzetti}, Daniela and {Sabbi}, Elena and {Schinnerer}, Eva and {{\"O}stlin}, G{\"o}ran and {Grasha}, Kathryn and {Cignoni}, Michele and {Fumagalli}, Michele},
        title = "{CLusters in the UV as EngineS (CLUES). I. Survey Presentation and FUV Spectral Analysis of the Stellar Light}",
	journal = {\aj},
         year = 2022,
        month = nov,
       volume = {164},
       number = {5},
          eid = {208},
        pages = {208},
          doi = {10.3847/1538-3881/ac9311},
archivePrefix = {arXiv},
       eprint = {2209.09914},
 primaryClass = {astro-ph.GA},
       adsurl = {https://ui.adsabs.harvard.edu/abs/2022AJ....164..208S}
}

@ARTICLE{2017ApJ...840..113G,
       author = {{Grasha}, K. and {Calzetti}, D. and {Adamo}, A. and {Kim}, H. and {Elmegreen}, B.~G. and {Gouliermis}, D.~A. and {Dale}, D.~A. and {Fumagalli}, M. and {Grebel}, E.~K. and {Johnson}, K.~E. and {Kahre}, L. and {Kennicutt}, R.~C. and {Messa}, M. and {Pellerin}, A. and {Ryon}, J.~E. and {Smith}, L.~J. and {Shabani}, F. and {Thilker}, D. and {Ubeda}, L.},
        title = "{The Hierarchical Distribution of the Young Stellar Clusters in Six Local Star-forming Galaxies}",
      journal = {\apj},
         year = 2017,
        month = may,
       volume = {840},
       number = {2},
          eid = {113},
        pages = {113},
          doi = {10.3847/1538-4357/aa6f15},
archivePrefix = {arXiv},
       eprint = {1704.06321},
 primaryClass = {astro-ph.GA},
       adsurl = {https://ui.adsabs.harvard.edu/abs/2017ApJ...840..113G}
}

@PHDTHESIS{2018PhDT........36G,
       author = {{Grasha}, Kathryn},
        title = "{The Clustering Of Young Stellar Clusters In Nearby Galaxies}",
       school = {UMass Amherst},
         year = 2018,
        month = may,
       adsurl = {https://ui.adsabs.harvard.edu/abs/2018PhDT........36G}
}

@ARTICLE{2023MNRAS.518...87D,
       author = {{Dage}, Kristen C. and {Sun}, Yifan and {Kundu}, Arunav and {Zepf}, Stephen E. and {Haggard}, Daryl},
        title = "{Far ultra-violet insights into NGC 1399's globular cluster population}",
      journal = {\mnras},
         year = 2023,
        month = jan,
       volume = {518},
       number = {1},
        pages = {87-92},
          doi = {10.1093/mnras/stac3132},
archivePrefix = {arXiv},
       eprint = {2210.14637},
 primaryClass = {astro-ph.GA},
       adsurl = {https://ui.adsabs.harvard.edu/abs/2023MNRAS.518...87D}
}

@ARTICLE{2016MNRAS.457.4296W,
       author = {{Wofford}, A. and {Charlot}, S. and {Bruzual}, G. and {Eldridge}, J.~J. and {Calzetti}, D. and {Adamo}, A. and {Cignoni}, M. and {de Mink}, S.~E. and {Gouliermis}, D.~A. and {Grasha}, K. and {Grebel}, E.~K. and {Lee}, J.~C. and {{\"O}stlin}, G. and {Smith}, L.~J. and {Ubeda}, L. and {Zackrisson}, E.},
        title = "{A comprehensive comparative test of seven widely used spectral synthesis models against multi-band photometry of young massive-star clusters}",
      journal = {\mnras},
     year = 2016,
        month = apr,
       volume = {457},
       number = {4},
        pages = {4296-4322},
          doi = {10.1093/mnras/stw150},
archivePrefix = {arXiv},
       eprint = {1601.03850},
 primaryClass = {astro-ph.GA},
       adsurl = {https://ui.adsabs.harvard.edu/abs/2016MNRAS.457.4296W}
}

@ARTICLE{2023ApJ...954..136J,
       author = {{Jung}, Dooseok Escher and {Calzetti}, Daniela and {Messa}, Matteo and {Heyer}, Mark and {Sirressi}, Mattia and {Linden}, Sean T. and {Adamo}, Angela and {Chandar}, Rupali and {Cignoni}, Michele and {Cook}, David O. and {Dobbs}, Clare L. and {Elmegreen}, Bruce G. and {Evans}, Aaron S. and {Fumagalli}, Michele and {Gallagher}, John S. and {Hunter}, Deidre A. and {Johnson}, Kelsey E. and {Kennicutt}, Robert C. and {Krumholz}, Mark R. and {Schaerer}, Daniel and {Sabbi}, Elena and {Smith}, Linda J. and {Tosi}, Monica and {Wofford}, Aida},
        title = "{Universal Upper End of the Stellar Initial Mass Function in the Young and Compact LEGUS Clusters}",
	journal = {\apj},
         year = 2023,
        month = sep,
       volume = {954},
       number = {2},
          eid = {136},
        pages = {136},
          doi = {10.3847/1538-4357/aceb5c},
archivePrefix = {arXiv},
       eprint = {2307.15831},
 primaryClass = {astro-ph.GA},
       adsurl = {https://ui.adsabs.harvard.edu/abs/2023ApJ...954..136J}
}

@ARTICLE{Roman2025ApJ...985..109R,
       author = {{Roman-Duval}, Julia and {Fischer}, William J. and {Fullerton}, Alexander W. and {Taylor}, Jo and {Plesha}, Rachel and {Proffitt}, Charles and {Monroe}, TalaWanda and {Fischer}, Travis C. and {Aloisi}, Alessandra and {Bouret}, Jean-Claude and {Britt}, Christopher and {Calvet}, Nuria and {Carlberg}, Joleen K. and {Crowther}, Paul A. and {De Rosa}, Gisella and {Dixon}, William V. and {Espaillat}, Catherine C. and {Evans}, Christopher J. and {Fox}, Andrew J. and {France}, Kevin and {Garcia}, Miriam and {Fleming}, Scott W. and {Frazer}, Elaine M. and {G{\'o}mez de Castro}, Ana I. and {Herczeg}, Gregory J. and {Hernandez}, Svea and {Hirschauer}, Alec S. and {James}, Bethan L. and {Johns-Krull}, Christopher M. and {Leitherer}, Claus and {Lockwood}, Sean and {Najita}, Joan and {Oey}, M.~S. and {Oliveira}, Cristina and {Pauly}, Tyler and {Reid}, I. Neill and {Riedel}, Adric and {Rodriguez}, David R. and {Sahnow}, David and {Sankrit}, Ravi and {Sembach}, Kenneth R. and {Shaw}, Richard and {Smith}, Linda J. and {Sohn}, S. Tony and {Som}, Debopam and {{\'U}beda}, Leonardo and {Welty}, Daniel E.},
        title = "{The UV Legacy Library of Young Stars as Essential Standards (ULLYSES) Large Director's Discretionary Program with Hubble. I. Goals, Design, and Initial Results}",
      journal = {\apj},
         year = 2025,
        month = may,
       volume = {985},
       number = {1},
          eid = {109},
        pages = {109},
          doi = {10.3847/1538-4357/adc45b},
archivePrefix = {arXiv},
       eprint = {2504.05446},
 primaryClass = {astro-ph.SR},
       adsurl = {https://ui.adsabs.harvard.edu/abs/2025ApJ...985..109R}
}

@ARTICLE{2024ApJ...964...74S,
       author = {{Shvartzvald}, Y. and {Waxman}, E. and {Gal-Yam}, A. and {Ofek}, E.~O. and {Ben-Ami}, S. and {Berge}, D. and {Kowalski}, M. and {B{\"u}hler}, R. and {Worm}, S. and {Rhoads}, J.~E. and {Arcavi}, I. and {Maoz}, D. and {Polishook}, D. and {Stone}, N. and {Trakhtenbrot}, B. and {Ackermann}, M. and {Aharonson}, O. and {Birnholtz}, O. and {Chelouche}, D. and {Guetta}, D. and {Hallakoun}, N. and {Horesh}, A. and {Kushnir}, D. and {Mazeh}, T. and {Nordin}, J. and {Ofir}, A. and {Ohm}, S. and {Parsons}, D. and {Pe'er}, A. and {Perets}, H.~B. and {Perdelwitz}, V. and {Poznanski}, D. and {Sadeh}, I. and {Sagiv}, I. and {Shahaf}, S. and {Soumagnac}, M. and {Tal-Or}, L. and {Santen}, J. Van and {Zackay}, B. and {Guttman}, O. and {Rekhi}, P. and {Townsend}, A. and {Weinstein}, A. and {Wold}, I.},
        title = "{ULTRASAT: A Wide-field Time-domain UV Space Telescope}",
	journal = {\apj},
         year = 2024,
        month = mar,
       volume = {964},
       number = {1},
          eid = {74},
        pages = {74},
          doi = {10.3847/1538-4357/ad2704},
archivePrefix = {arXiv},
       eprint = {2304.14482},
 primaryClass = {astro-ph.IM},
       adsurl = {https://ui.adsabs.harvard.edu/abs/2024ApJ...964...74S}
}

@ARTICLE{2018PhyS...93b4001D,
       author = {{de Grijs}, Richard and {Li}, Chengyuan},
        title = "{Not-so-simple stellar populations in nearby, resolved massive star clusters}",
      journal = {\physscr},
         year = 2018,
        month = feb,
       volume = {93},
       number = {2},
          eid = {024001},
        pages = {024001},
          doi = {10.1088/1402-4896/aa9ae8},
archivePrefix = {arXiv},
       eprint = {1711.06079},
 primaryClass = {astro-ph.SR},
       adsurl = {https://ui.adsabs.harvard.edu/abs/2018PhyS...93b4001D}
}

@ARTICLE{2025arXiv250608951H,
       author = {{Hota}, Sipra and {de Grijs}, Richard and {Subramaniam}, Annapurni},
        title = "{UVIT Study of the Magellanic Clouds (U-SMAC). III. Hierarchical Star Formation in the Small Magellanic Cloud Regulated by Turbulence}",
      journal = {\apj},
         year = 2025,
        month = aug,
       volume = {989},
       number = {2},
          eid = {216},
        pages = {216},
          doi = {10.3847/1538-4357/adec84},
archivePrefix = {arXiv},
       eprint = {2506.08951},
 primaryClass = {astro-ph.GA},
       adsurl = {https://ui.adsabs.harvard.edu/abs/2025ApJ...989..216H}
}

@ARTICLE{2022MNRAS.512.1196M,
       author = {{Miller}, Amy E. and {Cioni}, Maria-Rosa L. and {de Grijs}, Richard and {Sun}, Ning-Chen and {Bell}, Cameron P.~M. and {Choudhury}, Samyaday and {Ivanov}, Valentin D. and {Marconi}, Marcella and {Oliveira}, Joana M. and {Petr-Gotzens}, Monika and {Ripepi}, Vincenzo and {van Loon}, Jacco Th},
        title = "{The VMC survey - XLVII. Turbulence-controlled hierarchical star formation in the Large Magellanic Cloud}",
      journal = {\mnras},
     year = 2022,
        month = may,
       volume = {512},
       number = {1},
        pages = {1196-1213},
          doi = {10.1093/mnras/stac508},
archivePrefix = {arXiv},
       eprint = {2202.09267},
 primaryClass = {astro-ph.GA},
       adsurl = {https://ui.adsabs.harvard.edu/abs/2022MNRAS.512.1196M}
}

@ARTICLE{2005HiA....13..363D,
       author = {{de Grijs}, Richard},
        title = "{`Super' Star Clusters}",
      journal = {Highlights of Astronomy},
         year = 2005,
        month = jan,
       volume = {13},
        pages = {363},
       adsurl = {https://ui.adsabs.harvard.edu/abs/2005HiA....13..363D}
}

@ARTICLE{2019MNRAS.487..364L,
       author = {{Li}, Hui and {Vogelsberger}, Mark and {Marinacci}, Federico and {Gnedin}, Oleg Y.},
        title = "{Disruption of giant molecular clouds and formation of bound star clusters under the influence of momentum stellar feedback}",
      journal = {\mnras},
         year = 2019,
        month = jul,
       volume = {487},
       number = {1},
        pages = {364-380},
          doi = {10.1093/mnras/stz1271},
archivePrefix = {arXiv},
       eprint = {1904.11987},
 primaryClass = {astro-ph.GA},
       adsurl = {https://ui.adsabs.harvard.edu/abs/2019MNRAS.487..364L}
}

@ARTICLE{Piotto2015AJ....149...91P,
       author = {{Piotto}, G. and {Milone}, A.~P. and {Bedin}, L.~R. and {Anderson}, J. and {King}, I.~R. and {Marino}, A.~F. and {Nardiello}, D. and {Aparicio}, A. and {Barbuy}, B. and {Bellini}, A. and {Brown}, T.~M. and {Cassisi}, S. and {Cool}, A.~M. and {Cunial}, A. and {Dalessandro}, E. and {D'Antona}, F. and {Ferraro}, F.~R. and {Hidalgo}, S. and {Lanzoni}, B. and {Monelli}, M. and {Ortolani}, S. and {Renzini}, A. and {Salaris}, M. and {Sarajedini}, A. and {van der Marel}, R.~P. and {Vesperini}, E. and {Zoccali}, M.},
        title = "{The Hubble Space Telescope UV Legacy Survey of Galactic Globular Clusters. I. Overview of the Project and Detection of Multiple Stellar Populations}",
      journal = {\aj},
         year = 2015,
        month = mar,
       volume = {149},
       number = {3},
          eid = {91},
        pages = {91},
          doi = {10.1088/0004-6256/149/3/91},
archivePrefix = {arXiv},
       eprint = {1410.4564},
 primaryClass = {astro-ph.SR},
       adsurl = {https://ui.adsabs.harvard.edu/abs/2015AJ....149...91P}
}

@ARTICLE{2012A&ARv..20...50G,
       author = {{Gratton}, Raffaele G. and {Carretta}, Eugenio and {Bragaglia}, Angela},
        title = "{Multiple populations in globular clusters. Lessons learned from the Milky Way globular clusters}",
      journal = {\aapr},
         year = 2012,
        month = feb,
       volume = {20},
          eid = {50},
        pages = {50},
          doi = {10.1007/s00159-012-0050-3},
archivePrefix = {arXiv},
       eprint = {1201.6526},
 primaryClass = {astro-ph.SR},
       adsurl = {https://ui.adsabs.harvard.edu/abs/2012A&ARv..20...50G}
}

@ARTICLE{Bastian2018ARA&A..56...83B,
       author = {{Bastian}, Nate and {Lardo}, Carmela},
        title = "{Multiple Stellar Populations in Globular Clusters}",
      journal = {\araa},
         year = 2018,
        month = sep,
       volume = {56},
        pages = {83-136},
          doi = {10.1146/annurev-astro-081817-051839},
archivePrefix = {arXiv},
       eprint = {1712.01286},
 primaryClass = {astro-ph.SR},
       adsurl = {https://ui.adsabs.harvard.edu/abs/2018ARA&A..56...83B}
}

@ARTICLE{Gratton2019A&ARv..27....8G,
       author = {{Gratton}, Raffaele and {Bragaglia}, Angela and {Carretta}, Eugenio and {D'Orazi}, Valentina and {Lucatello}, Sara and {Sollima}, Antonio},
        title = "{What is a globular cluster? An observational perspective}",
      journal = {\aapr},
         year = 2019,
        month = nov,
       volume = {27},
       number = {1},
          eid = {8},
        pages = {8},
          doi = {10.1007/s00159-019-0119-3},
archivePrefix = {arXiv},
       eprint = {1911.02835},
 primaryClass = {astro-ph.SR},
       adsurl = {https://ui.adsabs.harvard.edu/abs/2019A&ARv..27....8G}
}

@ARTICLE{Milone2022Univ....8..359M,
       author = {{Milone}, Antonino P. and {Marino}, Anna F.},
        title = "{Multiple Populations in Star Clusters}",
      journal = {Universe},
         year = 2022,
        month = jun,
       volume = {8},
       number = {7},
          eid = {359},
        pages = {359},
          doi = {10.3390/universe8070359},
archivePrefix = {arXiv},
       eprint = {2206.10564},
 primaryClass = {astro-ph.GA},
       adsurl = {https://ui.adsabs.harvard.edu/abs/2022Univ....8..359M}
}

@ARTICLE{Milone2017MNRAS.464.3636M,
       author = {{Milone}, A.~P. and {Piotto}, G. and {Renzini}, A. and {Marino}, A.~F. and {Bedin}, L.~R. and {Vesperini}, E. and {D'Antona}, F. and {Nardiello}, D. and {Anderson}, J. and {King}, I.~R. and {Yong}, D. and {Bellini}, A. and {Aparicio}, A. and {Barbuy}, B. and {Brown}, T.~M. and {Cassisi}, S. and {Ortolani}, S. and {Salaris}, M. and {Sarajedini}, A. and {van der Marel}, R.~P.},
        title = "{The Hubble Space Telescope UV Legacy Survey of Galactic globular clusters - IX. The Atlas of multiple stellar populations}",
      journal = {\mnras},
         year = 2017,
        month = jan,
       volume = {464},
       number = {3},
        pages = {3636-3656},
          doi = {10.1093/mnras/stw2531},
archivePrefix = {arXiv},
       eprint = {1610.00451},
 primaryClass = {astro-ph.SR},
       adsurl = {https://ui.adsabs.harvard.edu/abs/2017MNRAS.464.3636M}
}

@ARTICLE{2018ApJ...866...21C,
       author = {{Cummings}, Jeffrey D. and {Kalirai}, Jason S. and {Tremblay}, P. -E. and {Ramirez-Ruiz}, Enrico and {Choi}, Jieun},
        title = "{The White Dwarf Initial-Final Mass Relation for Progenitor Stars from 0.85 to 7.5 M $_{{\ensuremath{\odot}}}$}",
      journal = {\apj},
         year = 2018,
        month = oct,
       volume = {866},
       number = {1},
          eid = {21},
        pages = {21},
          doi = {10.3847/1538-4357/aadfd6},
archivePrefix = {arXiv},
       eprint = {1809.01673},
 primaryClass = {astro-ph.SR},
       adsurl = {https://ui.adsabs.harvard.edu/abs/2018ApJ...866...21C}
}

@ARTICLE{2007ApJ...671..380H,
       author = {{Hansen}, Brad M.~S. and {Anderson}, Jay and {Brewer}, James and {Dotter}, Aaron and {Fahlman}, Greg. G. and {Hurley}, Jarrod and {Kalirai}, Jason and {King}, Ivan and {Reitzel}, David and {Richer}, Harvey B. and {Rich}, R. Michael and {Shara}, Michael M. and {Stetson}, Peter B.},
        title = "{The White Dwarf Cooling Sequence of NGC 6397}",
      journal = {\apj},
         year = 2007,
        month = dec,
       volume = {671},
       number = {1},
        pages = {380-401},
          doi = {10.1086/522567},
archivePrefix = {arXiv},
       eprint = {astro-ph/0701738},
 primaryClass = {astro-ph},
       adsurl = {https://ui.adsabs.harvard.edu/abs/2007ApJ...671..380H}
}

@article{Jiang2017,
	title = {Contribution of {Primordial} {Binary} {Evolution} to the {Two} {Blue}-straggler {Sequences} in {Globular} {Cluster} {M30}},
	volume = {849},
	doi = {10.3847/1538-4357/aa8ee1},
	language = {en},
	number = {2},
	journal = {\apj},
	author = {Jiang, Dengkai and Chen, Xuefei and Li, Lifang and Han, Zhanwen},
	month = nov,
	year = {2017},
	pages = {100}
}
\end{document}